\documentclass[12pt]{article}

\usepackage{newtxtext,newtxmath}
\usepackage{ulem}

\usepackage{graphicx}

\usepackage[letterpaper,margin=1in]{geometry}

\renewenvironment{abstract}
	{\quotation}
	{\endquotation}

\date{}

\makeatletter
\renewcommand{\fnum@figure}{\textbf{Figure \thefigure}}
\renewcommand{\fnum@table}{\textbf{Table \thetable}}
\makeatother

\usepackage{scicite}

\usepackage{url}

\usepackage{hyperref}
\hypersetup{colorlinks=true}

\def\scititle{
 A first-principles binary neutron star merger model of GW170817, GRB170817A, and AT2017gfo
}

\title{\bfseries \boldmath \scititle}

\author{
	Kenta Kiuchi$^{1,2\ast}$,
    Sho Fujibayashi$^{3,4,1}$,
    Kota Hayashi$^{1}$,
    Kyohei Kawaguchi$^1$\and
    Quentin Pognan$^1$,
    Alexis Reboul-Salze$^1$,
    Yuichiro Sekiguchi$^{5,2}$,
    Masaru Shibata$^{1,2}$,\and
    Shinya Wanajo$^{2, 1}$\and
	\small$^{1}$Max Planck Institute for Gravitational Physics (Albert Einstein Institute), Potsdam \& D-14476, Germany. \and
	\small$^{2}$Center for Gravitational Physics and Quantum Information, Yukawa Institute for Theoretical Physics, \and
    \small Kyoto University, Kyoto \& 606-8502, Japan. \and
    \small$^3$ Frontier Research Institute for Interdisciplinary Sciences, Tohoku University, Sendai, 980-8578, Japan \and
    \small$^4$ Astronomical Institute, Graduate School of Science, Tohoku University, Sendai 980-8578, Japan \and
    \small$^5$ Department of Physics, Toho University, Chiba, 274-8510, Japan \and
	\small$^\ast$Corresponding author. Email: kenta.kiuchi@aei.mpg.de \and
}

\begin{document} 

\maketitle

\begin{abstract} \bfseries \boldmath
The multimessenger observation of the binary neutron star merger event GW170817, associated with its electromagnetic counterparts GRB170817A and AT2017gfo, marked a milestone in astrophysics, yet its unified physical explanation remains elusive. 
We conduct an end-to-end simulation based on a first-principles general-relativistic magnetohydrodynamics neutrino-radiation transfer merger simulation, followed by nucleosynthesis calculations and photon radiative transfer to generate kilonova light curves.  We show that the large-scale dynamo simultaneously produces a relativistic jet with an isotropic-equivalent luminosity of $\sim 10^{51}~{\rm erg~s^{-1}}$ and $\approx 0.08M_\odot$ of neutron-rich ejecta, reproducing the GRB170817A afterglow and the AT2017gfo kilonova light curves. 
Our results establish a unified first-principles framework for interpreting binary neutron star mergers across gravitational wave, gamma-ray burst, and kilonova observations. 
\end{abstract}

\noindent
{\it Introduction and unresolved issues}--The binary neutron star merger event GW170817, together with the associated gamma-ray burst GRB170817A and kilonova AT2017gfo, marked the beginning of the era of multimessenger astronomy~\cite{LIGOScientific:2017vwq,LIGOScientific:2017ync,LIGOScientific:2018hze,LIGOScientific:2018cki}. The gravitational wave inspiral signal placed stringent constraints on the nuclear equations of state through measurement of the tidal deformability~\cite{LIGOScientific:2017vwq,De:2018uhw}, excluding many previously proposed nuclear equation of state models. The associated gamma-ray emission was the first smoking-gun for the binary neutron star merger origin of short gamma-ray bursts~\cite{Eichler:1989,Goldstein:2017mmi,LIGOScientific:2017ync}. The kilonova observations provided the first direct photometric and spectroscopic evidence for the synthesis of heavy elements through rapid neutron capture ($r$-process) in binary neutron star mergers~\cite{Lattimer:1974,Metzger:2010,LIGOScientific:2017ync,Watson:2019xjv,Domoto:2022cqp}. 
Finally, it enabled an independent measurement of the Hubble constant through a gravitational-wave standard siren~\cite{LIGOScientific:2017adf,Hotokezaka:2018dfi}.

Since its discovery, extensive theoretical and observational efforts have been devoted to establishing a unified physical picture of this event (see, e.g., \cite{Shibata:2017xdx}). 
However, there are no first-principles models that reproduce the observations, the key challenge being that one must conduct an end-to-end simulation of a binary neutron star merger. 
This involves conducting a first-principles binary neutron star merger simulation, the data of which is then exported to nucleosynthesis calculations and ultimately used in a photon radiative transfer simulation in order to calculate the associated kilonova light-curve and spectrum in a self-consistent manner.
It is a challenging problem that spans several orders of magnitude in spatial scales and involves a wide variety of complex physics. Over the past years, many works have managed to draw a unified picture of the jet launching necessary to produce a short gamma-ray burst, the neutron-rich matter ejection required for the $r$-process nucleosynthesis, as well as the kilonova emission from the merger ejecta, on the basis of first principles simulations of binary neutron star mergers~\cite{Kiuchi:2023obe,Kiuchi:2022nin,Hayashi:2024jwt}. Such studies revealed that, in addition to general relativity and neutrino transport, resolving magnetic-field amplification through turbulence and dynamo action is essential for reproducing the jet launching and capturing the ejection of neutron-rich matter, demanding large-scale simulations at extremely high spatial resolution. 

However, the model of GW170817 is still a riddle in the sense that if the maximum mass of a non-rotating neutron (Tolmann-Oppenheimer-Volkoff, TOV) star $M^\mathrm{TOV}_\mathrm{max}$ is $\lesssim 2.06M_\odot$, a remnant massive neutron star collapses to a black hole within $\sim 0.01~{\rm s}$. This leads to a dim kilonova emission compared to AT2017gfo because the neutron-rich ejecta of $\approx 0.01M_\odot$ is not large enough to generate sufficient luminosity~\cite{Kiuchi:2022nin,Kawaguchi:2023zln}. If $M^\mathrm{TOV}_\mathrm{max} \gtrsim 2.42M_\odot$, the remnant massive neutron star survives for $\gtrsim 10~{\rm s}$. A jet with an isotropic-equivalent luminosity of $L_\mathrm{iso}\sim 10^{52}~{\rm erg/s}$, and a huge amount of neutron-rich ejecta $\gtrsim 0.1M_\odot$, are launched due to the large-scale dynamo activating inside the remnant massive neutron star~\cite{Kiuchi:2023obe}. This model predicts overbright kilonova emission compared to AT2017gfo~\cite{Pognan2026}. 
In this case, a part of the rotational energy of $\sim 10^{52}~{\rm erg}$ of the remnant massive neutron star could be injected into the ejecta via the magnetic dipole radiation, which is incompatible with the radio observations~\cite{Margalit:2017dij}. 
These previous works suggest that a sweet spot for the GW170817 model could lie somewhere between these remnant lifetimes. 
In this article, we demonstrate for the first time that a fully first-principles, end-to-end model of a binary neutron star merger approximately reproduces the full set of multimessenger observations of GW170817, GRB170817A, and AT2017gfo. We should stress that a comprehensive understanding of this particular event is desired for potential future events in the observing run 5 of the LIGO-VIRGO-KAGRA collaboration, and in the era of next-generation gravitational wave detectors such as the Einstein Telescope and Cosmic Explorer ~\cite{Punturo:2010zz,Reitze:2019iox}. 

{\it End-to-end binary neutron star merger model}--
We first conduct a fully general relativistic, high-resolution, long-term, neutrino-radiation transfer magnetohydrodynamics simulation for a symmetric binary neutron star merger with the in-house code {\tt NANASI}~\cite{Kiuchi:2022}. The binary neutron star chirp mass is exactly the same as GW170817 within the uncertainty, and we employ the nuclear equation of state BHB$\Lambda\phi$~\cite{Banik:2014qja}, which is compatible with GW170817 observations, predicts $M^\mathrm{TOV}_\mathrm{max}\approx 2.11M_\odot$, and tidal deformability $\Lambda_{135}\approx 854$ for a $1.35M_\odot$ neutron star. We initialize the magnetic field strength as $10^{15}~{\rm G}$ at maximum. The employed spatial resolution achieves $12.5~{\rm m}$ with a nested mesh-refinement technique. With this setup, the remnant massive neutron star is neither a very short-lived $\sim 0.01~{\rm s}$~\cite{Kiuchi:2022nin}, nor a very long-lived $\sim 10~{\rm s}$~\cite{Kiuchi:2023obe} case. 
The simulation duration is $\approx 0.3~{\rm s}$ for the bulk evolution, and we continue the simulation until $\approx 0.7~{\rm s}$ to track the evolution of the neutron-rich matter ejecta. 
The employed grid resolution and its duration make the simulation in this article the highest resolution-longest duration first-principles global binary neutron star merger simulation in the literature. 
On top of the numerical relativity simulation data, we conduct the {\it r}-process nucleosynthesis calculation~\cite{Wanajo:2014wha}. 
Then, we export the resultant {\it r}-process elemental abundances and radioactive decay heating rate as well as ejecta structure 
to the Monte Carlo photon radiative transfer simulation to calculate 
the kilonova light curve~\cite{Kawaguchi:2019nju} (see Materials \& Methods for details of each stage). Throughout this article, the origin of the time axis is the merger time $t_\mathrm{merger}$, which is defined by the gravitational wave peak amplitude time.

\begin{figure} 
	\centering
	\includegraphics[width=0.48\textwidth]{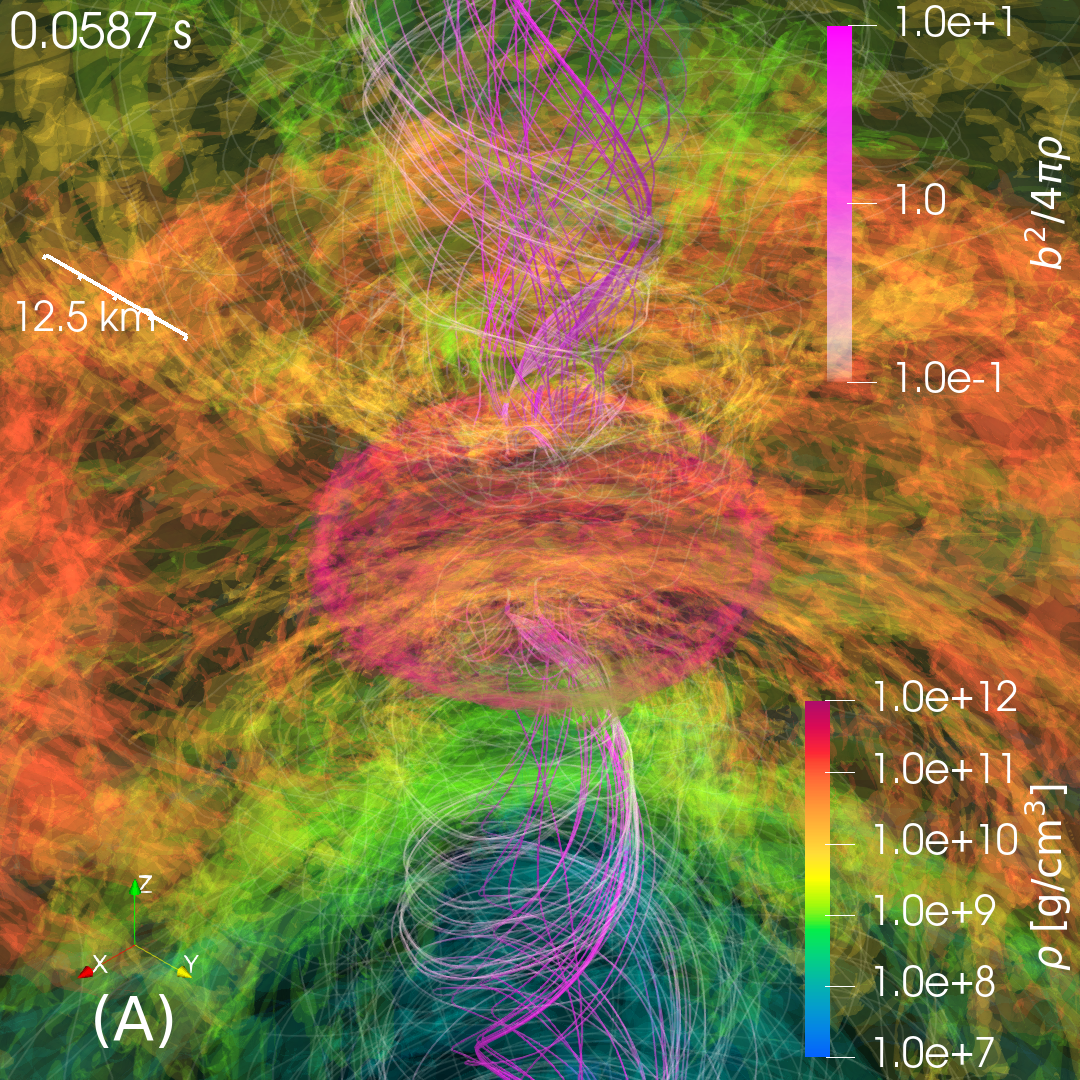} 
    \includegraphics[width=0.48\textwidth]{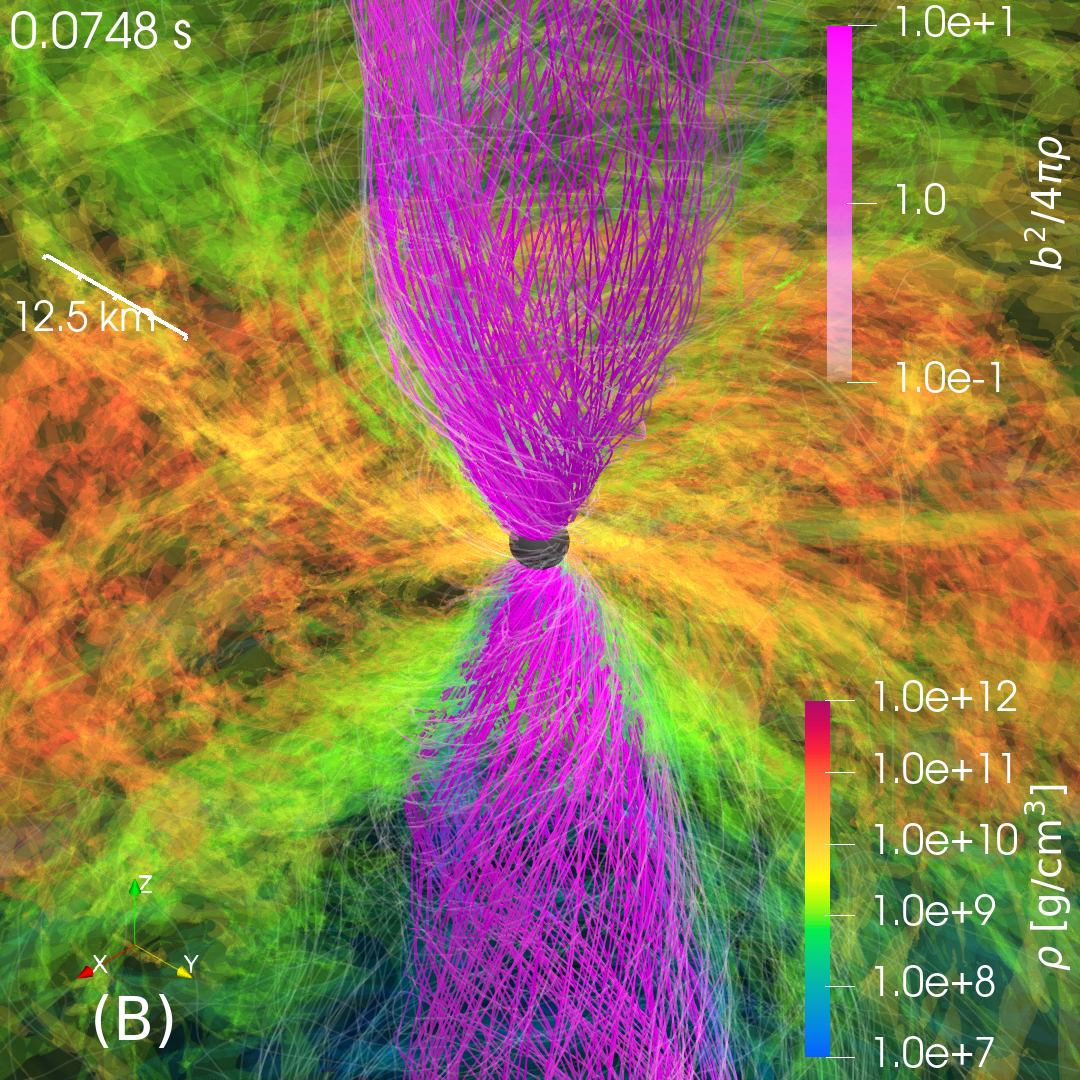}

	\caption{\textbf{Jet launching from binary neutron star merger remnant.}
    (\textbf{A}) The iso-surface of the rest-mass density (color) and the magnetic-field lines (magenta-curve) at $t-t_{\rm merger}\approx 0.059$~s inside the remnant massive neutron star formed after the merger. The large-scale magnetic field established via the magnetorotational instability-driven $\alpha\Omega$ dynamo drives a jet. 
    (\textbf{B}) The same as (\textbf{A}), but at $t-t_\mathrm{merger}\approx 0.075~{\rm s}$, i.e., after the black hole formation whose horizon is represented by the black sphere at the center. 
        }
	\label{fig:main_image1} 
\end{figure}

\begin{figure} 
	\centering
	\includegraphics[width=0.49\textwidth]{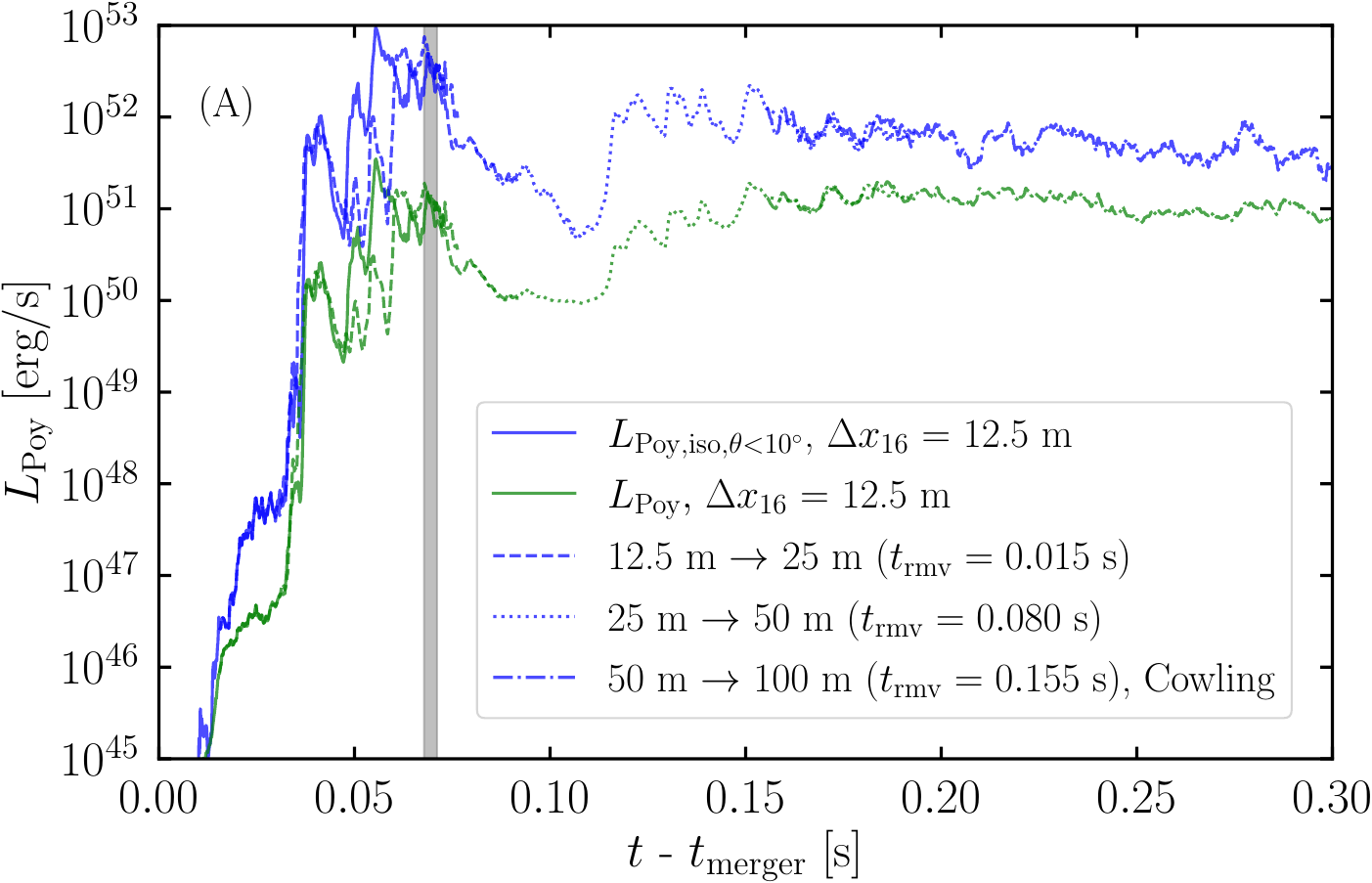}
    \includegraphics[width=0.49\textwidth]{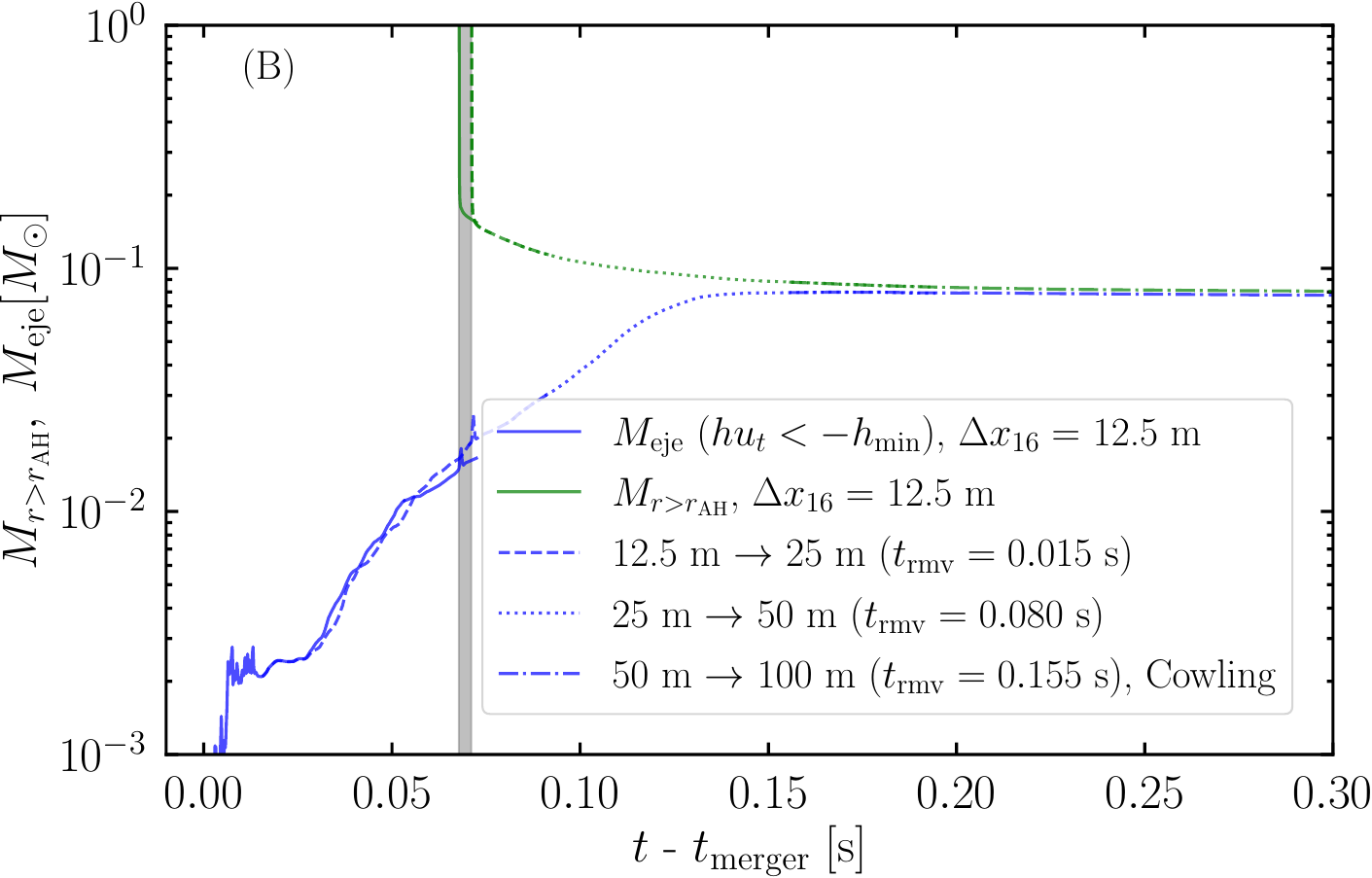}\\
    \includegraphics[width=0.49\textwidth]{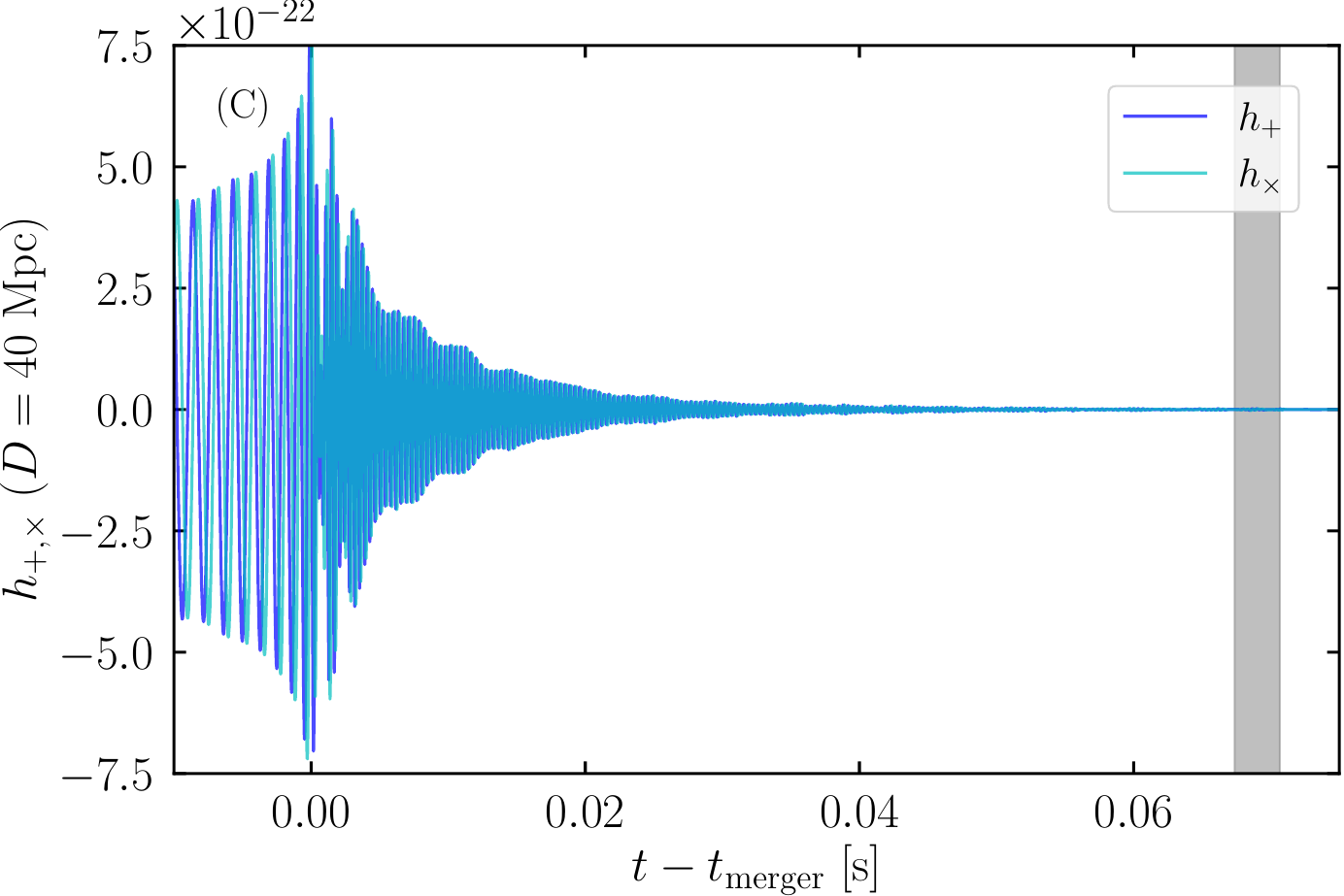}
	\caption{\textbf{Multimessenger signals from binary neutron star merger.}
    (\textbf{A}) Isotropic-equivalent (blue) and raw (green) jet luminosity. 
    (\textbf{B}) Neutron-rich matter ejecta (blue) and the baryonic mass outside of the black hole (green). 
    (\textbf{C}) Gravitational waves. 
    The horizontal axis is the post-merger time, and the gray-shaded region denotes the uncertainty in black hole formation timing. See Materials \& Method for the legend in detail. 
        }
	\label{fig:main_MMA} 
\end{figure}

{\it A comprehensive merger process}--
When the two neutron stars collide, the contact interface is subject to the Kelvin-Helmholtz instability, and the resultant small-scale vorticities efficiently amplify the magnetic field 
within $t-t_\mathrm{merger} \approx 0.005~{\rm s}$~\cite{Kiuchi:2015sga,Kiuchi:2017zzg,Kiuchi:2023obe,Aguilera-Miret:2024cor}. The electromagnetic-field energy saturates at an expected value of $\sim 10^{50}~{\rm erg}$ (see Supplementary Text)~\cite{Kiuchi:2023obe,Aguilera-Miret:2023qih,Aguilera-Miret:2024cor}. After the Kelvin-Helmholtz vortices are dissipated by the shock waves generated in the collision of the two neutron stars, the magnetorotational instability~\cite{Balbus-Hawley:1991} takes over as the main process for generating the turbulence inside the remnant massive neutron star. The resultant turbulence leads to large-scale magnetic field generation through the $\alpha\Omega$ dynamo inside the remnant massive neutron star (see Supplementary Text)~\cite{Brandenburg:2004jv,Kiuchi:2023obe,Reboul-Salze:2020mnw}. Since the remnant massive neutron star survives for $5$--$6$ $\alpha\Omega$ dynamo periods, the large-scale magnetic field is well established, and a first jet/outflow is launched by the magnetic-tower effect before the collapse to a black hole, which happens at $t-t_\mathrm{merger}\approx 0.07$~s. After the black hole formation, the large-scale magnetic field penetrates the black hole horizon, which provides the necessary conditions for the Blandford-Znajek mechanism~\cite{Blandford:1977ds} 
(see Fig.~\ref{fig:main_image1}, Supplementary Text, and Movie S1:~\url{http://www2.yukawa.kyoto-u.ac.jp/~kenta.kiuchi/anime/FUGAKU2025/out_yuv420p_3D.mp4} ). The isotropic-equivalent jet luminosity reaches up to $\sim 10^{52}$--$10^{53}~{\rm erg~s^{-1}}$ during the remnant massive neutron star phase. Then, the luminosity drops to $\sim 10^{51}~{\rm erg~s^{-1}}$ after the black hole formation since the central part of the remnant massive neutron star, sustaining the large-scale magnetic field, accretes onto the newly formed black hole. After the black hole and the surrounding massive torus relax to a quasi-stationary state at $t-t_{\rm merger}\approx 0.10$--$0.12~{\rm s}$, the luminosity again increases up to $\sim 10^{52}~{\rm erg~s^{-1}}$ due to the large-scale $\alpha\Omega$ dynamo activity inside the massive torus. The high luminosity state of $\sim 10^{51}$--$10^{52}~{\rm erg~s^{-1}}$ continues until $t-t_{\rm merger}\approx 0.3~{\rm s}$ as shown in Fig.~\ref{fig:main_MMA}~({\bf A}). The jet with an opening angle of $\approx 10^\circ$ has a terminal Lorentz factor of $\sim 30$. 
The large-scale magnetic field generation via the $\alpha\Omega$ dynamo also results in the Lorentz-force-driven post-merger ejecta~\cite{Kiuchi:2023obe}, and the total amount of the neutron-rich matter ejecta is $\approx 0.08M_\odot$, as shown in Fig.~\ref{fig:main_MMA}~(\textbf{B}). 
The post-merger gravitational wave amplitude quickly damps for $t-t_{\rm merger} \lesssim 0.02$--$0.03~{\rm s}$ as shown in Fig.~\ref{fig:main_MMA}~(\textbf{C}). 
Thus the present simulation demonstrates for the first time a self-consistent scenario of a binary neutron star merger applicable to GW170817, starting from the inspiral phase, and including the large-scale magnetic field generation from the remnant massive neutron star, through to the subsequent black hole and jet formation. 

\begin{figure} 
	\centering
    \includegraphics[width=0.49\textwidth]{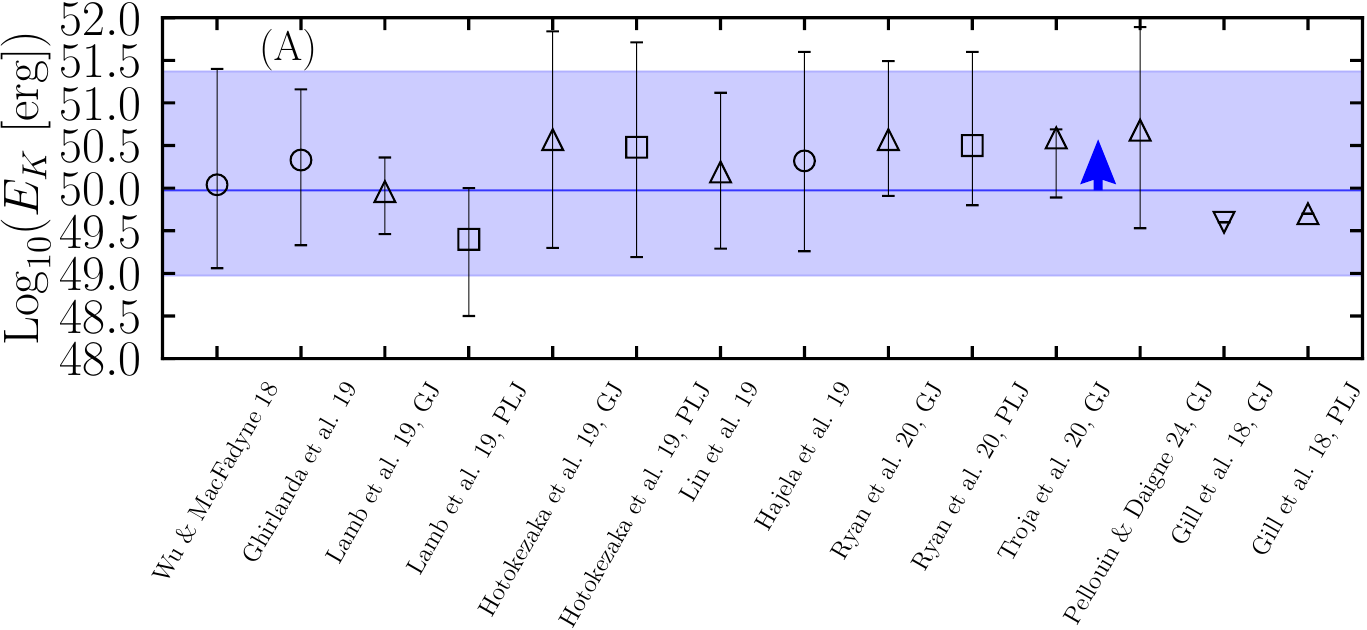}
	\includegraphics[width=0.49\textwidth]{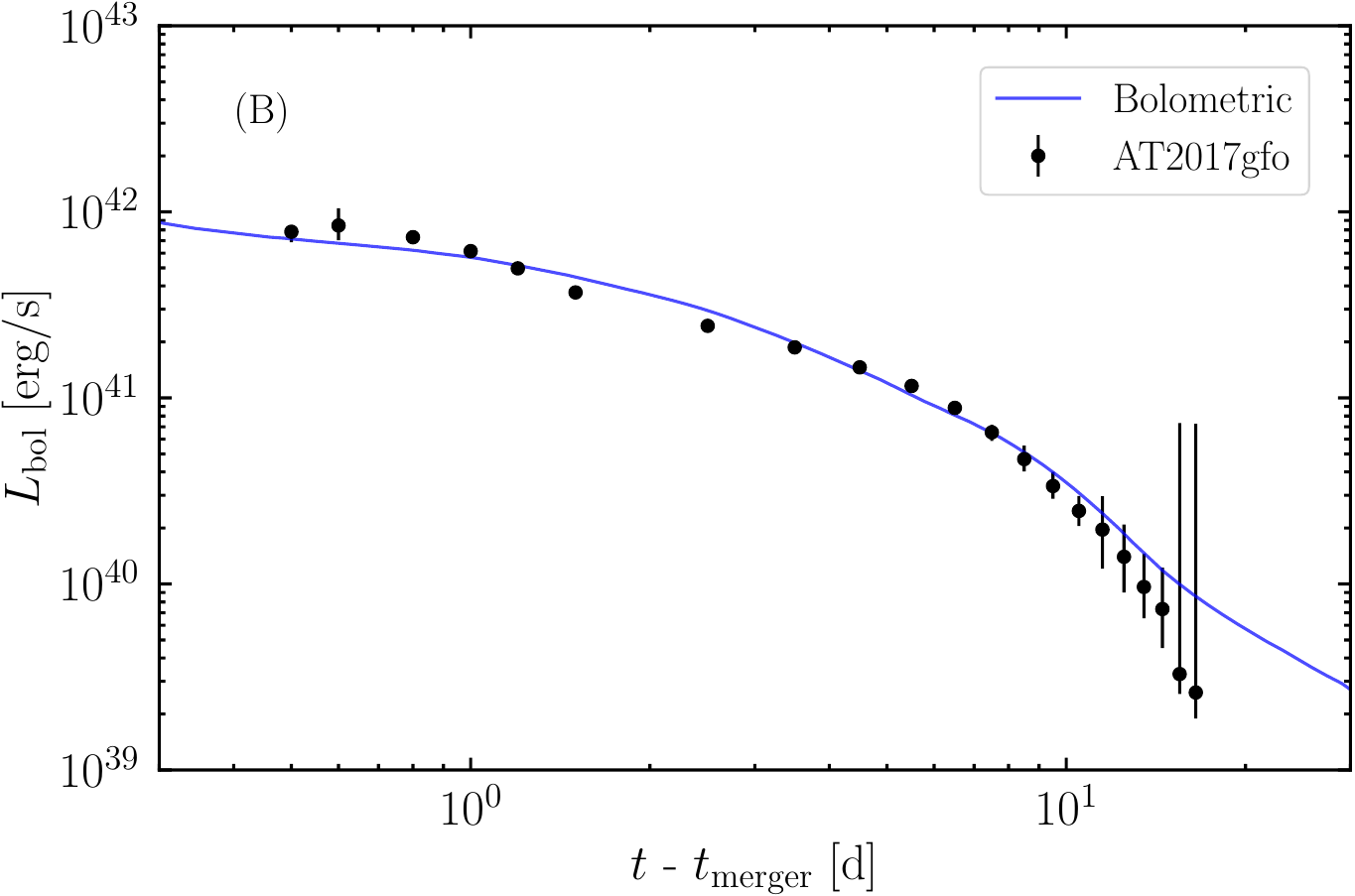}\\
	\includegraphics[width=0.49\textwidth]{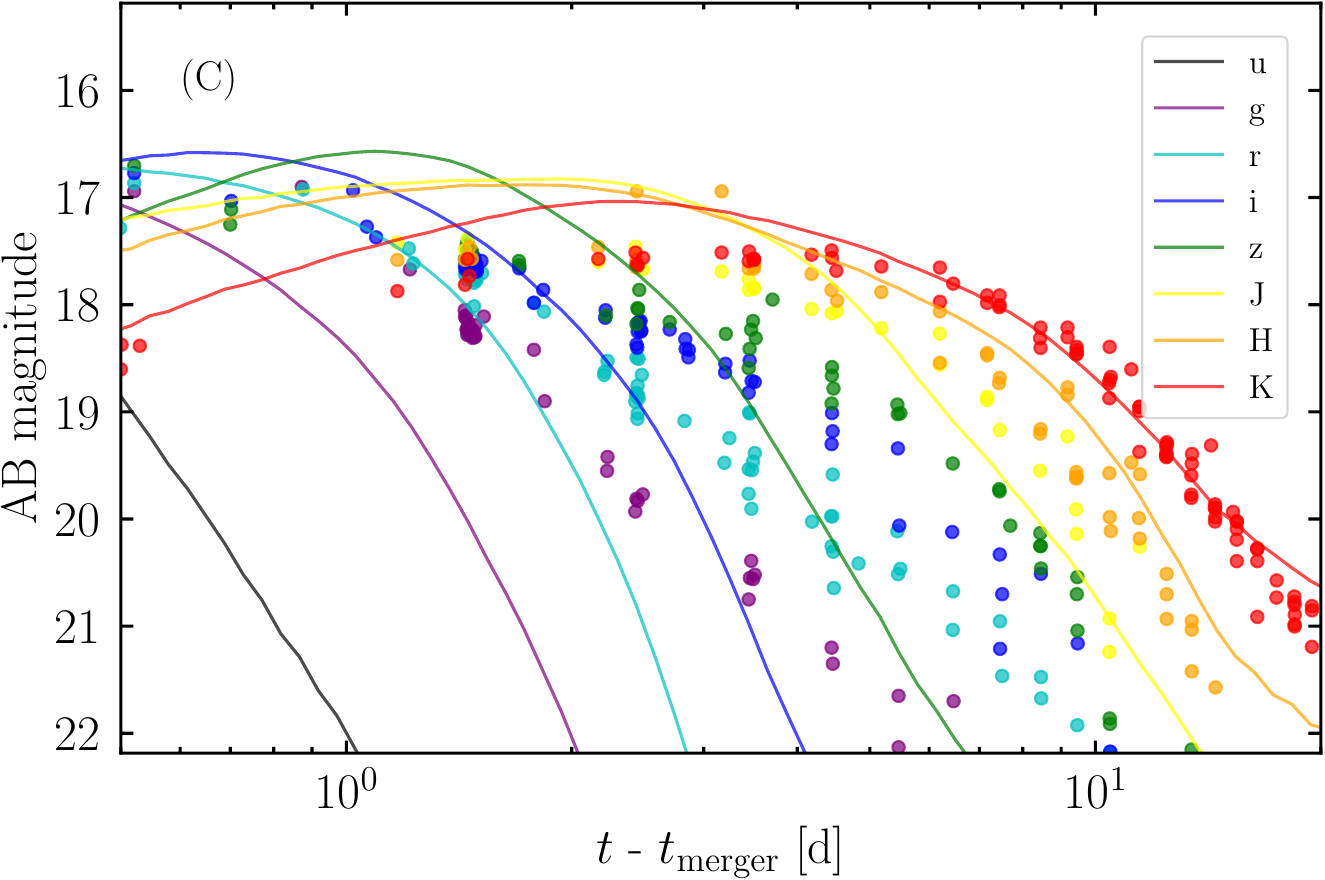}
	\caption{\textbf{Comparison to the observations.}
    (\textbf{A}) The required jet kinetic energy to fit the afterglow observations of GRB170817A in the radio, X-ray, and optical bands in the literature. GJ and PLJ are abbreviations for Gaussian Jet and Power-Law Jet, respectively. 
    The blue solid line is the estimated energy emitted as 
    of the jet until the end of the simulation $t-t_{\rm merger}\approx 0.3~{\rm s}$. The colored shaded region gives the estimated jet kinetic energy assuming $10\%$ conversion efficiency (lower-bound) and boosted by a factor $25$ due to the non-ideal magnetohydrodynamics effect (upper bound)~\cite{Reboul-Salze:2025eqt}. 
    (\textbf{B}) Total bolometric kilonova light curve.  
    The observed bolometric luminosity is
    taken from~\cite{Waxman:2017sqv}. (\textbf{C}) Spherically averaged ugrizJHK-band kilonova light curves at $40\,{\rm Mpc}$. The observed ugrizJHK data points are taken from \cite{Villar:2017wcc}.
    }
	\label{fig:main_knlc} 
\end{figure}

{\it Application to GRB170817A afterglow}--
Among many observables in GRB170817A, the jet kinetic energy inferred from the afterglow observation is the most robust. Figure~\ref{fig:main_knlc}~(\textbf{A}) summarizes the required jet kinetic energy to fit the afterglow observation for GRB170817A in the literature~\cite{Wu:2018bxg,Ghirlanda:2018uyx,Lamb:2018qfn,Hotokezaka:2018dfi,Lin:2018woe,Hajela:2019mjy,Ryan:2019fhz,Troja:2020pzf,Gill:2018kcw}, which has a $2$--$2.5$ order magnitude systematic uncertainty, due to many ad hoc assumptions, e.g., the jet opening angle and the jet energy angular distribution. 
The blue-horizontal line shows the jet kinetic energy in our simulation integrated up to $t-t_{\rm merger}=0.3~{\rm s}$ with $\theta < 10^\circ$ in our model assuming $100\%$ conversion efficiency~\cite{Drenkhahn:2002ug}. 
The lower bound of the blue-shaded region assumes $10\%$ conversion efficiency. The upper bound of the blue-shaded region is a jet kinetic energy hypothetically accelerated by a factor of $25$ by the increase of magnetic fields due to non-ideal magnetohydrodynamics effects~\cite{Reboul-Salze:2025eqt}. 
Our model can reproduce the required jet kinetic energy to fit the afterglow.

{\it Application to AT2017gfo}--
The massive post-merger ejecta being larger than those previously reported makes the bolometric kilonova light curve in close agreement with the observation, 
as shown in Fig.~\ref{fig:main_knlc} (\textbf{B}).
The peak times and magnitudes of the UV, optical, and NIR broad-band light curves reasonably agree with the observation as shown in Fig.~\ref{fig:main_knlc} (\textbf{C}). 
Several discrepancies between the broad-band light curves and the observations, such as the rapid fading in the optical {\it gri} bands, or the early excess at $t-t_\mathrm{merger}\lesssim 2~{\rm d}$ in the {\it z} band, could stem from non-local thermodynamic equilibrium effects, which our kilonova model does not take into account. As demonstrated in \cite{Brethauer:2025plw}, suppressing the populations in neutral and singly-ionized states tends to decrease
the opacity in the optical bands, which could potentially solve these discrepancies (see also Supplementary Text for more discussion). Hence, while these remaining discrepancies likely reflect limitations in current kilonova radiative transfer modeling, our self-consistent ejecta model provides a physically motivated foundation for future improvements in kilonova simulations.

The helium (He) and strontium (Sr) abundance constraints have been proposed through detailed analyses of the absorption feature observed at $\sim 8000$~{\AA} in the photospheric spectra of AT2017gfo~\cite{Watson:2019xjv,Perego:2020evn,Sneppen:2024jch}. Specifically, previous nucleosynthesis studies of binary neutron star mergers, which do not take into account the magnetohydrodynamics effect, resulted in a severe overproduction of He compared to this constraint when the remnant neutron star survives longer than $\sim 0.02$--0.03~s \cite{Sneppen:2024jch}. In our model, we find that the spherically averaged mass fractions of He and Sr at velocities of $v\approx 0.15$--$0.2~{\rm c}$ are $7\times 10^{-3}$ and $7\times 10^{-2}$, respectively, which are consistent with the abundance constraints despite the longer lifetime of the remnant neutron star ($\sim 0.07$~s) compared to the previously proposed lifetime constraint~\cite{Sneppen:2024jch}. This is a consequence of the fact that the ejecta in our model are more neutron-rich than those in previous studies that lack the magnetohydrodynamics effect, and thus, He is less favorably produced (see Supplementary Text). This agreement highlights the importance of self-consistent end-to-end modeling for establishing a direct connection between the merger physics and kilonova observables.

A previous study \cite{Jacobi:2025eak} also has suggested that, when the remnant neutron star survives longer than $\sim 0.1$~s, about $10^{-3}\,M_\odot$ of $^{56}$Ni is ejected, resulting in flattening of the kilonova light curve owing to its decay on timescales of days for polar angles. We should note that their model also lacks the magnetohydrodynamics effects. However, our model predicts a negligible amount of $^{56}$Ni owing to the absence of ejecta with the electron fraction of $\gtrsim 0.48$. Therefore, our result indicates that the contribution of radioactive heating from $^{56}$Ni is unimportant in kilonova light curves.

{\it Application to GW170817 post merger}-- 
Although our equation of state passes the constraint from the inspiral gravitational waves in GW170817~\cite{LIGOScientific:2018hze}, it does not necessarily mean the equation of state fulfills the constraint coming from the post-merger gravitational waves because the remnant massive neutron star is a strong gravitational wave emitter.
The spectral energy density peak amplitude is $dE_\mathrm{gw}/df \approx 1.8\times 10^{-4}M_\odot c^2{\rm Hz}^{-1}$ at $f\approx 2650~{\rm Hz}$, which is well below the upper limit at this frequency in GW170817~\cite{LIGOScientific:2018hze}. We also confirm that the spectral energy density of our model does not conflict with the upper limit for GW170817 in $1500~{\rm Hz} \leq f \leq 4000~{\rm Hz}$. 

{\it Remaining challenges}--
With these results, we conclude that our end-to-end model of a binary neutron star merger can account for the broad features of GW170817, GRB170817A, and AT2017gfo. 
Our work establishes a unified first-principles framework for the multimessenger event from the binary neutron star merger. 
The next key challenges to address are (\textbf{A}) solving the $1.7$~s time-lag mystery in GW170817 and GRB170817A, (\textbf{B}) reproducing the detailed observed kilonova spectra with improved radiative transfer simulations, (\textbf{C}) narrowing down the nuclear equation of state parameter space on the $M^{\rm TOV}_{\rm max}$--$\Lambda_{135}$ plane, which is consistent with GW170817, GRB170817A, and AT2017gfo, 
and (\textbf{D}) predicting features of binary neutron star mergers in future gravitational wave events. 


\clearpage 

%
\bibliography{science_template} 
\bibliographystyle{sciencemag}

%
%
%
%
%
%


\section*{Acknowledgments}
This work used computational resources of the supercomputer FUGAKU provided by RIKEN through the HPCI System Research Project (Project ID: hp240532, hp250570,hp250066, hp260063). The test simulations were also performed on Sakura, Momiji, Raven, and Viper clusters at the Max Planck Computing and Data Facility. 
\paragraph*{Funding:}
This work was in part supported by the Grant-in-Aid for Scientific Research (grant Nos. 23K25869, 23H04900 and 26K21721) of Japan MEXT/JSPS. 
\paragraph*{Author contributions:}
KK (Kiuchi) is the primary driver of this article, developing {\tt NANASI}, conducting a numerical relativity simulation, analyzing simulation data, conducting the tracer particle, and writing the article. 
SF extends the equation of state table to lower density and temperature ranges and  develops the post-processing tracer particle code. KH is responsible for generating the initial data and for the 3D visualization. 
KK (Kawaguchi) conducted a long-term ejecta hydrodynamics simulation as well as a Monte Carlo photon radiative transfer simulation for the kilonova light-curve calculation, and wrote the article.
QP conducted tests and additional calculations for the long-term ejecta hydrodynamic evolution and radiative transfer simulations, participated in the analysis of the kilonova light curves, and contributed to writing the article.
ARS conducts a $\alpha\Omega$ dynamo analysis to support the findings in this article and is responsible for writing the article. 
YS is a primary developer of {\tt NANASI}. MS is a primary developer of {\tt NANASI}, and also prepares a $3$PB storage server to store all the data supporting the findings. 
SW is responsible for the $r$-process nucleosynthesis calculation and for writing the article. 
\paragraph*{Competing interests:}
There are no competing interests to declare.

\paragraph*{Data and materials availability:}
All the simulation data and the post-processing data supporting the findings in this article are openly available upon request. 
The prerequisite of sharing the numerical relativity simulation raw data is to prepare a storage server large enough to transfer the data, ideally $>100$TB, for those who make the request. 
The numerical relativity code~{\tt NANASI} is available for a reasonable request. 
\subsection*{Supplementary materials}
Materials and Methods\\
Supplementary Text\\
Figs. S1 to S17\\
Tables S1 to S2\\
References \textit{(51-\arabic{enumiv})}\\ 
Movie S1\\


\newpage


\renewcommand{\thefigure}{S\arabic{figure}}
\renewcommand{\thetable}{S\arabic{table}}
\renewcommand{\theequation}{S\arabic{equation}}
\renewcommand{\thepage}{S\arabic{page}}
\setcounter{figure}{0}
\setcounter{table}{0}
\setcounter{equation}{0}
\setcounter{page}{1} 


\begin{center}
\section*{Supplementary Materials for\\ \scititle}

Kenta Kiuchi$^{\ast}$,
Sho Fujibayashi,
Kota Hayashi,
Kyohei Kawaguchi,
Quentin Pognan,\\
Alexis Reboul-Salze,
Yuichiro Sekiguchi,
Masaru Shibata,
Shinya Wanajo\\ 
\small$^\ast$Corresponding author. Email: kenta.kiuchi@aei.mpg.de
\end{center}



\subsection*{Materials and Methods}
\subsubsection*{Numerical relativity simulation}
We employ our in-house numerical-relativity neutrino-radiation transport magnetohydrodynamics code {\tt NANASI}~\footnote{Numerical relativity code for Astrophysical and Non-Astrophysical object SImulation}~\cite{Kiuchi:2022}. The Einstein solver implements the Baumgarte-Shapiro-Shibata-Nakamura-puncture formulation~\cite{Shibata:1995,Baumgarte:1998te,Campanelli:2005dd,Baker:2005vv} augmented by the Z4c constraint propagation prescription~\cite{Hilditch:2012fp}. The fourth-order accurate finite difference scheme is applied for the spatial derivative of the geometrical variables, and the sixth-order accurate Kreiss-Oliger dissipation is added to suppress the high-frequency noise. The relativistic magnetohydrodynamics solver implements the second-order accurate finite-volume Harten-Lax-van Leer-Discontinuity Riemann solver~\cite{MUB:2009} combined with the advanced constrained transport scheme~\cite{Evans:1988,Gardiner:2007nc}. The third-order accurate piecewise parabolic method is employed for the cell reconstruction~\cite{Colella:1984}. The neutrino radiation transport solver implements the M1+GR-Leakage scheme to take into account the neutrino heating and cooling~\cite{Sekiguchi:2015dma}. The fourth-order accurate Runge-Kutta scheme with the Courant-Friedrichs-Levy (CFL) number 0.45 is employed for the time update. 

We employ the public spectral library FUKA~\cite{Papenfort:2021hod} to generate a quasi-equilibrium configuration of a non-spinning equal-mass binary neutron star with $m_0=2.725M_\odot$, which corresponds to the chirp mass ${\mathcal M}_c \approx 1.186 M_\odot$ consistent with GW170817 ${\mathcal M}^\text{GW170817}_c =1.186^{+0.001}_{-0.001}M_\odot$~\cite{LIGOScientific:2018hze}. 
The initial coordinate separation of the binary is $\approx 44.4$~km. As an equation of state of the neutron star, we assume the BHB$\Lambda\phi$~\cite{Banik:2014qja}, which predicts the tidal deformability $\Lambda_{1.35} \approx 854$ for the $1.35M_\odot$ neutron star, and the maximum  mass of Tolman-Oppenheimer-Volkoff star $M^\text{TOV}_\text{max} \approx 2.11M_\odot$. 
This equation of state satisfies the astronomical observation constraints as well as the nuclear physics experimental constraints~\cite{Li:2021thg}.
We also employ the initial orbital eccentricity reduction procedure to reduce the residual initial orbital eccentricity down to $O(10^{-3})$~\cite{Papenfort:2021hod}.

To cover a wide dynamic range of binary neutron star mergers, {\tt NANASI} implements a Cartesian nested grid structure with a 2:1 refinement rule and employs the Berger-Oliger type time marching algorithm~\cite{Berger:1984zza}. Each concentric nested domain has a cell-centered grid structure with $x^i_{\rm (lv)}\in [(-N-1/2)\Delta x^i_{\rm (lv)},(N+1/2)\Delta x^i_{\rm (lv)}]$ where $i=1$--$3$ denotes the $x$, $y$, and $z$-direction, respectively, and $\rm lv=1$--$16$ denotes the depth of the nested domains. We employ $\Delta x_{16}=12.5$~m and $N=385$. With this set-up, the size of 
the finest nested domain is $L_{16} \approx 9.64 {\rm km}$. We impose orbital-plane symmetry to reduce computational cost. 
Since the coordinate radius of the neutron star is $\approx 10.6$~km, the fourth finest domain with $L_{13}\approx 77.1~{\rm km}$ covers the entire binary neutron star. We also employ the conservative mesh-refinement techniques, such as the reflux prescription, the conservative prolongation, and divergence-free-magnetic-flux-conservative prolongation~\cite{Kiuchi:2012qv,Kiuchi:2022}. With these techniques, the violation of the baryonic mass conservation stays around $O(10^{-7})\%$ until the black hole formation~\cite{Kiuchi:2023obe}. The divergence-free condition and the magnetic-flux conservation are satisfied in machine precision~\cite{Kiuchi:2012qv,Neuweiler:2024jae}.

This grid set up enables us to properly capture the Kelvin-Helmholtz instability at the contact interface of the two neutron stars when they collide~\cite{Price:2006fi,Kiuchi:2014hja,Kiuchi:2017zzg,Kiuchi:2022,Kiuchi:2023obe,Rasio:1999}, and the magnetorotational instability-driven $\alpha\Omega$ dynamo, which subsequently sets in inside the remnant massive neutron star~\cite{Kiuchi:2017zzg,Kiuchi:2023obe,Balbus-Hawley:1991}. We use 110,592 cores on FUGAKU, and the total CPU time for the simulations reported in this article is $\approx 240$ million core hours. 

The Kelvin-Helmholtz instability phase continues only for a particular timescale, $\approx 0.002$--
$0.003$~s in our model, after the merger, because the shock wave dissipates the shear layers~\cite{Kiuchi:2023obe}. Also, a region deep inside the remnant massive neutron star is unlikely to be subject to the magnetorotational instability because of the positive radial gradient of the angular velocity~\cite{Shibata:2005ss,Kiuchi:2023obe,Balbus-Hawley:1991}. Furthermore, after a black hole formation, the maximum rest-mass density typically decreases by two or three orders of magnitude. This implies that we may not have to keep the finest nested domain(s) after the Kelvin-Helmholtz instability phase, and after the black hole formation. Therefore, we remove the finest nested domains step by step at certain moments denoted by $t_{\rm rmv}$ as follows to accelerate the simulation. 

Specifically, we remove the finest domain with $\Delta x_{16}=12.5$ m at $t-t_{\rm merger}=0.015$~s, the second finest domain with $\Delta x_{15}=25$~m at $t-t_{\rm merger}=0.08$~s, and the third finest domain with $\Delta x_{14}=50$~m at $t-t_{\rm merger}=0.155$~s. Also, we apply the Cowling approximation at the final moment of removal. Furthermore, to follow the evolution of the ejecta, we reduce the grid number from $N=385$ to $217$ while keeping $\Delta x_{13}=100$~m at $t-t_{\rm merger}=0.32$~s. 
Thanks to the concentric nested grid structure, this removal of the nested domain(s) does not require any interpolations or extrapolations. To validate this strategy, we continue the original run up to a certain moment and check the consistency of relevant quantities. Specifically, we keep running the simulation with $\Delta x_{16}=12.5$ m up to the black hole formation, which happens at $t-t_{\rm merger}\approx 0.07$~s. Also, we continue the simulation with $\Delta x_{15}=25$ m up to $t-t_{\rm merger}\approx 0.09$~s, that with $\Delta x_{14}=50$ m up to $t-t_{\rm merger}\approx 0.19$~s, and that with $\Delta x_{13}=100$~m up to $t-t_{\rm merger}\approx 0.36$~s, respectively.

During the evolution, we employ an atmospheric prescription where we assume a constant atmospheric density of $10^3~{\rm g~cm^{-3}}$ for $r\le 38.6~{\rm km}$, and a power-law profile with $\rho \propto 10^3 (38.6~{\rm km}/r)^3~{\rm g~cm^{-3}}$ for $r>38.6~{\rm km}$. Once this profile reaches the equation of state table limit of $0.167~{\rm g~cm^{-3}}$, where we stitch the original BHB$\Lambda\phi$ nuclear equation of state with the Helmholtz-type equation of state~\cite{Timmes} to extend the low-density and temperature regime, we assume a constant rest-mass density profile with this floor value. 
We also assume an atmospheric temperature of $10^{-3}~{\rm MeV}$. If the rest-mass density touches the atmosphere profile after the time update, we reset the momentum to be zero, and the rest-mass density and temperature to be the atmospheric values. Inside the magnetosphere formed in the polar regions after the black hole formation, we monitor the magnetization parameter $\sigma_B\equiv b^\mu b_\mu/(4\pi\rho {\rm c}^2)$ where $b^\mu$ denotes the magnetic field measured in the fluid rest frame. If $\sigma_B$ exceeds a critical value $\sigma_{B,\rm crit}$, we reset the fluid elements by $\rho\to\rho (\sigma_B/\sigma_{B,\rm crit})$. We found the typical maximum value of $\sigma_B$ is $O(10^2)$ inside the magnetosphere region. We employ $\sigma_{B,\rm crit}=10^2$. Like the atmospheric prescription widely used in the numerical relativity community, this prescription introduces the violation of the baryonic mass conservation. By monitoring its violation, we confirm it is much below the post-merger ejecta mass. The mass conservation violation at the saturation of the post-merger ejecta at $t-t_{\rm merger}\approx 0.12$~s is $\approx 4\times10^{-6}M_\odot$ (see the main text). 

Finally, we remark on the initialization of the magnetic field. Following Ref.~\cite{Kiuchi:2023obe}, we initialize it by the vector potential:
\begin{align}
A_\varphi \propto \max\left(P-0.0002 P_{\rm max},0\right)^2, \label{eq:vector}
\end{align}
where $P$ and $P_{\rm max}$ denote the pressure and its maximum respectively. The overall normalization is set to be the initial maximum field strength $B_{0,{\rm max}}=10^{15}~{\rm G}$. We should note that the binary neutron star merger global simulations with the subgrid-scale modeling~\cite{Aguilera-Miret:2023qih,Aguilera-Miret:2024cor,Campanelli:2005dd}, in which the magnetic field is initialized by several different field topologies, and $B_{0,{\rm max}}=10^{11}$--$10^{12}$~G reported a very similar electromagnetic saturation energy at the end of the Kelvin-Helmholtz instability phase to the direct simulation with the grid resolution of $12.5$ m and $B_{0,{\rm max}}=10^{15.5}$ G~\cite{Kiuchi:2023obe}. 
Furthermore, our very recent binary neutron star merger simulation with $\Delta x_{16}=6.25$~m and $B_{0,\mathrm{max}}=10^{12.5}~{\rm G}$ demonstrates that the stellar-scale magnetic field with the strength of $10^{15}~{\rm G}$ is developed via the Kelvin-Helmholtz instability within $5~{\rm ms}$ after the merger~\cite{Kiuchi:2026pgb}. These findings justify the initialization of the unrealistically strong magnetic field compared to the observed binary pulsars in this article.

\subsubsection*{Tracer particle}

To perform the {\it r}-process nucleosynthesis calculation as a post-process, a \textit{tracer particle method} is employed to extract the Lagrangian time evolution of the thermodynamic quantities from the numerical relativity simulation data. Suppose that we have a time series of three-velocity field $v^{(n)i}$ derived by the numerical relativity, where $i$ is the index of the spatial coordinates, at a time slice $t=t^{(n)}$. The velocity field is defined on the Cartesian grid points $(x^1_j,x^2_k,x^3_l)$ discretely, where $j$, $k$, $l$ denote the grid points in the three-dimensional space. For a given spatial position of a particle $x^{(n)i}_\mathrm{p}$ at the $n$th time slice, the position of the particle in the $(n+1)$th time slice is solved with the implicit-trapezoidal rule.
In this method, we solve
\begin{align}
\frac{x^{(n+1)i}_\mathrm{p}-x^{(n)i}_\mathrm{p}}{t^{(n+1)}-t^{(n)}} = \frac{1}{2}\big(v^{(n)i}_\mathrm{p} + v^{(n+1)i}_\mathrm{p}\big)
\end{align}
for $x^{(n+1)i}_\mathrm{p}$ where $v^{(n)i}_\mathrm{p} := v_\mathrm{p}^{(n)i}(x_\mathrm{p}^{(n)})$ is the velocity of the particle at the $n$th time slice tri-linearly interpolated from $v^{(n)i}$ to the particle position. 
We apply this tracer particle method to a given set of numerical relativity simulation data. Thus, the time step $\Delta t = t^{(n+1)}-t^{(n)}$ is limited by the numerical relativity simulation data output cadence. The numerical accuracy of this tracer particle method may be diagnosed with a time step defined by
\begin{align}
\Delta t' = \min\bigg(\frac{x^1_{j+1}-x^1_{j}}{v_\mathrm{p}^{(n)1}}, \frac{x^2_{k+1}-x^2_{k}}{v_\mathrm{p}^{(n)2}}, \frac{x^3_{l+1}-x^3_{l}}{v_\mathrm{p}^{(n)3}}\bigg).
\end{align}

If $\Delta t'<\Delta t$, the particle position changes significantly within the time step $\Delta t$. In such a case, we perform several sub-steps between the $(n+1)$th and $n$th time slices. The velocity fields at a time $t$ between $t^{(n+1)}$ and $t^{(n)}$ are linearly interpolated from the two time slices.

The particles are injected on an extraction sphere of radius $r_\mathrm{ext}=3\times10^8$\,cm. At each particle-injection time, we evaluate the ejecta criterion at angular grid points on the extraction sphere. If the ejecta criterion is satisfied at a given angular grid point, a tracer particle is placed at $r=r_\mathrm{ext}$ in that angle and then traced backward in time. We employ the Bernoulli criterion $h u_t < -h_{\rm min}$ for the ejecta where $h$, $h_{\rm min}$, and $u_t$ denote the specific enthalpy, its minimum value, and the time component of the four velocity, respectively. 
The polar angles in the set $\{(\theta_i, \varphi_i)\}$ are on a uniform grid $\theta_i \in [0,\pi/2]$ with a spacing $\Delta \theta$. The azimuthal angles in the set are also on a uniform grid $\varphi_i \in [0,2\pi]$ but with a $\theta$-dependent spacing $\Delta \varphi$, which decreases, i.e., the grid number in the azimuthal direction increases, with the polar angle. The function of $\Delta \varphi$ is determined to distribute the particles in an approximately uniform manner on the sphere.

The particle injection is repeated with a time interval
\begin{align}
\Delta t_\textrm{p-set} = \frac{r_\mathrm{ext}\Delta \theta}{\langle v^r\rangle},
\end{align}
where $\langle v^r\rangle$ is the average radial velocity of the particles located at the particle injection time. This choice makes the radial separation between successive particle sets comparable to the lateral spacing on the extraction sphere, i.e.,
$\langle v^r\rangle \Delta t_\textrm{p-set} = r_\mathrm{ext}\Delta\theta$.


The mass of each particle is assigned as
\begin{align}
m = {r_\mathrm{ext}}^2 \Delta\Omega \rho u^r \sqrt{-g}\Delta t_\textrm{p-set},
\end{align}
where $u^r$ is the radial component of the fluid four-velocity, $\Delta\Omega$ is the solid angle that is assigned to each angular grid point, and $g$ is the determinant of the metric. This definition of the particle mass is consistent with the conserved mass flux at $r=r_\mathrm{ext}$. Therefore, the total mass of the particles converges to the ejecta mass of the numerical relativity simulation for increasing the number of particles set on the sphere (i.e., $\Delta \theta \rightarrow0$ and thus $\Delta t_\textrm{p-set}\rightarrow0$).

We evolve the tracer particle backward in time from $t-t_\mathrm{merger}=0.29$\,s to the time slightly before the merger. 
However, at $t-t_\mathrm{merger}=0.29$\,s, we find that there is still a significant amount of matter inside the extraction radius $r_\mathrm{ext}$. A part of the matter at the time is diagnosed to be unbound, but such a matter is not flagged by the tracer particles when they are initialized in the way described above, because the matter does not cross the radius $r_\mathrm{ext}$ yet at the time. In addition, the matter gravitationally bound at the time could become unbounded in the subsequent evolution. Therefore, we place another set of particles inside the radius $r_\mathrm{ext}$ at $t-t_\mathrm{merger}=0.29$\,s. They are traced backward in time from $t-t_\mathrm{merger}=0.29$ to the time before the merger in the same manner. In addition, they are also traced forward in time until the time of the termination of the simulation ($t-t_\mathrm{merger}\approx 0.68$\,s).
In this way, we capture the Lagrangian motion of all unbound matter at the end of the simulation.

\subsubsection*{Nucleosynthesis}
Nucleosynthesis calculations are performed for a limited number of tracer particles in order to reduce the computational cost (see justification in Supplementary Text). 
For each tracer particle, the nucleosynthetic yields are obtained by using the nuclear reaction network code \texttt{rNET}~\cite{Wanajo:2018wra,Wanajo:2022mhf}. 
The code adopts experimentally evaluated reaction rates when available, and theoretical rates otherwise (the latter dominates for {\it r}-process nucleosynthesis). Specifically, theoretical neutron-capture and $\beta$-decay rates are taken from those of TALYS~\cite{Goriely:2008zu} and GT2~\cite{Tachibana1990}, respectively, both based on the HFB-21 mass prediction~\cite{Goriely:2010bm}. Each calculation for a given tracer particle starts when the temperature decreases to 10~GK, where GK~$\equiv 10^9$~K, with the initial composition of $Y_\mathrm{e}$ and $1 - Y_\mathrm{e}$ for free protons and neutrons, respectively. Here, $Y_\mathrm{e}$ is the electron fraction (or the number of protons per nucleon) evaluated at 5~GK. At such a high temperature, the composition immediately relaxes to that determined by nuclear statistical equilibrium. Each calculation is terminated at 1~Gyr.

We then construct a table of radioactive heating rates and elemental abundances for the subsequent long-term hydrodynamics simulation. The table is defined on a grid of the extraction time and angular coordinates, $(t_\mathrm{ext},\theta_\mathrm{ext},\varphi_\mathrm{ext})$, on the extraction sphere. For each grid point, we select the four tracer particles closest to that point at $t=t_\mathrm{ext}$. The nucleosynthetic yields and heating-rate histories of these particles are interpolated with weights determined by their distances from the point on the sphere. The resulting weighted averages are stored in the table as the mass fraction $X(Z;t_\mathrm{ext},\theta_\mathrm{ext},\varphi_\mathrm{ext})$ and history of specific heating rate, $q(\tau;t_\mathrm{ext},\theta_\mathrm{ext},\varphi_\mathrm{ext})$, for matter crossing the extraction sphere at the given time and angular position. $Z$ is the atomic number of elements, and $\tau$ is the time after the beginning of the nucleosynthesis calculation.
 In the long-term hydrodynamics simulation, the elemental abundances and heating rate for each fluid element are assigned from this table (see the following subsection for details). 


\subsubsection*{Long-term ejecta evolution}

We employ a 3D code developed in our previous studies~\cite{Kawaguchi:2024hdk} for the HD simulation. This code solves the relativistic Euler equations under a spherical coordinate system. In order to incorporate the effect of gravity, a fixed background metric for a non-rotating black hole expressed in isotropic coordinates is used (see Appendix A of~\cite{Kawaguchi:2024hdk} for the formulation of the basic equations). We note that the orbital plane symmetry is imposed for the HD simulation following the setup of the numerical relativity simulation. For the equation of state, we consider the contributions from gas and radiation: the total pressure $P$ is given by $P=P_{\rm gas}+P_{\rm rad}$ with $P_{\rm gas}=n_{\rm B} k_{\rm B} T/\mu$ and $P_{\rm rad}=a_{\rm rad} T^4/3$, where $n_{\rm B}$, $T$, $ k_{\rm B}$, $\mu$, and $a_{\rm rad}$ are the baryon number density, temperature, Boltzmann constant, mean molecular weight, and radiation density constant, respectively. We simplify the gas pressure assuming that atoms are fully ionized with $\mu=2$, and the gas pressure is dominated by the contribution from electrons, given the fact that the average atomic mass number is expected to be much larger than unity.

The uniform grid with $N_\theta$ and $N_\varphi$ grid points is employed for the angular direction. 
For the radial direction, the following non-uniform grid structure is employed; for a given $j$-th radial grid-node
\begin{align}
	{\rm ln}\,r_j={\rm ln}\left(\frac{r_{\rm out}}{r_{\rm in}}\right)\frac{j-1}{N_r}+{\rm ln}\,r_{\rm in},\,j=1\cdots N_r+1,\label{eq:grid}
\end{align} 
where $r_{\rm in}$ and $r_{\rm out}$ denote the inner and outer radii of the computational domain, respectively, and $N_r$ denotes the total number of the grid along the radial direction. In the present work, we employ $(N_r,N_\theta,N_\varphi)=(1024,64,128)$, and $r_{\rm in}$ and $r_{\rm out}$ are initially set to be $3,000\,{\rm km}$ and $10^3\,r_{\rm in}$, respectively. We confirm that this grid resolution is sufficiently high for our purpose of the study by 
conducting the convergence study with $(N_r,N_\theta,N_\varphi)=(512,32,64)$. 
To relax the CFL condition, we average over the conservative variables of hydrodynamics in the direction of $\varphi$ for all the grid located in $\theta\leq\theta_{\rm c}$ for each sub-step of the evolution (see~\cite{Hirai:2022hbf} for a similar prescription). For the present study, we choose $\theta_{\rm c}$ to be $\pi/24$, while we confirm that the resulting light curves are essentially unchanged even if we employ $\theta_{\rm c}=\pi/12$.

The hydrodynamic properties of the outflow are extracted at $r=r_{\rm ext}$ in the numerical relativity simulation, and the time-sequential data are employed as the inner boundary condition of the HD simulations. The outflow data obtained from the numerical relativity simulation run out at $t-t_{\rm merger}\gtrsim 0.7$\,s, and after then, the HD simulation is continued by setting a floor value to the rest-mass density of the inner boundary. To follow the evolution of ejecta even after the high-velocity edge of the outflow reaches the outer boundary of our HD simulation, the radial grid points are added to the outside of the original outer boundary, while at the same time, the innermost radial grid points are removed so as to keep the total number of radial grid points. The radial regridding is performed so that the region of $r/(ct)\approx$ $10^{-3}$--$1$ is always covered with the computational domain up to $t-t_{\rm merger}=0.1\,{\rm d}$.

The effect of radioactive heating is incorporated into the HD simulation in the same manner as in our previous studies~\cite{Kawaguchi:2020vbf,Kawaguchi:2022bub,Kawaguchi:2024hdk}. The spatial distribution of the radioactive heating rates of the ejecta is determined by referring to a table constructed from the nucleosynthesis calculation performed along the tracer particle trajectories as described above. To this end, we label each fluid element with the time ($t_{\rm ext}$) and angles ($\theta_{\rm ext}$ and $\varphi_{\rm ext}$) at which it reaches at $r=r_{\rm ext}$ and advects these quantities with the three velocity $v^i$. 
At each time step of the evolution, the radioactive heating rate is determined from these quantities by linearly interpolating the heating-rate table.



\subsubsection*{Kilonova light curve}

The light curves of kilonovae are calculated using a wavelength-dependent radiative transfer simulation code~\cite{Tanaka:2013ana,Tanaka:2017qxj,Tanaka:2017lxb,Kawaguchi:2019nju,Kawaguchi:2020vbf}. In this code, the photon transfer is simulated by a Monte Carlo method for given ejecta profiles composed of the density, velocity, and elemental abundance under the assumption of homologous expansion. The time-dependent thermalization efficiency is taken into account following an analytic formula derived by~\cite{Barnes:2016umi}. We employ $0.03\ {\rm cm^2\ g^{-1}}$ for the effective gamma ray opacity as a fiducial setup following \cite{Guttman:2024bxl} (see "Uncertainties and systematic errors in kilonova light curve modeling" for the impact of the uncertainty in the effective gamma ray opacity on the light curves). In our radiative transfer code, the local gas temperature at each time step is calculated as $T_{\rm gas}=(u_{\rm rad}/a_{\rm rad})^{1/4}$, i.e., under the assumption of local thermodynamic equilibrium (LTE) using the Stefan-Boltzmann law with the radiation energy density, $u_\mathrm{rad}$, obtained by the radiative transfer simulation.
Then, the ionization and excitation states are determined from the gas density and temperature by solving Saha's ionization and Boltzmann excitation equations.

For the photon-matter interaction, bound-bound, bound-free, and free-free transitions, and electron scattering are taken into account for the transfer of optical and infrared photons~\cite{Tanaka:2013ana,Tanaka:2017qxj,Tanaka:2017lxb}. The formalism of the expansion opacity~\cite{1983ApJ...272..259F,1993ApJ...412..731E,Kasen:2006ce} and the line list derived in~\cite{Domoto:2022cqp} are employed for the bound-bound transitions. In this line list, the atomic data of VALD~\cite{1995A&AS..112..525P,1999A&AS..138..119K,2015PhyS...90e4005R} or Kurucz's database~\cite{1995all..book.....K} is used for $Z=20$–$29$, while the results of atomic calculations from~\cite{Tanaka:2019iqp} are used for $Z=30$--$88$. For Sr II, Y I, Y II, Zr I, Zr II, Ba II, La III, and Ce III, which are the ions producing strong lines, the line data are replaced with those calibrated with the atomic data of VALD and NIST databases~\cite{NIST}. Note that, since our atomic data include only up to the triple ionization for all the ions, the early phase of the light curves ($t-t_{\rm merger}< 1\,{\rm d}$) may not be very reliable due to high ejecta temperature (see~\cite{Banerjee:2020myd,Banerjee:2022doa,Banerjee:2023gye} for the work accounting for the higher ionization atoms).

The radiative transfer simulations are performed from $t-t_{\rm merger}=0.1\,{\rm d}$ to $30\,{\rm d}$ employing the density and internal energy profiles of the HD simulations at $t-t_{\rm merger}=0.1\,{\rm d}$ and assuming the homologous expansion for $t-t_{\rm merger} > 0.1$\,d. The spatial distributions of the heating rate and elemental abundances are determined by the table obtained by the nucleosynthesis calculations, referring to the injected time and angle of the fluid elements. Note that, as an approximation, the elemental abundances at $t-t_{\rm merger}=1\,{\rm d}$ are used during the entire time evolution in the radiative transfer simulations to reduce the computational cost. We confirm that this simplified prescription gives only a minor systematic error on the resultant light curves, as also illustrated in~\cite{Kawaguchi:2020vbf}. 

A three-dimensional cylindrical grid is applied for storing the local elemental abundances and radioactive heating rate as well as for solving the temperature and opacity. The $50$, $50$, and $32$ grid points are set to the cylindrical radius, vertical, and longitudinal directions, which cover the domain with the coordinate ranges of $(0,0.7\,ct)$, $(0,0.7\,ct)$, and $(0,2\pi)$, respectively. For the time grid, we use a logarithmically-spaced time step with a time step of $\Delta {\rm log}((t-t_{\rm merger})/{\rm d})=0.025$. We confirm that the resulting light curves are modified only within the systematic uncertainties discussed in the bellow by changing each grid numbers from $50$, $50$, and $32$ to $40$, $40$, and $28$, changing the time stepping from $\Delta {\rm log}((t-t_{\rm merger})/{\rm d})=0.025$ to 0.0125 as well as changing the maximum cylindrical radius and vertical coordinate ranges from $0.7\,ct$ to $0.8\,ct$.


\subsection*{Supplementary Text}

In this Supplementary Text, we describe the relevant evolution stages such as the Kelvin-Helmholtz instability phase, the magnetorotational-instability driven $\alpha\Omega$ dynamo phase, and the subsequent black-hole-accretion torus phase. The link (\url{http://www2.yukawa.kyoto-u.ac.jp/~kenta.kiuchi/anime/FUGAKU2025/out_yuv420p_1600.mp4}) visualizes the entire evolution on a meridional plane for the rest-mass density, the magnetic-field strength, the magnetization parameter, the unboundness of the fluid elements with the Bernoulli criteria, the electron fraction, the temperature, the specific entropy, and the Shakura-Sunyaev parameter from the top-left to the bottom-right panel, respectively.
Subsequently, we describe the {\it r}-process nucleosynthesis calculation and the kilonova light curve calculation. 

\subsubsection*{Kelvin-Helmholtz instability saturation}
\begin{figure} 
	\centering
	\includegraphics[width=0.65\textwidth]{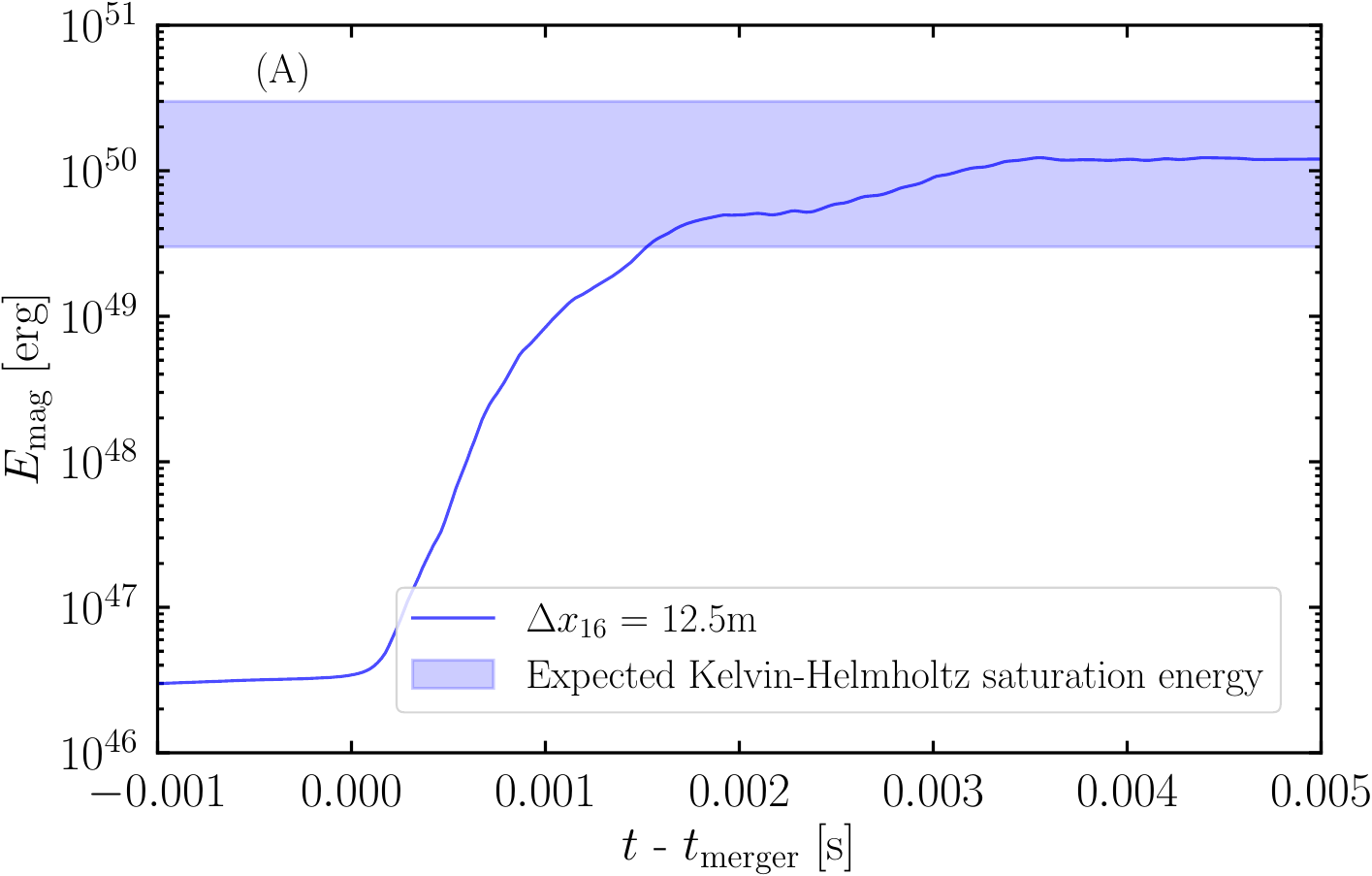} 
	\caption{\textbf{Kelvin-Helmholtz instability-driven amplification of magnetic field.}
    (\textbf{A}) Electromagnetic energy as a function of the post-merger time. The shaded color region presents the expected saturation energy estimated in the high-resolution direct simulations and the subgrid-scale modeling of binary neutron star mergers~\cite{Kiuchi:2023obe,Aguilera-Miret:2021fre,Aguilera-Miret:2023qih,Aguilera-Miret:2024cor,Palenzuela:2021gdo,Kiuchi:2026pgb}. 
		}
	\label{fig:SM_KH} 
\end{figure}

Figure~\ref{fig:SM_KH} plots the electromagnetic energy as a function of the post-merger time during the Kelvin-Helmholtz instability phase. 
The shaded color region presents the expected electromagnetic saturation energy due to the Kelvin-Helmholtz instability, which is estimated in the high-resolution direct binary neutron star merger simulation~\cite{Kiuchi:2023obe}, and the subgrid-scale modeling of binary neutron star merger simulations~\cite{Aguilera-Miret:2021fre,Aguilera-Miret:2023qih,Aguilera-Miret:2024cor,Palenzuela:2021gdo}. We note that we select the non-spinning binary neutron star models with the total mass of $2.700$--$2.724M_\odot$. The equations of state of the neutron star are the APR4~\cite{Akmal:1998cf} or DD2~\cite{Hempel:2009mc}. Also, the magnetic-field topology is initialized by the confined dipole field, the multipole field, or the misaligned dipole field. These differences make variance in the expected saturation energy. We should note that the very recent unprecedented high-resolution simulation of the magnetized binary neutron star merger is consistent with this estimation~\cite{Kiuchi:2026pgb}. 
The plot shows that the magnetic field is amplified up to the expected saturation energy before the shear layer is dissipated by the shock waves at $t-t_{\rm merger}\approx 0.002$--$0.003$~s. 

\subsubsection*{Magnetorotational instability}
\begin{figure} 
	\centering
	\includegraphics[width=0.495\textwidth]{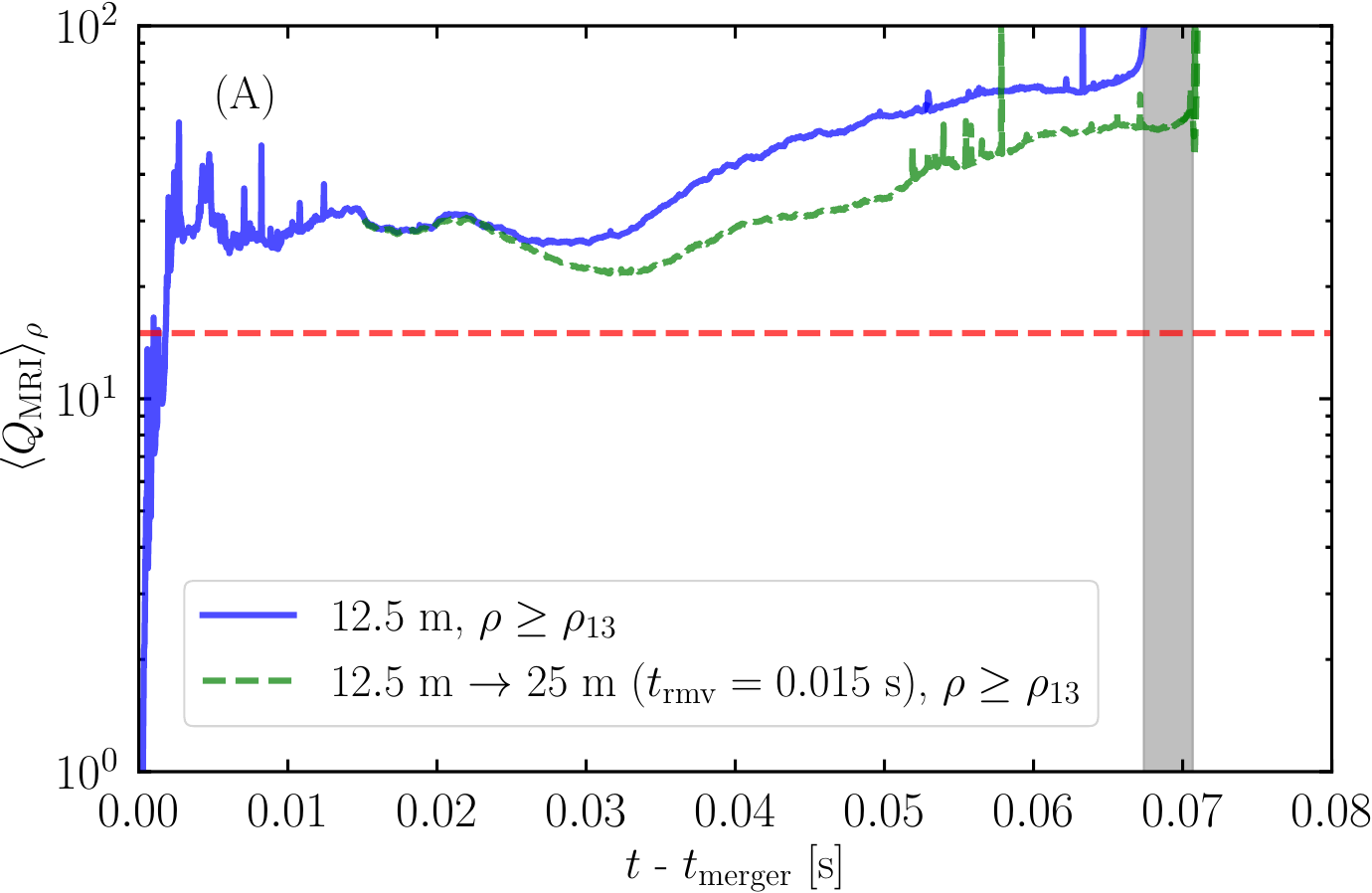}
    \includegraphics[width=0.495\textwidth]{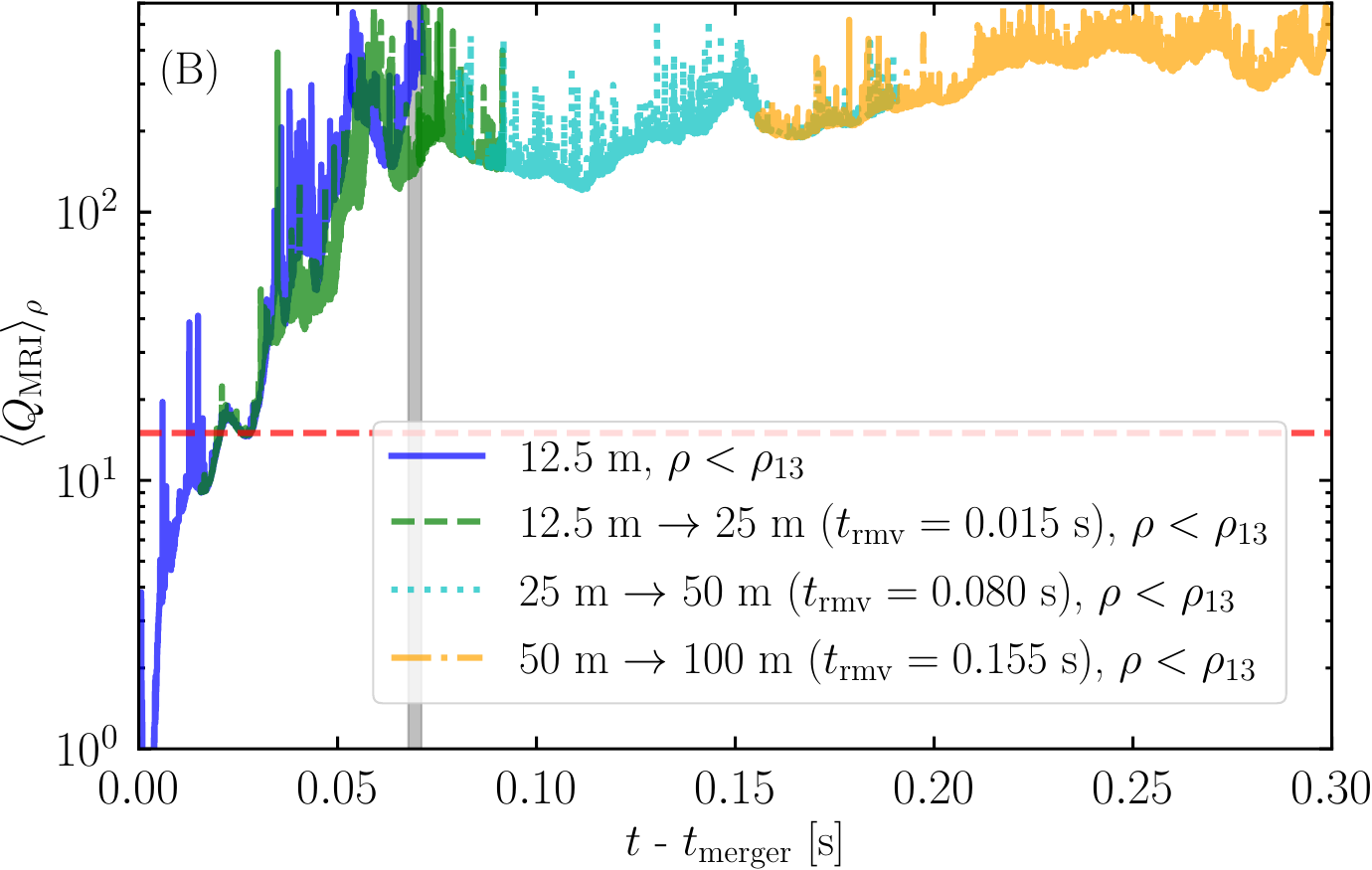}\\ 
    \includegraphics[width=0.495\textwidth]{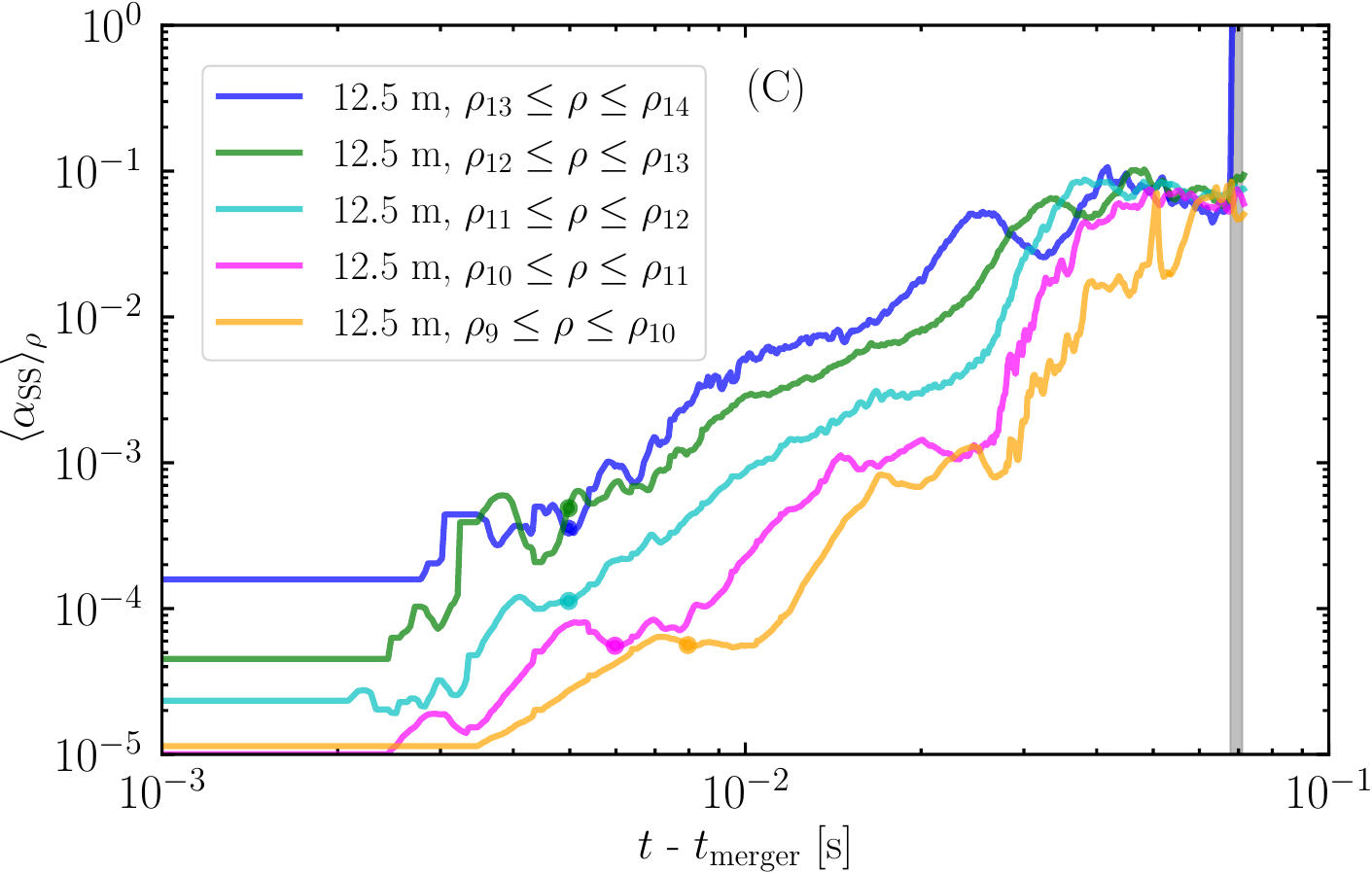}
    \includegraphics[width=0.495\textwidth]{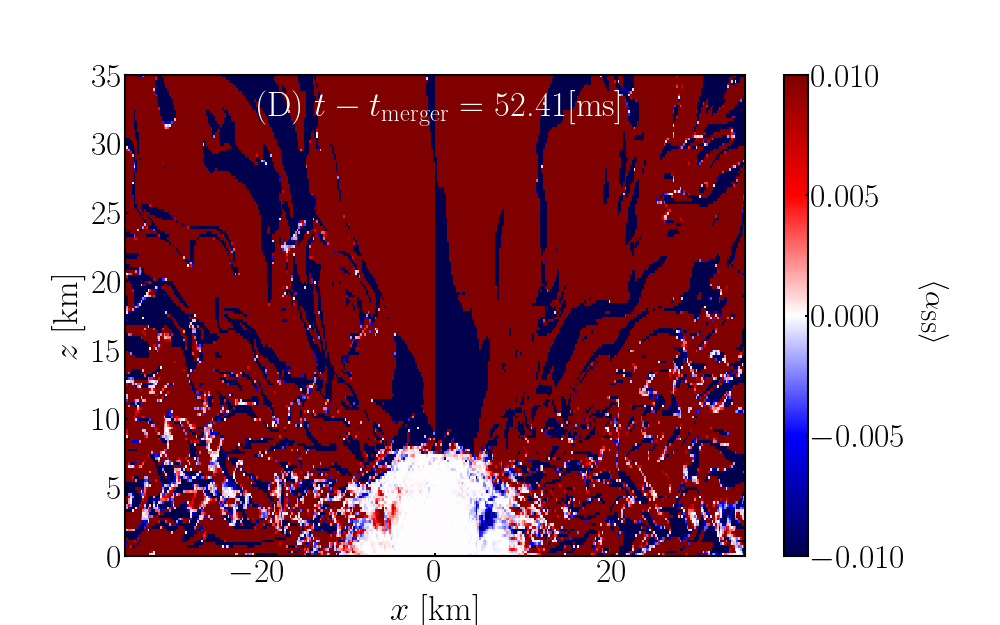}    
	\caption{\textbf{Magnetorotational instability inside the merger remnant.}
    ({\bf A}) Magnetorotational-instability quality factor in the core region with $\rho \ge 10^{13}~{\rm g~cm^{-3}}$. The blue curve denotes the simulation with $\Delta x_{16}=12.5$ m, and the green curve denotes the simulation in which the finest nested domain with $\Delta x_{16}=12.5$ m is removed at $t-t_{\rm merger}=0.015$~s. The vertical gray-shaded region denotes the uncertainty of the black hole formation estimated by the simulations with $\Delta x_{16}=12.5$~m and  $\Delta x_{15}=25$~m. 
    ({\bf B}) Magnetorotational-instability quality factor in the disk region with $\rho < 10^{13}~{\rm g~cm^{-3}}$. The cyan (orange) curve presents the simulation, in which the second (third) finest nested domain with $\Delta x_{15}=25$ m $(\Delta x_{14}=50~{\rm m})$ is removed at $t-t_{\rm merger}=0.08~(0.155)$~s. 
    The horizontal red-dashed line denotes the critical quality factor of $15$~\cite{Hawley:2011tq}. 
    ({\bf C}) The rest-mass density conditioned Shakura-Sunyaev parameter as a function of the post-merger time. 
    The filled circles on top of each curve at the very beginning indicate the timing when the magnetorotational-instability quality factor exceeds the critical value. 
    ({\bf D}) The profile of the Shakura-Sunyaev parameter on a meridional plane at $t-t_{\rm merger}\approx 0.052$~s. 
    The link to the visualization:~\url{http://www2.yukawa.kyoto-u.ac.jp/~kenta.kiuchi/anime/FUGAKU2025/out_yuv420p_Maxwell.mp4}. 
		}
	\label{fig:SM_Qfac} 
\end{figure}

We monitor the magnetorotational instability quality factor~\cite{Balbus-Hawley:1991} defined by
\begin{align}
Q_{\rm MRI} \equiv \frac{\lambda_\text{MRI}}{\Delta x_{\rm (lv)}} = \frac{2\pi B^z}{\Delta x_{\rm (lv)} \sqrt{4\pi \rho} \Omega},
\end{align}
where $B^z$, $\rho$, and $\Omega$ denote the $z$-component of the magnetic field, the rest-mass density, and the angular velocity, respectively. Following Refs.~\cite{Kiuchi:2017zzg,Kiuchi:2022nin,Kiuchi:2023obe}, we evaluate the rest-mass density conditioned quality factor:
\begin{align}
\langle Q_{\rm MRI} \rangle_\rho  = \frac{\int_{\text{condition for }\rho} Q_{\rm MRI} {\rm d}^3x}{\int_{\text{condition for }\rho} {\rm d}^3x}.
\end{align}
As a condition for the rest-mass density, we apply the condition that the rest-mass density is greater (smaller) than $10^{13}~{\rm g~cm^ {-3}}$. Figures~\ref{fig:SM_Qfac} (\textbf{A}) and (\textbf{B}) plot the rest-mass density conditioned quality factor as a function of the post-merger time. It shows that the fastest growing mode of the magnetorotational instability is well resolved both in the core and disk throughout the simulation. 

References~\cite{Aguilera-Miret:2023qih,Aguilera-Miret:2024cor,Palenzuela:2021gdo} argue that the conventional magnetorotational-instability quality factor may not be an appropriate diagnostic for the magneto-turbulent state because the background flow, on which the linear perturbation calculation relies~\cite {Balbus-Hawley:1991}, is highly non-axisymmetric. On the other hand, Ref.~\cite{Hawley:2011tq} proposed a couple of diagnostics for the magneto-turbulent state in the non-linear phase of the magnetorotational instability. As a representative of their diagnostics, we estimate the Shakura-Sunyaev parameter defined by
\begin{align}
\alpha_{\rm SS} = \left\langle -\frac{B^r B_\varphi}{4\pi P}\right\rangle,
\end{align}
where $\langle\cdot\rangle$ denotes a time ensambling. Figure~\ref{fig:SM_Qfac} (\textbf{C}) plots the rest-mass density conditioned Shakura-Sunyaev parameter as a function of the post-merger time. As the rest-mass density condition, we choose the rest-mass density foliation :~$10^ {i} \le \rho < 10^{i+1}~{\rm g~cm^{-3}}$ with $i=9,10,11,12,$ and $13$, and $\rho_i=10^i~{\rm g~cm^{-3}}$. 
The filled circles on top of each curve in the plot indicate the timing when the magnetorotational-instability quality factor exceeds the critical value. It shows that once the quality factor exceeds the critical value, the Shakura-Sunyaev parameter steeply increases and saturates around $O(10^{-2}-10^{-1})$ depending on the foliation density (see also fig.~\ref{fig:SM_Qfac} (\textbf{D})). 
It indicates that the magneto-turbulent state due to the magnetorotational instability is established. 

\subsubsection*{Magnetorotational instability-driven $\alpha\Omega$ dynamo}
\begin{figure} 
	\centering
	\includegraphics[width=0.65\textwidth]{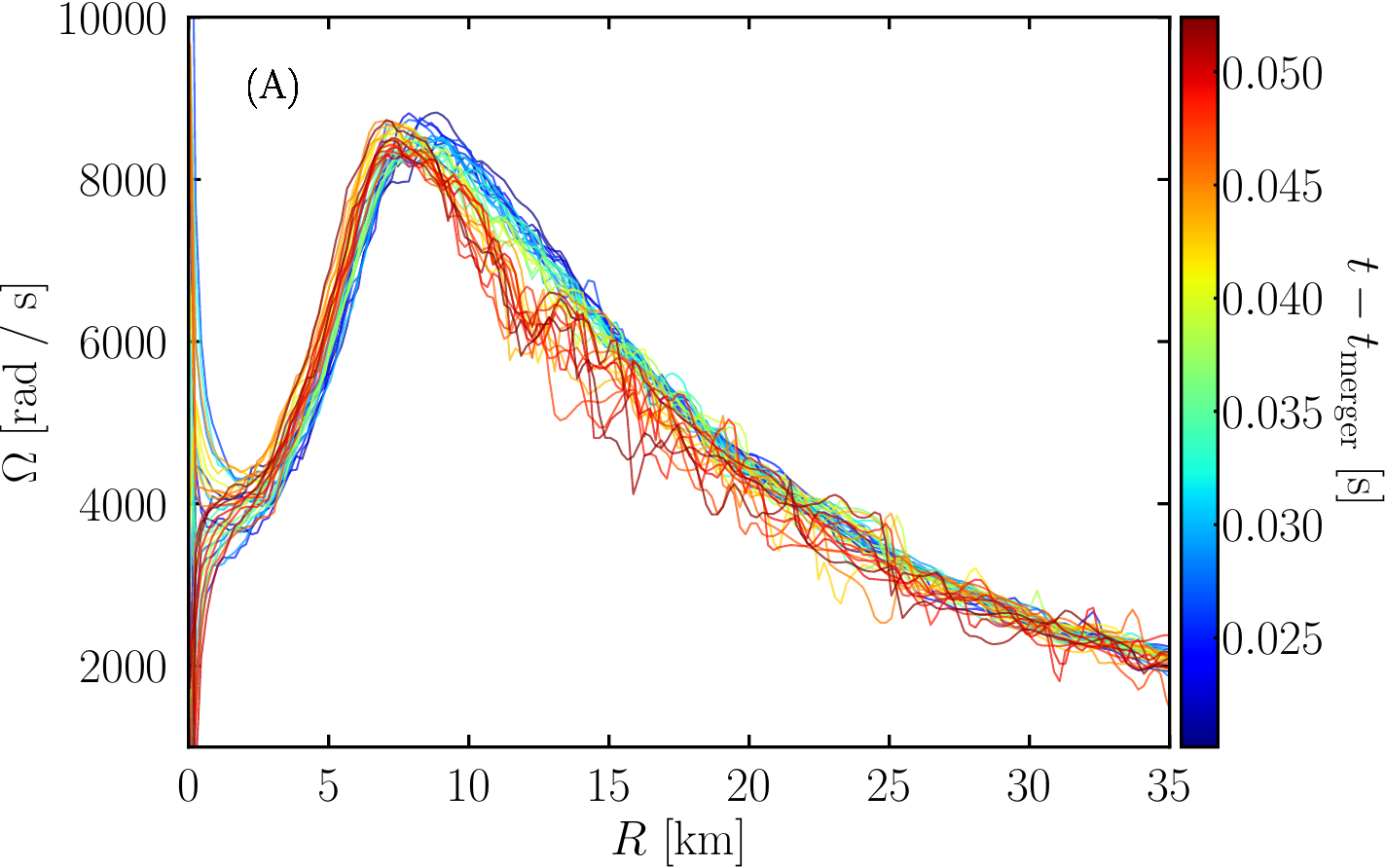}\\
    \includegraphics[width=0.65\textwidth]{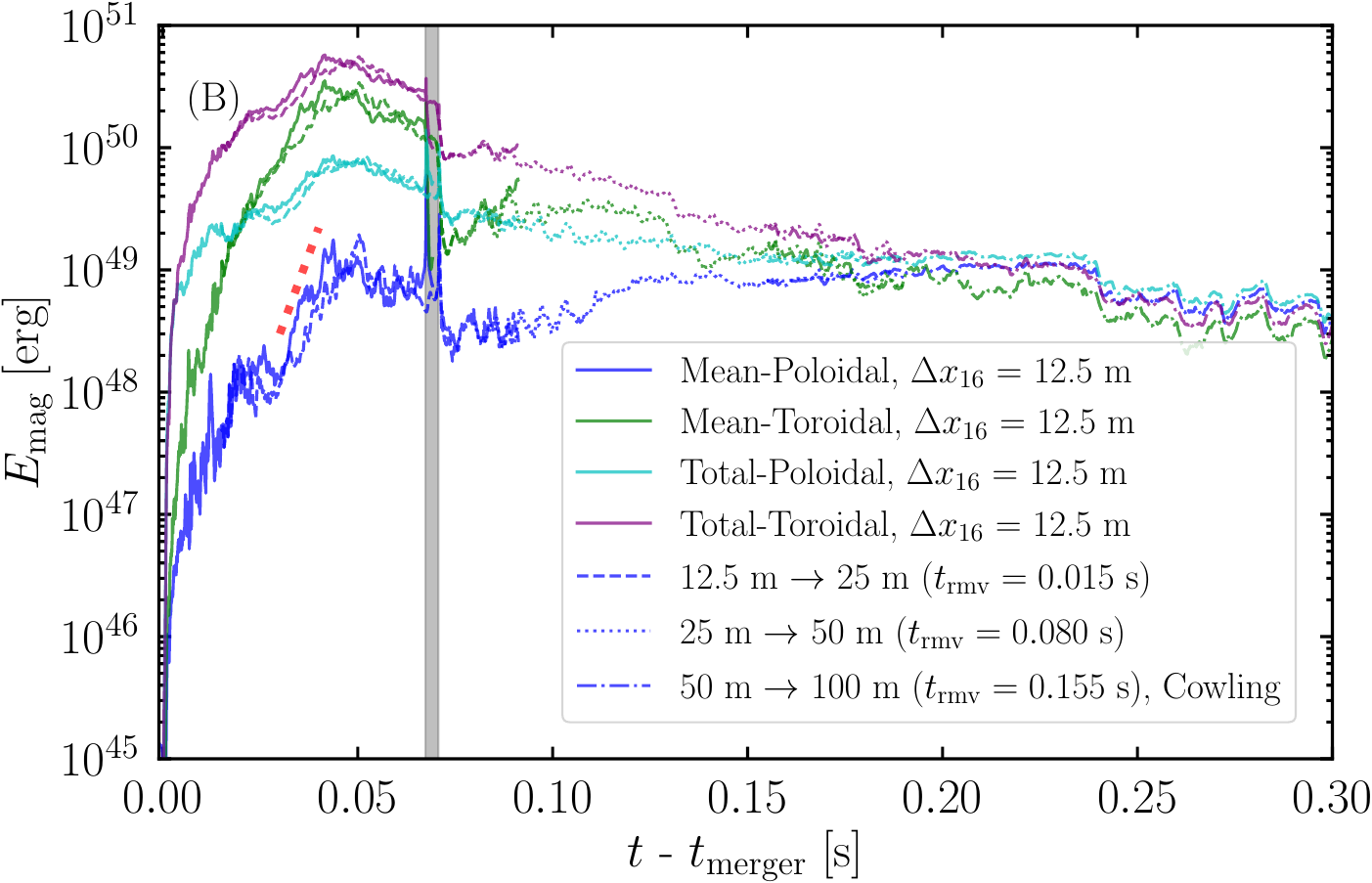} 
	\caption{\textbf{Mean magnetic field generation in a magnetorotational-instability active region.}
    (\textbf{A}) The time sequence of the angular velocity radial profile of the remnant massive neutron star on the orbital plane. The employed grid resolution is $\Delta x_{16}=12.5$ m. 
    (\textbf{B}) Mean- and total-magnetic field energy in the magnetorotational-instability active region as a function of the post-merger time. Blue, green, cyan, and magenta colors denote the mean-poloidal, mean-toroidal, total-poloidal, and total-toroidal components, respectively. 
    The vertical, gray-shaded region denotes the uncertainty in the timing of black hole formation estimated from simulations with $\Delta x_{16}=12.5$ m and  $\Delta x_{15}=25$ m. 
    The legends for the dashed, dotted, and dashed-dotted curves are the same as fig.~\ref{fig:SM_Qfac} (\textbf{B}), where we apply the Cowling approximation after the final removal of the nested domain. 
		}
	\label{fig:SM_Mean} 
\end{figure}

In the mean-field $\alpha\Omega$ dynamo theory, the induction equation for the mean field and the electromotive force due to the turbulent field are described by
\begin{align}
&\partial_t \bar{\bf B} = {\bf \nabla} \times \left(\bar{\bf V}\times \bar{\bf B} + \bar{\bf \mathcal E}\right),\\
&\bar{\mathcal E}_i \equiv \overline{\left(\bf v \times b\right)}_i = \alpha_{ij} \bar{B}_j + \beta_{ij}\overline{\left({\bf \nabla \times \bar{B}}\right)}_j \approx \alpha_{ii} \bar{B}_i, \label{eq:aOdynamo}
\end{align}
where $\bar{\bf B}$ and $\bar{\bf V}$ denote the mean magnetic field and the mean velocity field, respectively. We define the mean field by the axisymmetric average in our context. 
$\bf b$ and $\bf v$ denote the fluctuation components for the magnetic field and velocity field, where the magnetorotational instability is responsible for generating them. The first and second terms in the right-hand side of Eq.~(\ref{eq:aOdynamo}) are the dynamo $\alpha$ and the turbulent resistivity~\cite{Brandenburg:2004jv}. In our context, the off-diagonal component in the dynamo $\alpha$ and the turbulent resistivity are subdominant~\cite{Kiuchi:2023obe}. Therefore, if we assume the purely azimuthal flow $\bar{\bf V}=R\Omega {\bf e}_\varphi$ in the cylindrical coordinate, the above equation is cast into
\begin{align}
&\partial_t \bar{B}_R \approx - \partial_{z} \left(\alpha_{\varphi\varphi} \bar{B}_\varphi \right), ~
\partial_t \bar{B}_z \approx \partial_R \left(\alpha_{\varphi\varphi} \bar{B}_\varphi \right),~\partial_t \bar{B}_\varphi = R \bar{B}_A \nabla_A \Omega,
\end{align}
where $A=R,z$. Therefore, the mean poloidal field is generated only through the dynamo $\alpha$ term. 
Figure~\ref{fig:SM_Mean} (\textbf{A}) plots the time sequence of the angular velocity radial profile on the orbital plane. It shows that, in the region with $R \gtrsim (\lesssim) 7$--$9$~km, the magnetorotational instability is active (inactive). Therefore, we define the magnetorotational-instability active region by $\rho \le 10^{14.5}~\rm{g~cm^{-3}}$ to take into account the highly flattened structure of the remnant massive neutron star due to the strong and rapid differential rotation~\footnote{The negative radial gradient around the center is presumably an artifact due to the misalignment of the coordinate origin and the center of mass.}~\footnote{It corresponds to $R\ge 8$ km on the orbital plane.}. 
Figure~\ref{fig:SM_Mean} (\textbf{B}) plots the electromagnetic field energy in the magnetorotational-instability active region for both the mean field and total field. It indicates the mean poloidal-field energy (blue curve) exponentially increases for $0.03 \lesssim t-t_{\rm merger} \lesssim 0.05$~s (see the red-dashed guide), and the growth rate does not significantly depend on the resolution (see the blue-solid curve for $\Delta x_{16}=12.5$~m, and the blue-dashed curve for $\Delta x_{15}=25$~m). 

\begin{figure} 
	\centering
	\includegraphics[width=0.495\textwidth]{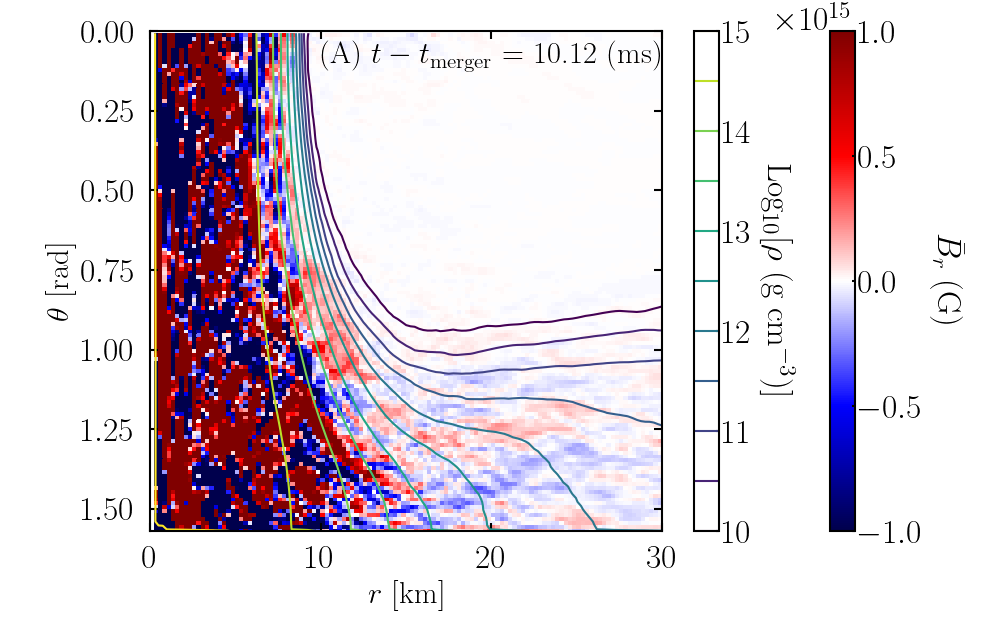}
    \includegraphics[width=0.495\textwidth]{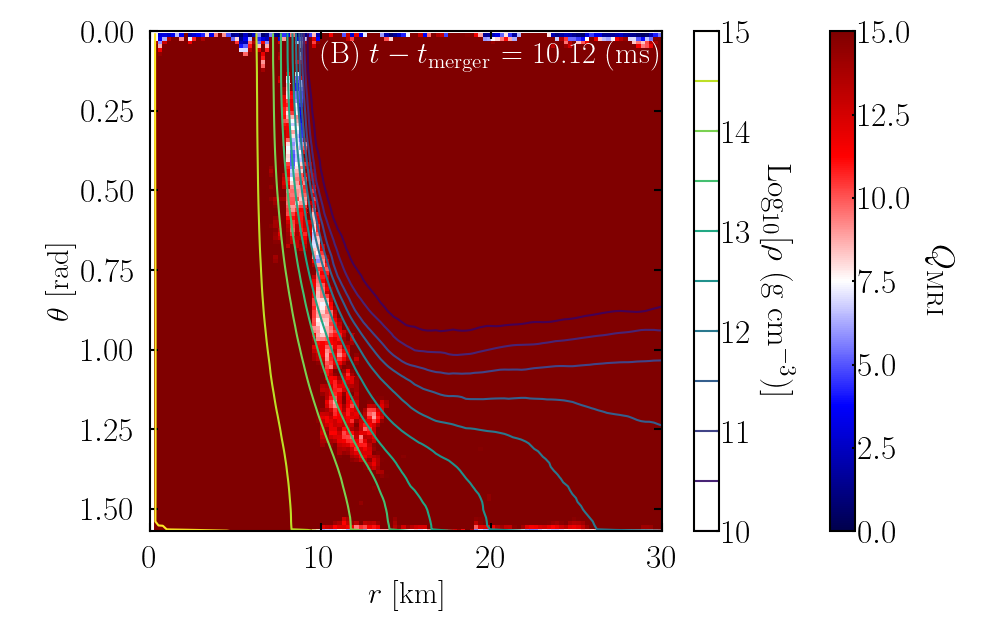}\\
    \includegraphics[width=0.495\textwidth]{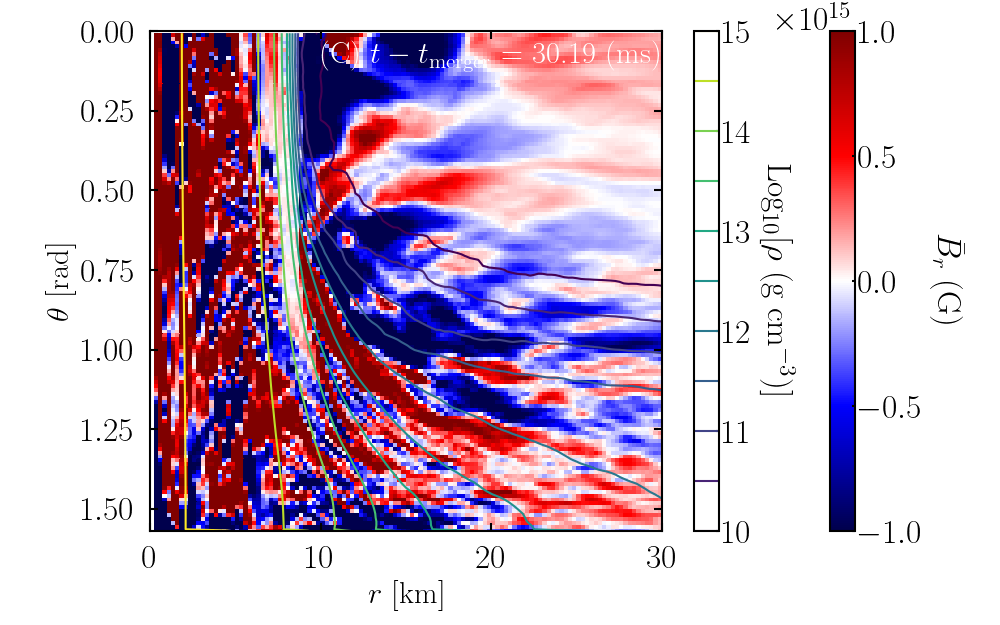}
    \includegraphics[width=0.495\textwidth]{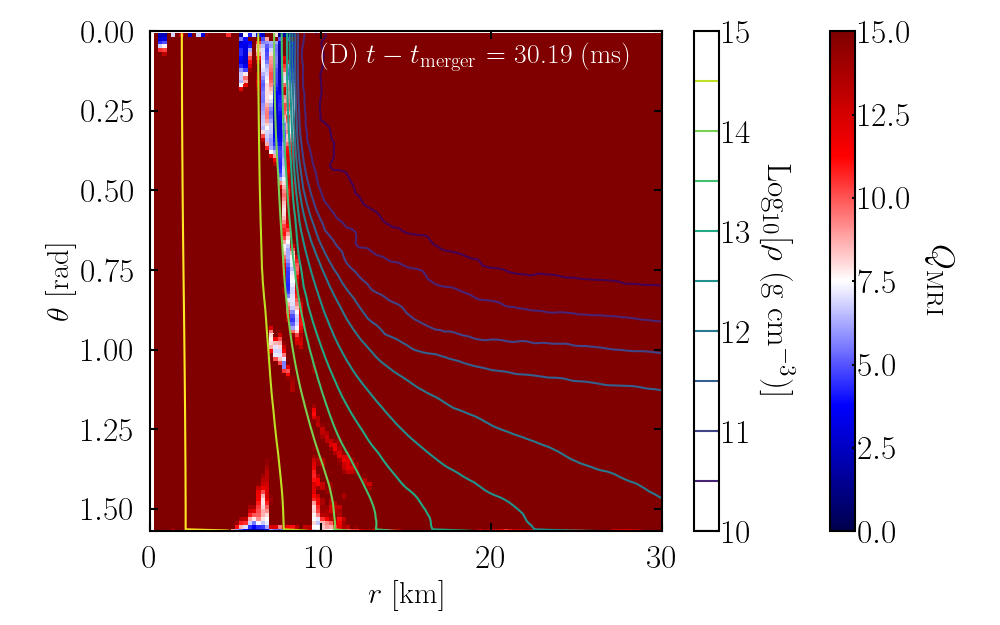}\\
    \includegraphics[width=0.495\textwidth]{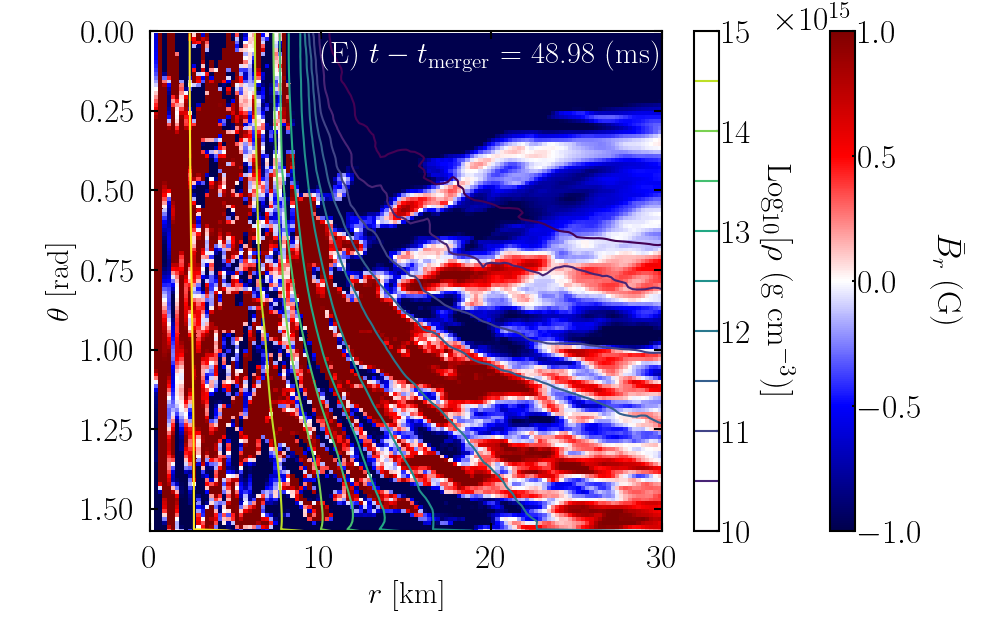}
    \includegraphics[width=0.495\textwidth]{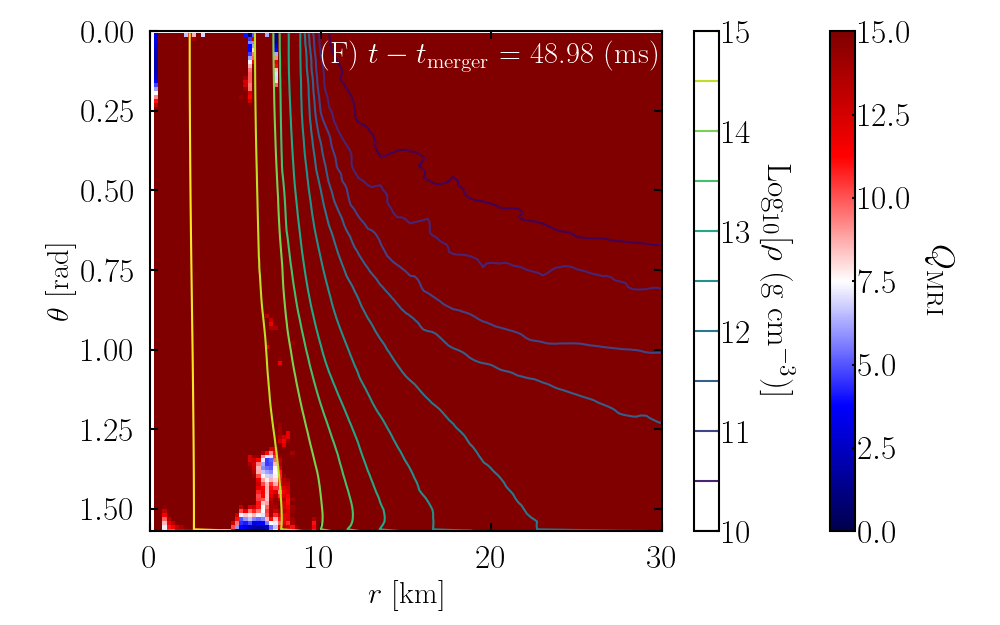} 
	\caption{\textbf{Mean poloidal magnetic field generation.}
    (\textbf{A}) The radial mean magnetic field on top of the rest-mass density contour on the $r$--$\theta$ plane at $t-t_{\rm merger}\approx 0.01$~s.
    (\textbf{B}) The magnetorotational-instability quality factor on top of the rest-mass density contour on the $r$--$\theta$ plane at $t-t_{\rm merger}\approx 0.01$~s.
    (\textbf{C}) The same as (\textbf{A}), but at $t-t_{\rm merger}\approx 0.03$~s.
    (\textbf{D}) The same as (\textbf{B}), but at $t-t_{\rm merger}\approx 0.03$~s.
    (\textbf{E}) The same as (\textbf{A}), but at $t-t_{\rm merger}\approx 0.05$~s.
    (\textbf{F}) The same as (\textbf{B}), but at $t-t_{\rm merger}\approx 0.05$~s. 
    The employed grid resolution is $\Delta x_{16}=12.5$ m. 
    The link to the visualizations:~\url{http://www2.yukawa.kyoto-u.ac.jp/~kenta.kiuchi/anime/FUGAKU2025/out_yuv420p_Mean_Field.mp4} for the mean poloidal magnetic field and ~\url{http://www2.yukawa.kyoto-u.ac.jp/~kenta.kiuchi/anime/FUGAKU2025/out_yuv420p_Qfac.mp4} for the magnetorotational-instability quality factor.
		}
	\label{fig:SM_Mean_Qfac} 
\end{figure}

Figure~\ref{fig:SM_Mean_Qfac} represents the generation of the radial mean magnetic field $\bar{B}_r$ on the $r$--$\theta$ plane in the magnetorotational-instability active region, which is characterized by the region with $\rho \le 10^{14.5}~{\rm g~cm^{-3}}$. We also find that the polarity of the radial mean magnetic field around the rotational axis after the development of the large-scale field for $t-t_{\rm merger}\gtrsim 0.05$~s has the opposite sign in the simulation with $\Delta x_{15}=25$ m. Thus, the polarity is stochastically determined (see the link for the visualization:~\url{http://www2.yukawa.kyoto-u.ac.jp/~kenta.kiuchi/anime/FUGAKU2025/out_yuv420p_Mean_Field_lv16to15.mp4}). It strongly indicates that the large-scale poloidal field does not originate from the initial magnetic field. Otherwise, the polarity should not exhibit the stochastic property. 

\begin{figure}
    \centering
    \includegraphics[width=\linewidth]{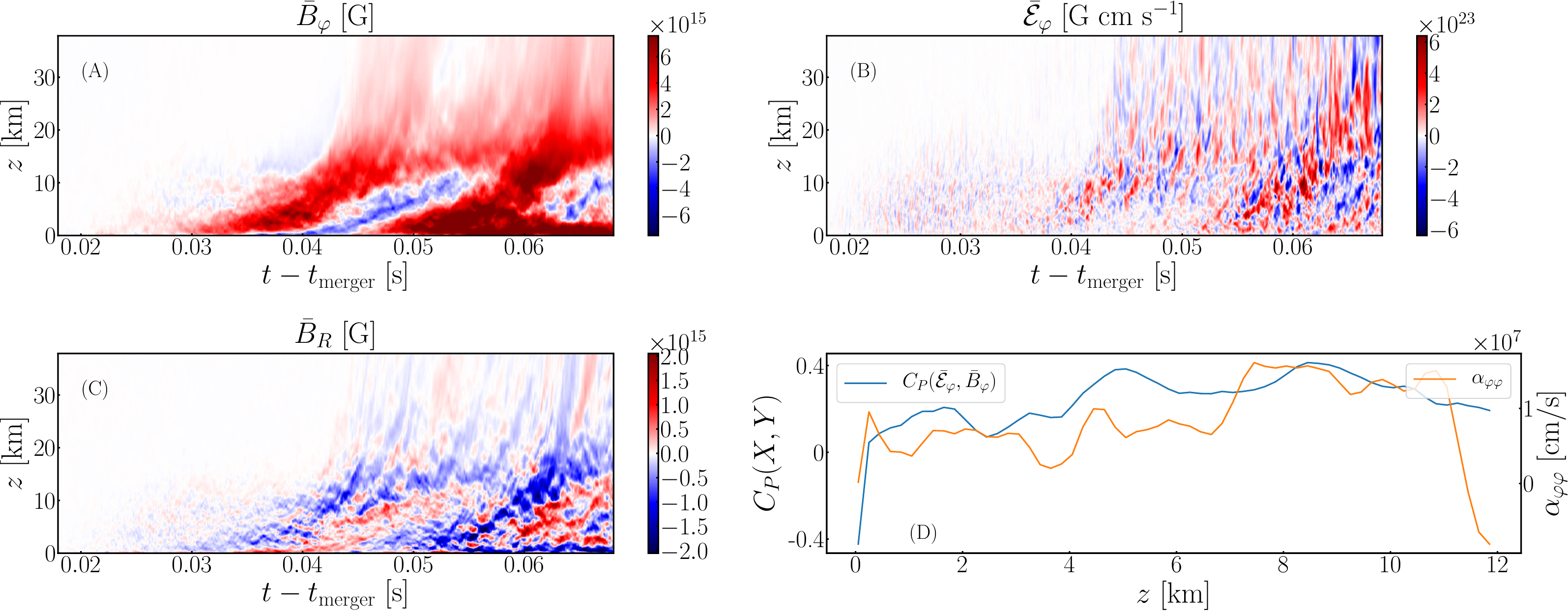}
    \caption{{\bf $\alpha\Omega$ dynamo inside the binary neutron star merger remnant}. Butterfly diagram at R = 20 km: (\textbf{A}) Mean toroidal magnetic field $\bar{B}_\varphi$. (\textbf{B}) Toroidal electromotive force $\bar{\mathcal{E}}_\varphi$. (\textbf{C}) Mean radial magnetic field $\bar{B}_R$. (\textbf{D}) $\alpha_{\varphi\varphi}$ parameters (orange) and correlation between $\bar{\mathcal{E}}_\varphi$ and $\bar{B}_\varphi$ (blue).}
    \label{fig:BF_diagram_20km}
\end{figure}

Figure~\ref{fig:BF_diagram_20km} shows the butterfly diagram for the mean toroidal field, poloidal field, and electromotive force at radius $R=20~{\rm km}$. 
The top-left and bottom-left panels show that the mean toroidal field $\bar{B}_\varphi$ is anti-correlated with $\bar{B}_R$, which indicates the $\Omega$ effect. To quantify the correlation between $\bar{B}_\varphi$ and $\mathcal{\bar{E}}_\varphi$, we compute the Pearson correlation $ C_P (X, Y )$  between the two quantities, X and Y, in the bottom-right panel of fig.~\ref{fig:BF_diagram_20km}. This figure shows that $\bar{B}_\varphi$ and $\mathcal{\bar{E}}_\varphi$ correlate for $z \leq 12$ km, where the density scale height at $R=20~{\rm km}$ is $H=11.1$~km. 
We also computed the diagonal dynamo coefficient $\alpha_{\varphi\varphi}$ with the Singular Value Decomposition method \cite{Racine:2011ApJ} 
and find that $\max(\alpha_{\varphi\varphi})=1.5\times10^7 \rm{cm\ s^{-1}}$ and averaged $\alpha_{\varphi\varphi}=8.6\times 10^6 \rm cm\ s^{-1}$ over a scale height. The period of the butterfly diagram is $~0.017$ s and can be explained by the theoretical $\alpha\Omega$ dynamo period $P_{\alpha\Omega}=2\pi\sqrt{\frac{2}{\alpha_{\varphi\varphi} q\Omega k_z} }=0.018~{\rm s}$, where $q=d\ln\Omega/d\ln R$ and $k_z = 2\pi/H$.  We compare the pattern periods in the simulation and the theoretical predictions at different radii in table \ref{tab:SM_aO} and find a good agreement between theoretical and simulation values. We also investigate the impact of resolution on the dynamo parameters in a simulation with $\Delta x_{15}=25$~m and $t_{\rm rmv}=0.015$~s. We find that the dynamo coefficient slightly decreased overall except at $R=30$ km, where the decrease is significant. 
The small decrease in dynamo efficiency in the region $\rho>\rho_{13}$ corresponding to $R \lesssim 15~{\rm km}$ on the orbital plane is due to the decrease of how well the magnetorotational instability is resolved as seen on the quality factor in fig.~\ref{fig:SM_Qfac}. The steeper decrease at $R=30$ km is of a different origin, as the finest nested domain does not cover this radius. Most likely, this difference comes from the different thermodynamic evolution and contraction of the remnant massive neutron star. It leads to some stochasticity in the dynamo period due to the turbulence and due to the small number of periods evolved in the simulation before the remnant massive neutron star collapses to a black hole. 
Overall, we obtain consistent results for the $\alpha\Omega$ dynamo with our previous study \cite{Kiuchi:2023obe}. 

\begin{table} 
	\centering
	\caption{\textbf{Remnant neutron star $\alpha\Omega$ dynamo.}
    Radius, the dynamo $\alpha$, the wavenumber in the z-direction, the shear rate, $\alpha\Omega$-dynamo period, the period measured in the butterfly diagram until the remnant massive neutron star collapses. Values in parentheses correspond to the simulation with $\Delta x_{15}=25$~m and $t_\mathrm{rmv}=0.015$~s.
    }
	\label{tab:SM_aO} 

	\begin{tabular}{cccccc} 
		\\
		\hline\hline
		$R$ [km] & $\alpha_{\varphi\varphi}$ $[10^6~\rm cm~s^{-1}]$ & $k_z~[10^{-6}{\rm cm^{-1}}]$ & $d\Omega/d\ln R~[{\rm s}^{-1}]$ & $P_{\alpha\Omega}~[{\rm s}]$ & $P_{\rm BF}$ [s]\\
		\hline\hline
        $10$ & $2.6~(1.4)$ & $24~(26)$ & $5359~(5397)$ & $0.015~(0.02)$ & $0.012~(0.02)$ \\
        $20$ & $8.6~(6.2)$ & $5.6~(5.9)$ & $4911~(4534)$ & $0.018~(0.027)$ & $0.017~(0.024)$ \\
        $30$ & $9.6~(3.8)$ & $3.4~(3.4)$ & $3505~(3326)$ & $0.026~(0.043)$ & $0.017~(\geq0.03)$ \\
        \hline
	\end{tabular}
\end{table}

\subsubsection*{Convergence study in numerical relativity simulation}
Because we initialize the magnetic field with an unrealistic strength, it is necessary to show that the large-scale magnetic field generation is not a consequence of the magnetic winding of the initial field. 
For this, we conduct a convergence study because, contrary to the Kelvin-Helmholtz instability and magnetorotational instability, we expect the magnetic winding not to have a steep dependence on the employed grid resolution. To demonstrate it, we run a non-magnetized binary neutron star merger simulation with the same mass, same equation of state, and $\Delta x_{13}=100$~m with $N=385$ until $t-t_{\rm merger}\approx 0.008$~s. Subsequently, we embed the poloidal magnetic field loop with $B_{0,\rm max}=10^{11}~{\rm G}$ inside the remnant massive neutron star (see also Eq.~(\ref{eq:vector})). With this setup, neither the Kelvin-Helmholtz instability nor the magnetorotational instability is responsible for the magnetic field amplification. After initialization of the magnetic field, we run two simulations with $\Delta x_{13}=100$~m with $N=385$ and $\Delta x_{12}=200$~m with $N=217$.

\begin{figure} 
	\centering
	\includegraphics[width=0.495\textwidth]{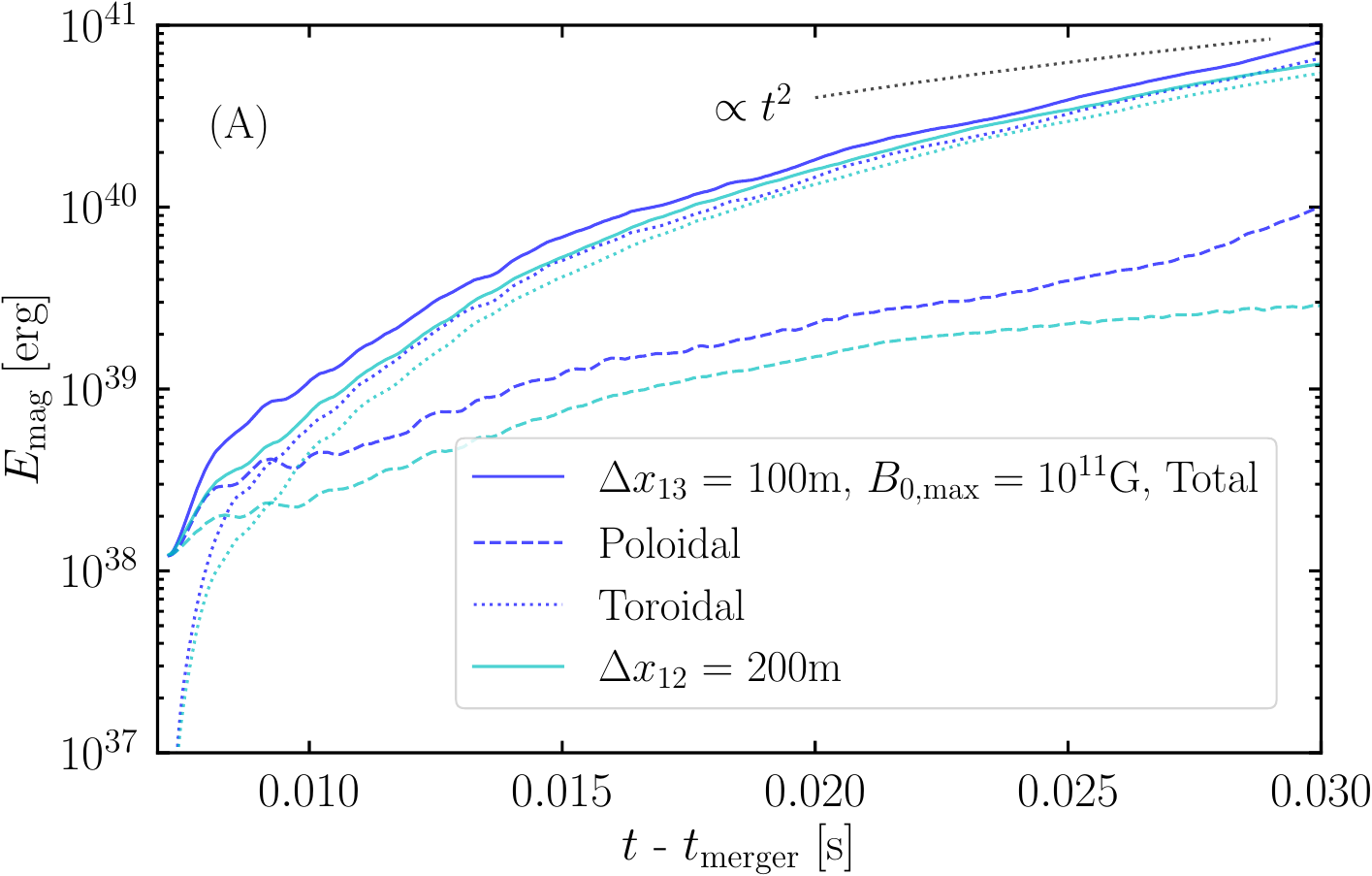}
    \includegraphics[width=0.495\textwidth]{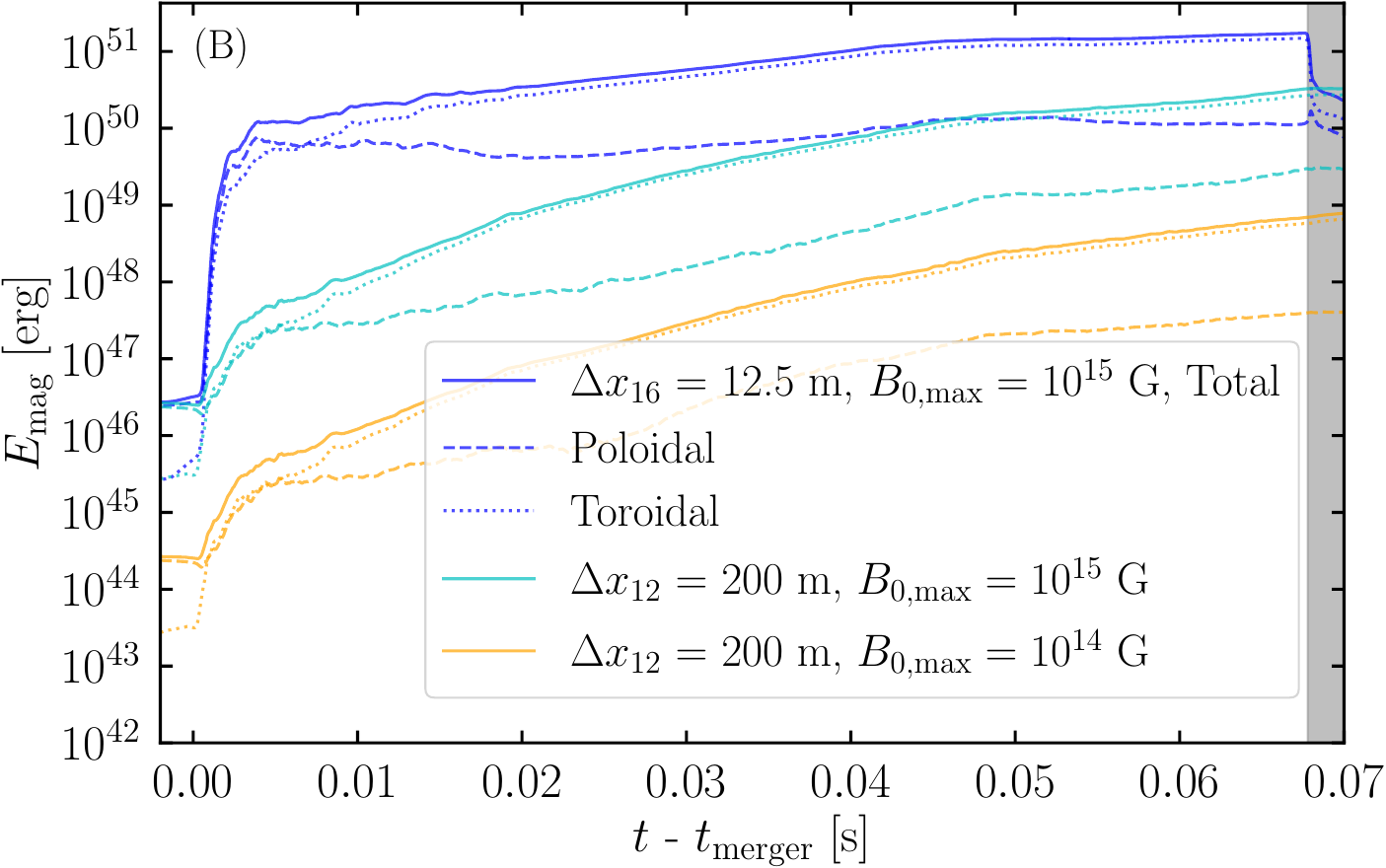}\\
    \includegraphics[width=0.495\textwidth]{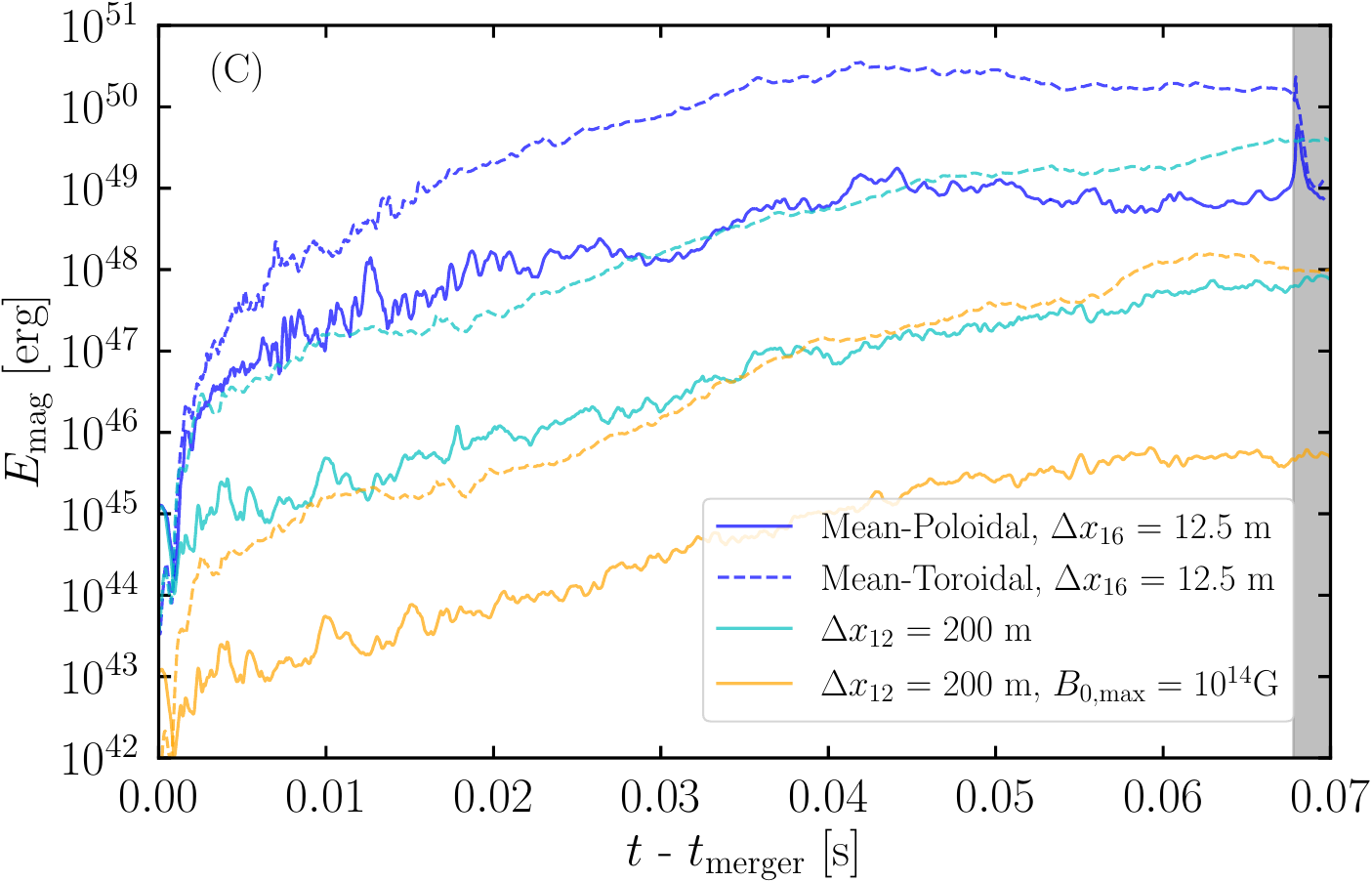} \includegraphics[width=0.495\textwidth]{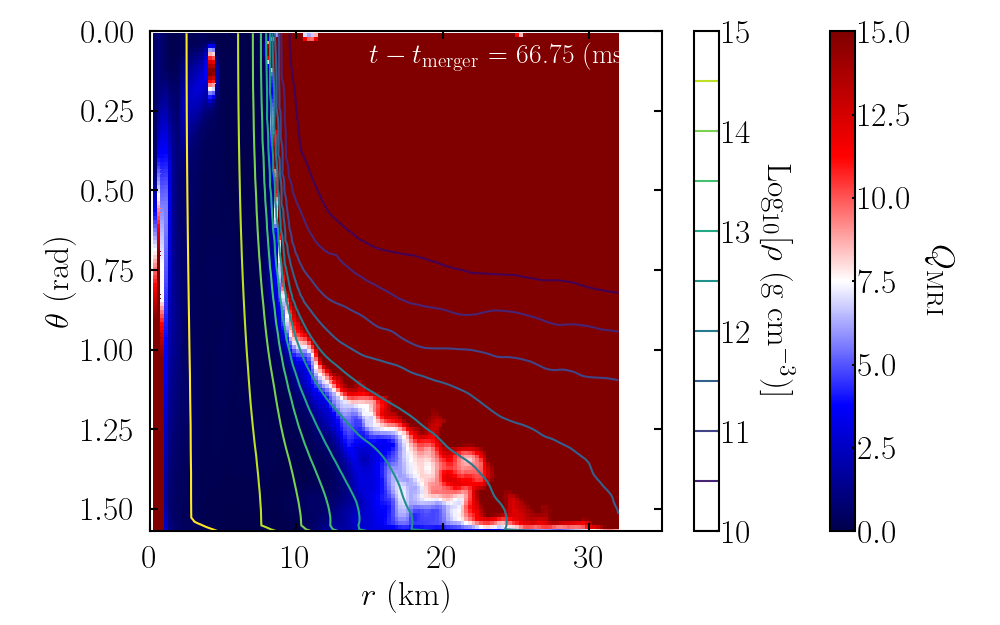}
	\caption{\textbf{Convergence test.}
    ({\bf A}) Electromagnetic energy as a function of the post-merger time in the magnetic winding problem. The magnetic field is initialized inside the merger remnant at $t-t_\mathrm{merger}\approx 0.008$ s. 
    The solid, dashed, and dotted curves present the total, poloidal, and toroidal components, respectively. The blue and cyan colors represent the resolution with $\Delta x_{13}=100$~m with $N=385$ and $\Delta x_{12}=200$~m with $N=217$, respectively. The black dashed curve denotes the guide with $\propto t^2$.
    ({\bf B}) The same as the panel (\textbf{A}), but the magnetic field is initialized during the inspiral phase. The blue and cyan curves denote the simulation with $\Delta x_{16}=12.5$~m with $N=385$ (main paper) and $\Delta x_{12}=200$~m with $N=217$, respectively. The orange curve represents the simulation with $\Delta x_{12}=200$~m, $N=217$ and $B_\mathrm{0,max}=10^{14}$~G. 
    ({\bf C}) Mean magnetic-field energy in the magnetorotational instability-active region as a function of the post-merger time. The solid and dashed curves denote the mean poloidal and toroidal field components, respectively. The legend is the same as the panel (\textbf{B}). 
    ({\bf D}) The magnetorotaional instability quality factor on the $r$--$\theta$ plane at $t-t_\mathrm{merger}\approx 0.07$ s. The employed resolution is $\Delta x_{12}=200$ m and $B_{0,\rm max}=10^{15}~{\rm G}$. 
		}
	\label{fig:Convergence} 
\end{figure}

Figure~\ref{fig:Convergence} (\textbf{A}) plots the electromagnetic energy as a function of the post-merge time in this magnetic winding test. The toroidal magnetic field energy asymptotically approaches the expected magnetic winding, i.e., the energy increases as $t^2$. It should be noted that the perfect match to $\propto t^2$ is not expected because the remnant massive neutron star has a non-axisymmetric structure, particularly for the angular velocity profile. It also demonstrates that the growth due to the magnetic winding is not severely sensitive to the employed grid resolution. For the toroidal-field energy, a $30\%$ difference is observed between the $100$~m and $200$~m simulations at $t-t_{\rm merger} \approx 0.03$~s. For the poloidal-field energy, it is a factor of three. Therefore, we conclude that the grid resolution of $200$~m is sufficient to capture the magnetic winding. 

Then, we conduct a convergence study with $\Delta x_{12}=200$~m with $N=217$ and $B_{0,\rm max}=10^{15}~{\rm G}$ starting from the inspiral phase. We also initialize a weak magnetic field with $B_{\rm 0,max}=10^{14}~{\rm G}$ and $\Delta x_{12}=200$~m. Figure~\ref{fig:Convergence} (\textbf{B}) plots the electromagnetic energy as a function of the post-merger time. Because the energy increase exhibits a steep dependence on the employed grid resolution, i.e., a factor of twenty in the toroidal-field energy and a factor of forty in the poloidal-field energy at $t-t_{\rm merger} \approx 0.03$~s, it suggests that a process other than the winding of the initial magnetic field is responsible for the electromagnetic-field energy increase. We repeat the analysis shown in figs.~\ref{fig:SM_Mean} and \ref{fig:SM_Mean_Qfac}, and fig.~\ref{fig:Convergence} (\textbf{C}) plots the mean-poloidal and toroidal magnetic-field energy contained in the magnetorotational-instability active region. Importantly, the mean-poloidal magnetic-field energy increases even in the low resolution run, and the magnetorotational instability quality factor exceeds the critical value in the region with $\rho \le 10^{12}$--$10^{12.5}~{\rm g~cm^{-3}}$ as shown in the panel (\textbf{D}) (see also the visualization for $\Delta x_{12}=200$~m:~\url{http://www2.yukawa.kyoto-u.ac.jp/~kenta.kiuchi/anime/FUGAKU2025/out_yuv420p_Mean_Field_200m.mp4} and ~\url{http://www2.yukawa.kyoto-u.ac.jp/~kenta.kiuchi/anime/FUGAKU2025/out_yuv420p_Qfac_200m.mp4}). 
We observe that the generated mean poloidal field propagates towards the polar region, and it shows the polarity flip even in the $200$~m resolution. These indicate that the magnetic winding due to the initial field is not responsible for the large-scale field generation. 

\begin{figure} 
	\centering
	\includegraphics[width=0.65\textwidth]{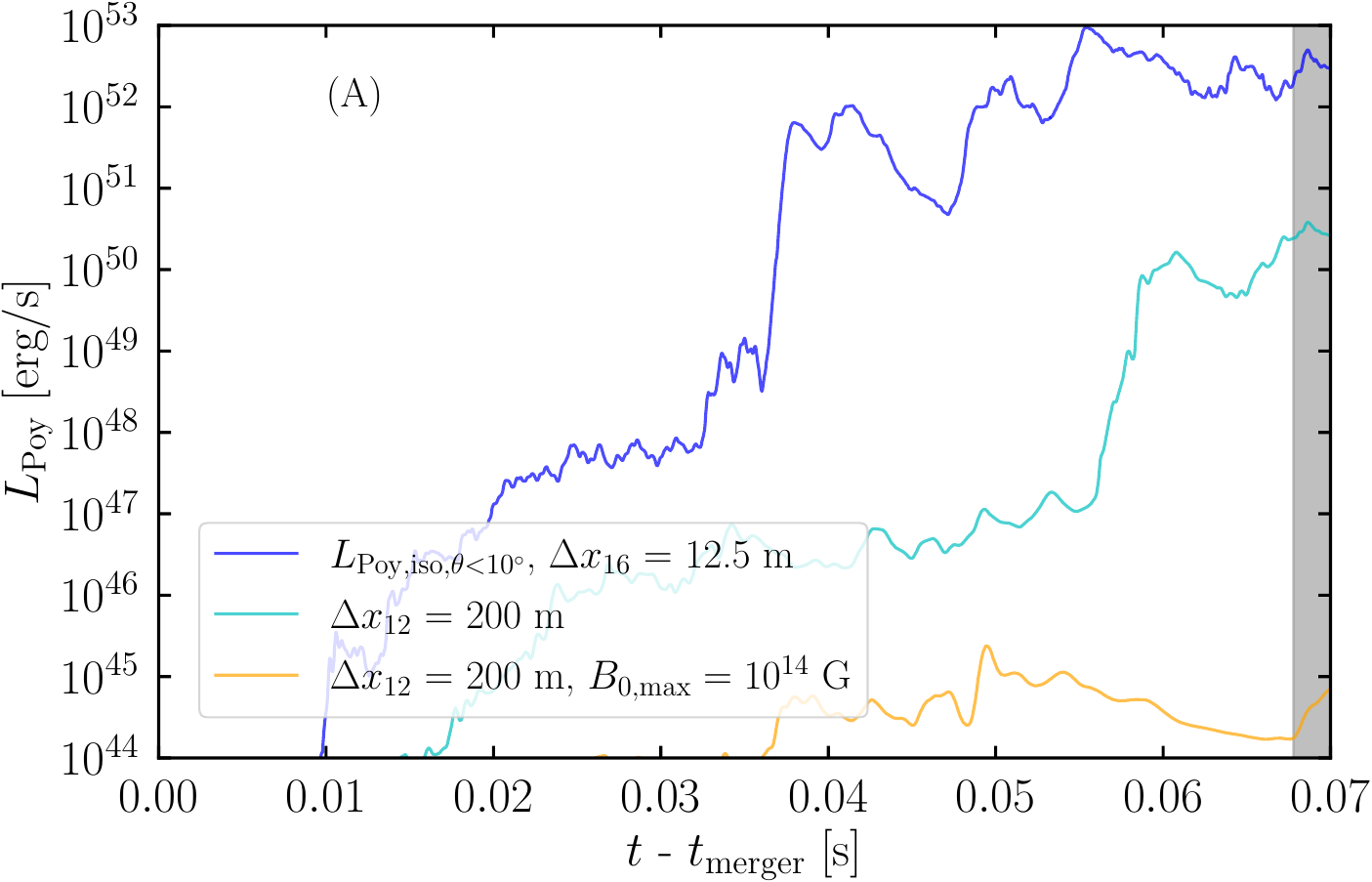}
	\caption{\textbf{Convergence test for Poynting flux.} (\textbf{A}) Isotoropic equivalent Poynting flux luminosity as a function of the post-merger time for the simulation with $\Delta x_{16}=12.5$ m (blue), $\Delta x_{12}=200$ m (cyan), and $\Delta x_{12}=200$ m with $B_{0,\rm{max}}=10^{14}~{\rm G}$ (orange). 
		}
	\label{fig:Convergence2} 
\end{figure}

Finally, we observe a launch of the Poynting-flux dominated outflow in the low resolution simulation as shown in fig.~\ref{fig:Convergence2} (see also the visualization for $\Delta x_{12}=200$~m:~\url{http://www2.yukawa.kyoto-u.ac.jp/~kenta.kiuchi/anime/FUGAKU2025/out_yuv420p_200m.mp4}). 
The caveat for the low resolution run is that the electromagnetic energy is one order of magnitude smaller than the simulation with $\Delta x_{16}=12.5$ m at the formation of the black hole (see figs.~\ref{fig:Convergence} (\textbf{B}) and (\textbf{C})). As a result, the Poynting flux luminosity is two orders of magnitude smaller than the simulation with $\Delta x_{16}=12.5$ m at the formation of the black hole. 

In figs.~\ref{fig:Convergence} and \ref{fig:Convergence2}, we show the result in which we initialize the magnetic field to be $10^{14}~{\rm G}$ and employ the resolution $\Delta x_{12}=200$~m by the orange curves. Inefficient growth of the electromagnetic field energy and lack of the Poynting flux launch during the simulation time are merely caused by the low resolution, not by the weak initial field strength. 

\subsubsection*{Assessment for the removal of the nested domains and Cowling approximation}
\begin{figure} 
	\centering
	\includegraphics[width=0.495\textwidth]{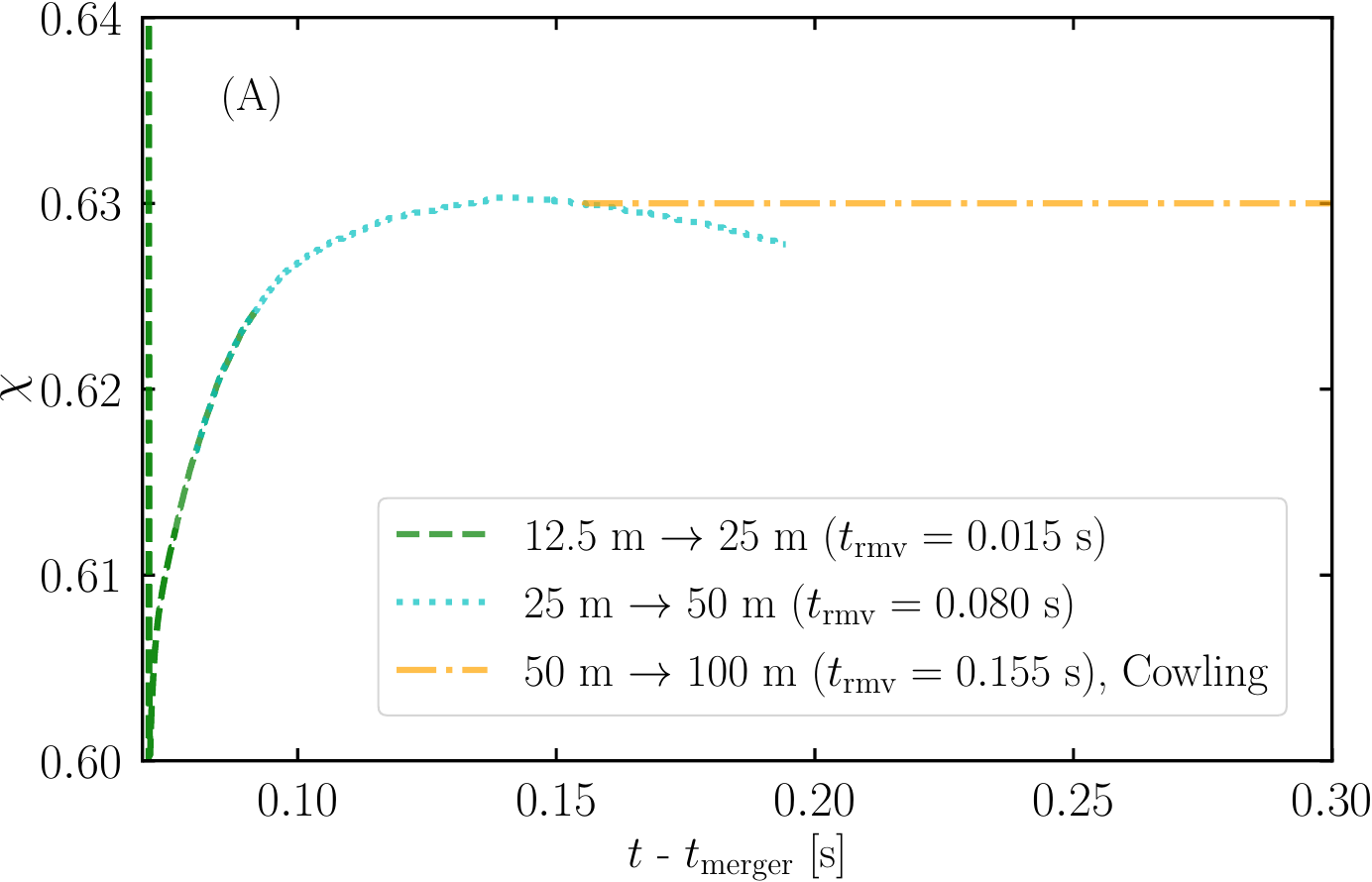}
    \includegraphics[width=0.495\textwidth]{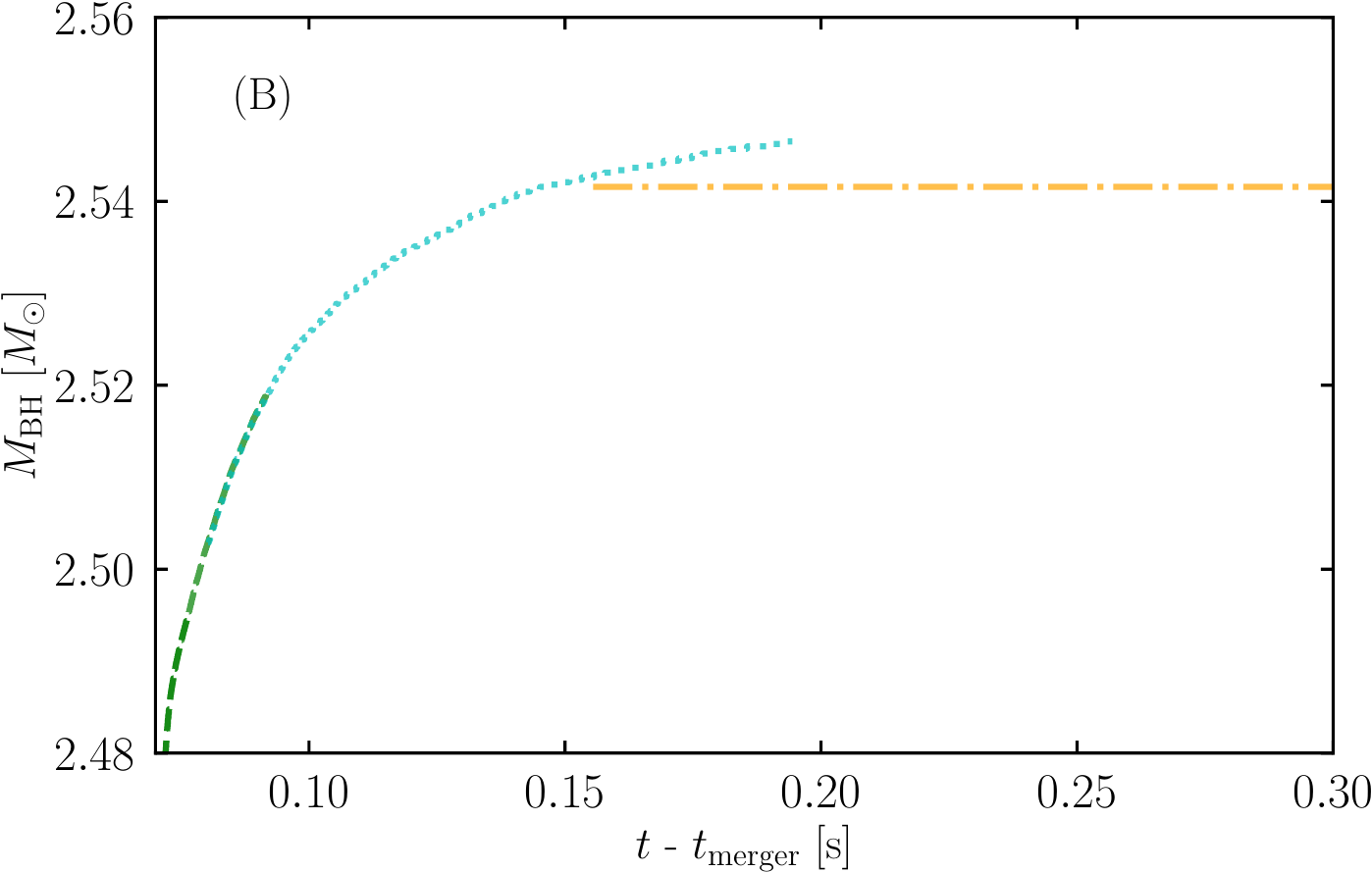}\\
    \includegraphics[width=0.495\textwidth]{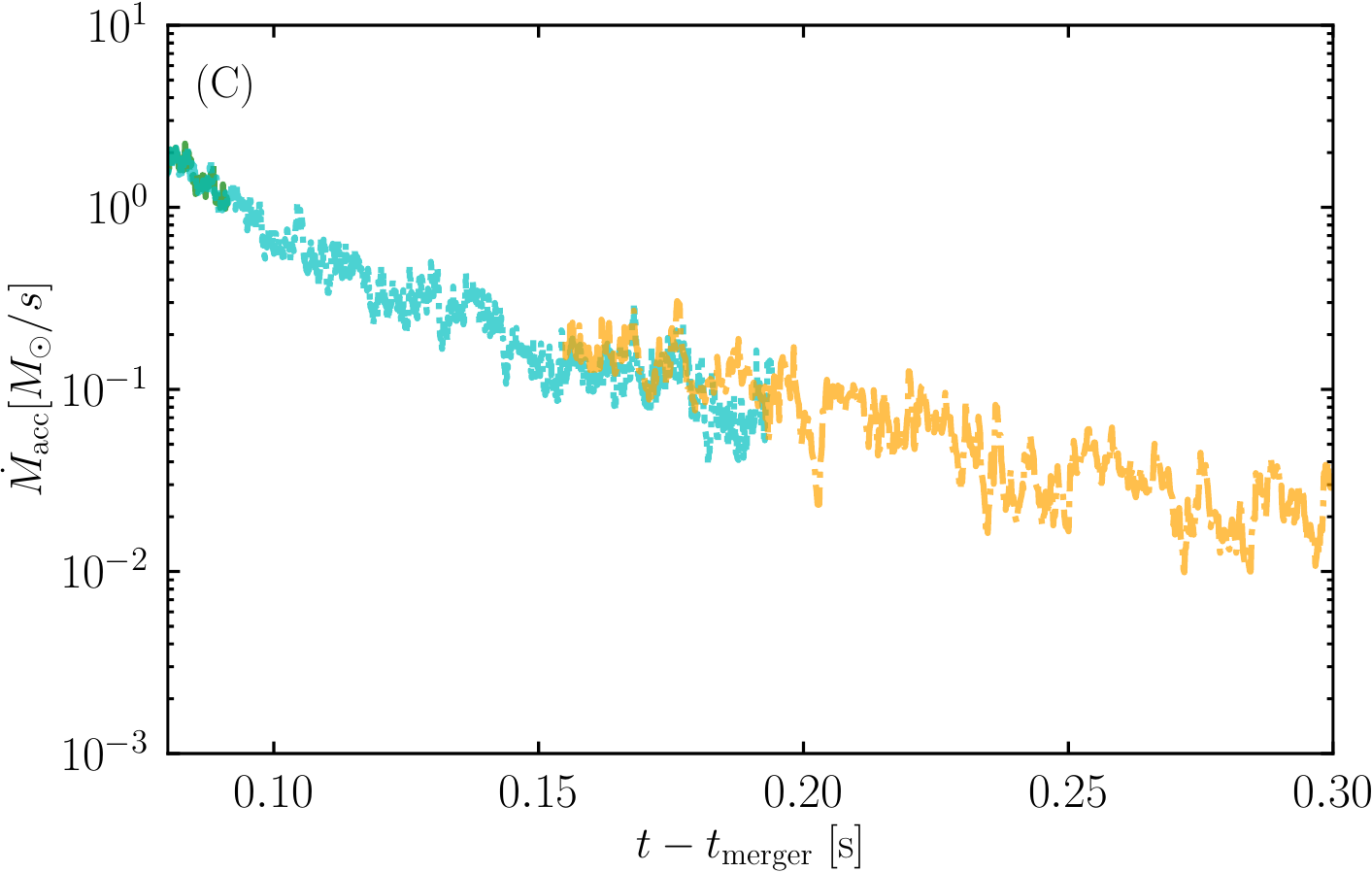}
    \includegraphics[width=0.495\textwidth]{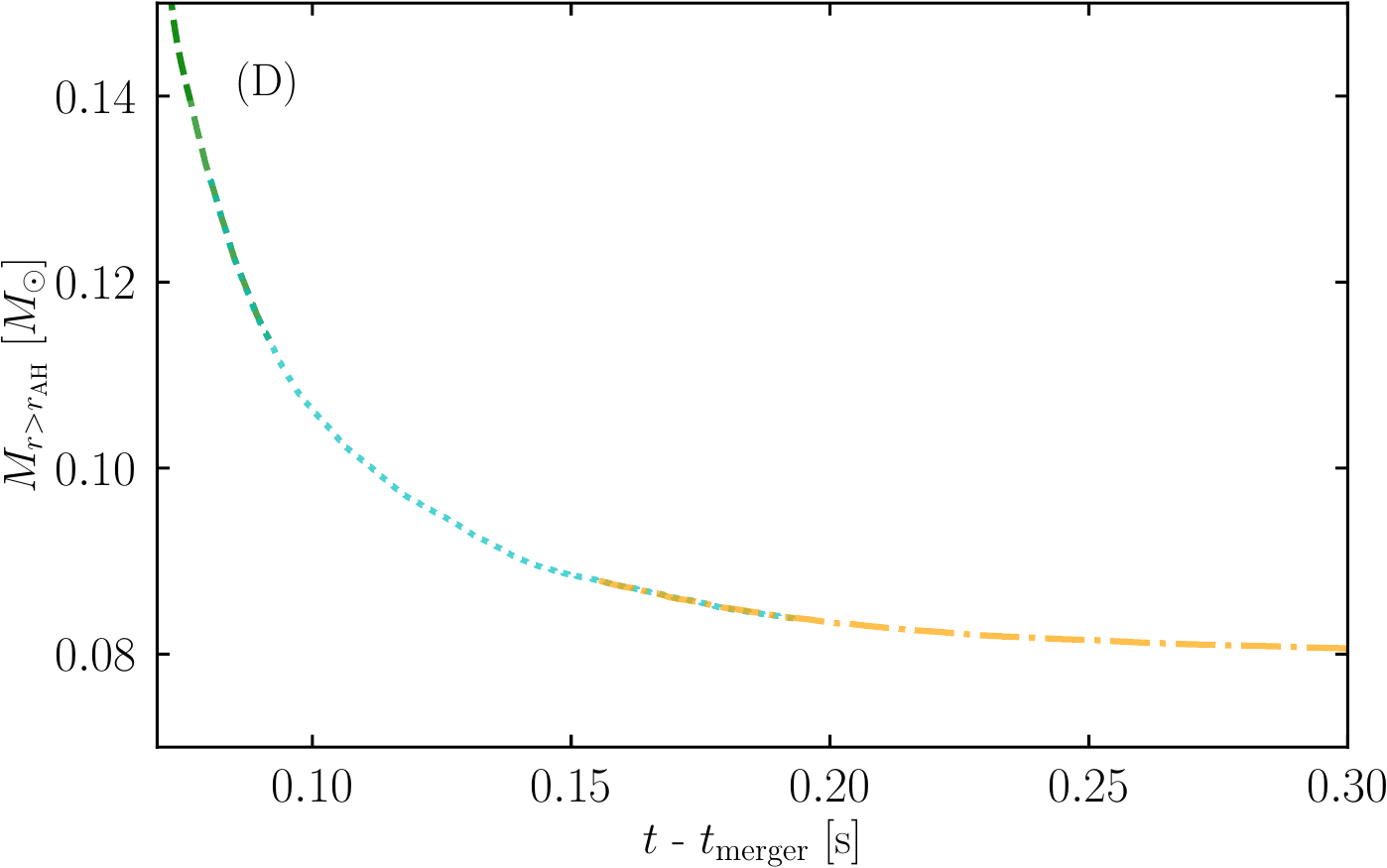}\\
    \includegraphics[width=0.495\textwidth]{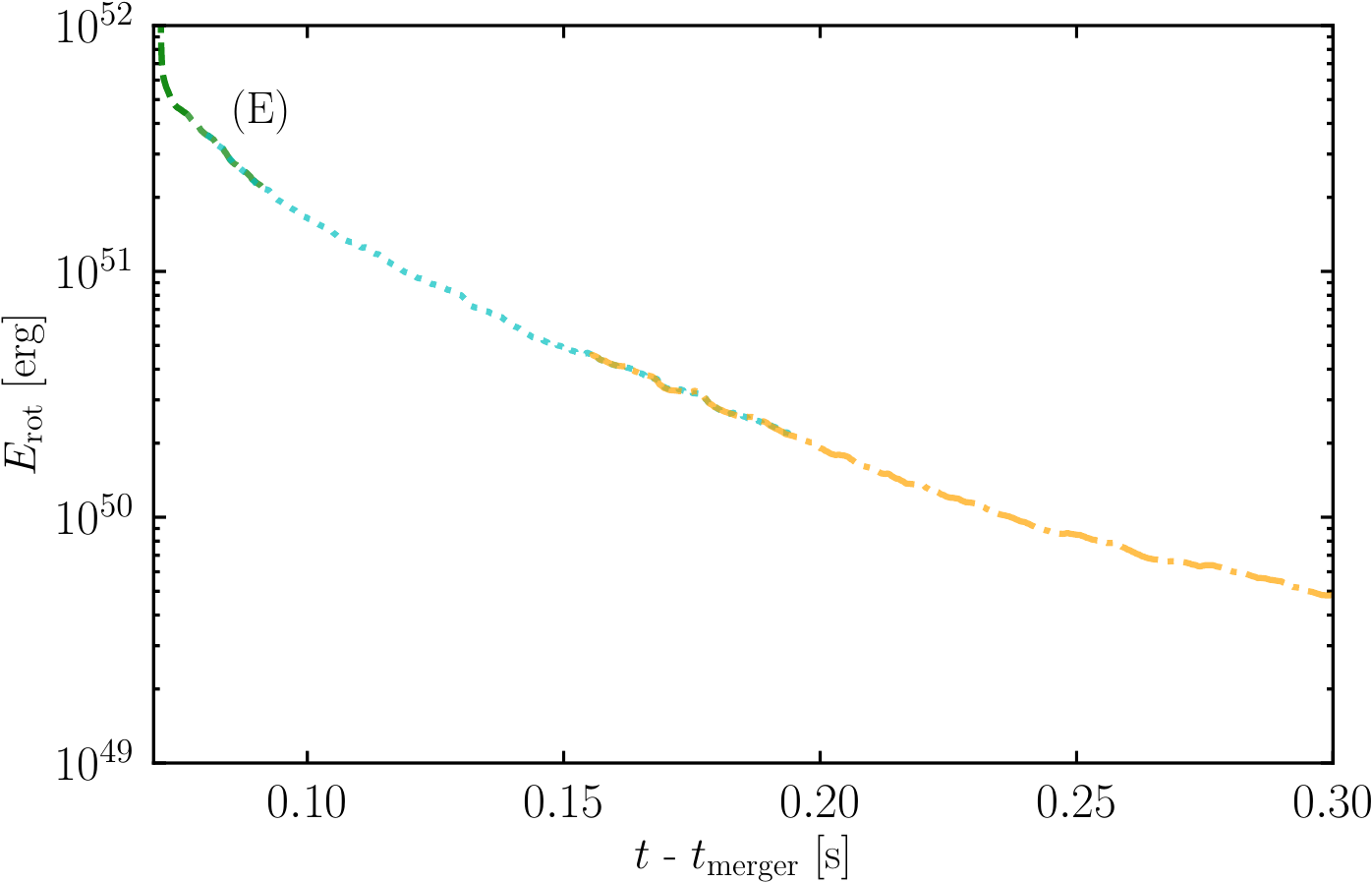}
    \includegraphics[width=0.495\textwidth]{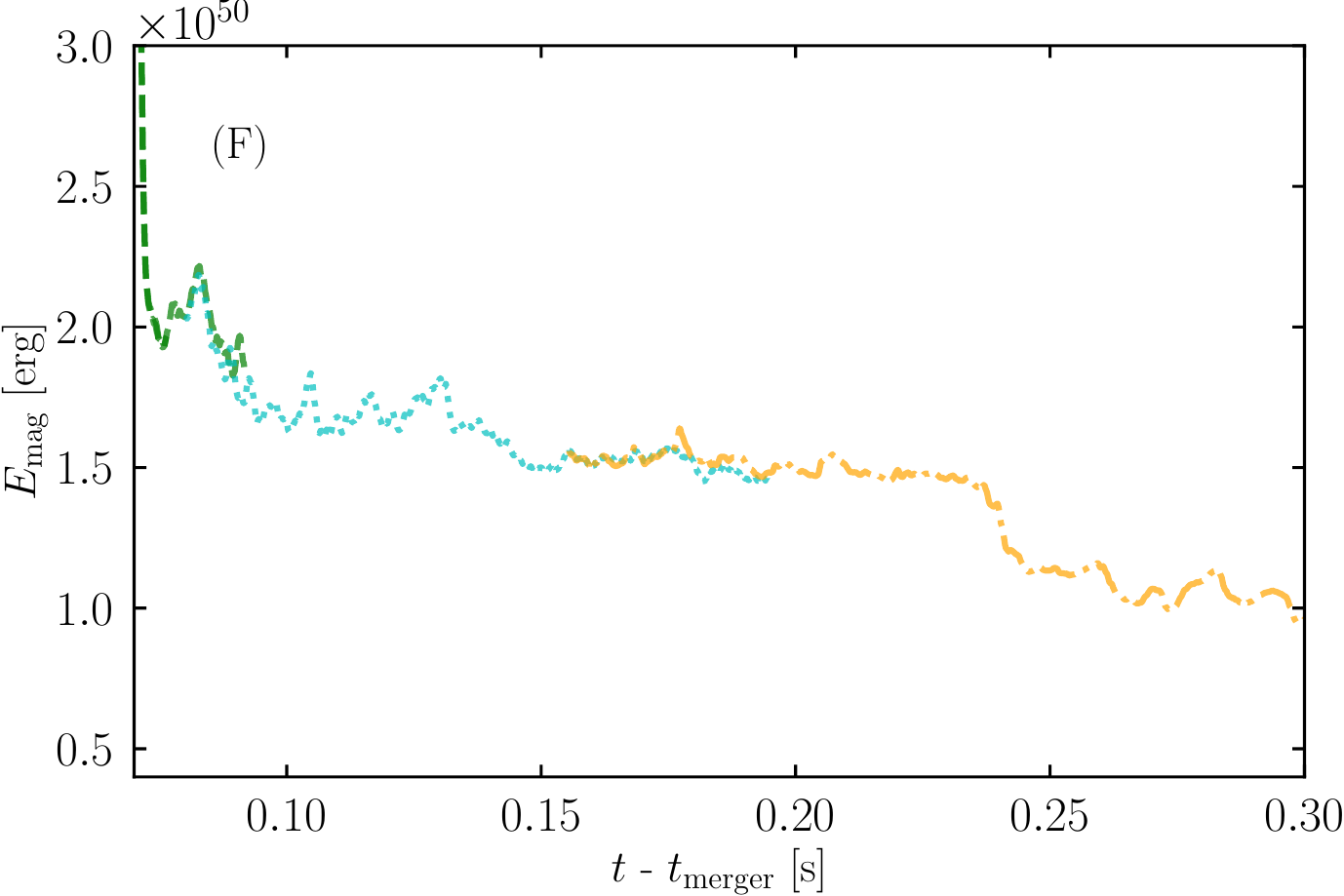} 
	\caption{\textbf{Properties of the black hole and accretion torus.}
    (\textbf{A}) Non-dimensional black hole spin. 
    (\textbf{B}) Black hole mass.
    (\textbf{C}) Mass accretion rate onto the black hole. 
    (\textbf{D}) Baryonic mass outside of the black hole. 
    (\textbf{E}) Rotational energy of the accretion torus. 
    (\textbf{F}) Electromagnetic energy.
    The horizontal axis is the post-merger time. 
    The legends for the dashed, dotted, and dashed-dotted curves are the same as fig.~\ref{fig:SM_Qfac} (\textbf{B}).
		}
	\label{fig:SM_BH_torus} 
\end{figure}

Figure~\ref{fig:SM_BH_torus} plots several relevant quantities describing the black hole and massive torus properties as a function of the post-merger time. The differences are barely observed after the removal of the nested domain(s) and after applying the Cowling approximation, except for the black hole spin and mass. By construction, both quantities stay constant in the Cowling approximation. These observables validate the removal strategy. We also estimate the energy conservation error by
\begin{align}
\Delta E_{\rm error} = M_{\rm ADM,0}c^2 - \left(M_{\rm BH}c^2+M_{r>r_{\rm AH}}c^2+E_{\rm GW}+E_{\nu} + M_{\rm acc,t>t_{Cowling}}c^2\right),
\end{align}
where $M_{\rm ADM,0},E_{\rm GW}, E_\nu$, and $M_{\rm acc,t>t_{Cowling}}$ denote the initial Arnowitt-Deser-Misner mass of the binary, the energy carried by gravitational waves for $l=|m|=2$ mode, the energy carried by the neutrino, and the baryonic mass accreted onto the black hole after the Cowling approximation is applied, respectively. We note that the gravitational binding energy of the accretion torus and ejecta is neglected. 

Similarly, the angular momentum conservation error is estimated by
\begin{align}
\Delta J_{\rm error} = J_{\rm ADM-like,0} - \left(J_{\rm BH}+J_{r>r_{\rm AH}}+J_{\rm GW}+J_{\nu} + J_{\rm acc,t>t_{Cowling}}\right),
\end{align}
Table~\ref{tab:SM_conservation} shows that energy and angular momentum conservation are satisfied with $\approx 0.3\%$ and $1.8\%$ error, respectively, at the end of the simulation $t-t_{\rm merger}\approx 0.7$~s. 

\begin{figure} 
	\centering
	\includegraphics[width=0.65\textwidth]{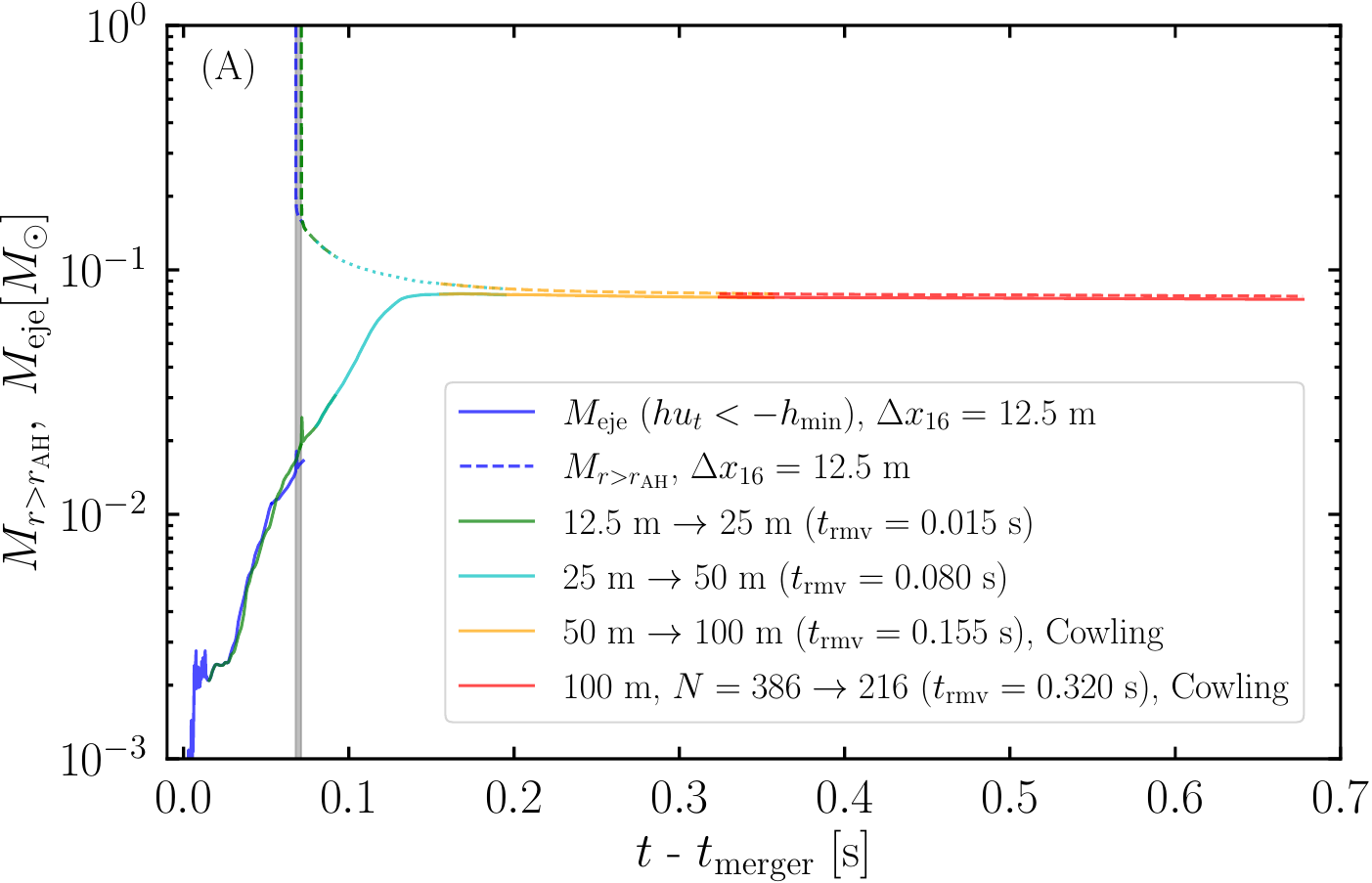}
	\caption{\textbf{Long-term evolution of ejecta}
    (\textbf{A}) Ejecta (solid) and baryonic mass outside the apparent horizon (dashed). The legend is the same as fig.~\ref{fig:SM_Qfac} (\textbf{B}), but red for $\Delta x_{13}=100$~m with $N=216$. 
		}
	\label{fig:SM_eje} 
\end{figure}

Figure~\ref{fig:SM_eje} plots the long-term evolution up to $t-t_{\rm merger}\approx 0.7$~s for the ejecta and baryonic mass outside the apparent horizon. It shows that there is no noticeable difference after the nested domain removal and/or the grid-point  reduction.

Finally, we remark on the convergence study in the numerical relativity simulation in the main text. Strictly speaking, the nested domain removal is not a convergence study; nonetheless, it gives a sense of the extent to which the result is sensitive to the employed grid resolution. The mean-field generation during the remnant massive neutron star, particularly, the poloidal component,  and the resultant Poynting flux are not very sensitive to the grid resolution except for the time variability (see Fig.~\ref{fig:main_MMA} (\textbf{A}) and fig.~\ref{fig:SM_Mean} (\textbf{B})). The collapsing time to the black hole is $t-t_{\rm merger} \approx 0.068~{\rm s}$ for $\Delta x_{16}=12.5~{\rm m}$ and $t-t_{\rm merger} \approx 0.071~{\rm s}$ for $\Delta x_{15}=25~{\rm m}$, respectively. The post-merger ejecta mass at the black hole formation is $\approx 0.015 M_\odot$ for $\Delta x_{16}=12.5~{\rm m}$ and $\approx 0.019 M_\odot$ for $\Delta x_{15}=25~{\rm m}$, respectively. Overall, the result is not very sensitive to the employed resolution. 

\begin{table} 
	\centering
	\caption{\textbf{Energy and angular momentum conservations.}
    The first and second columns denote the initial ADM mass, black hole mass, the baryonic mass outside the black hole, the energy emitted by gravitational waves and neutrinos, and the baryonic mass accreted onto the black hole after the Cowling approximation is applied. The final column gives the energy conservation error. The unit is $M_\odot c^2$. They are evaluated at $t-t_{\rm merger}\approx 0.7$~s. 
    The third and fourth columns correspond to the angular momentum counterpart, and the unit is ${\rm G} M_\odot^2/c$.
    }

	\begin{tabular}{ccccccc} 
		\\
		\hline\hline
		$M_{\rm ADM,0}{\rm c}^2$& $M_{\rm BH}{\rm c^2}$ & $M_{r>r_{\rm AH}}{\rm c^2}$&$E_{\rm GW}$ & $E_\nu$ & $M_{\rm acc,t>t_{Cowling}}{\rm c^2}$ & $\Delta E_{\rm error}$\\
		\hline\hline
 		$2.7030$ & $2.5415$ & $0.0787$ & $0.0494$ & $0.0137$ & $0.0105$ & $0.0092$ \\       
        \hline\hline
        $J_{\rm ADM-like,0}$ & $J_{\rm BH}$ & $J_{r>r_{\rm AH}}$&$J_{\rm GW}$ & $J_\nu$ & $J_{\rm acc,t>t_{Cowling}}$ & $\Delta J_{\rm error}$\\
        \hline\hline
        $7.3855$ & $4.0695$ & $1.1966$ & $1.8876$ & $0.0296$ & $0.0724$ & $0.1305$\\
		\hline
	\end{tabular}
    \label{tab:SM_conservation}
\end{table}



\subsubsection*{Nucleosynthesis}

\begin{figure} 
	\centering
	\includegraphics[width=0.495\textwidth]{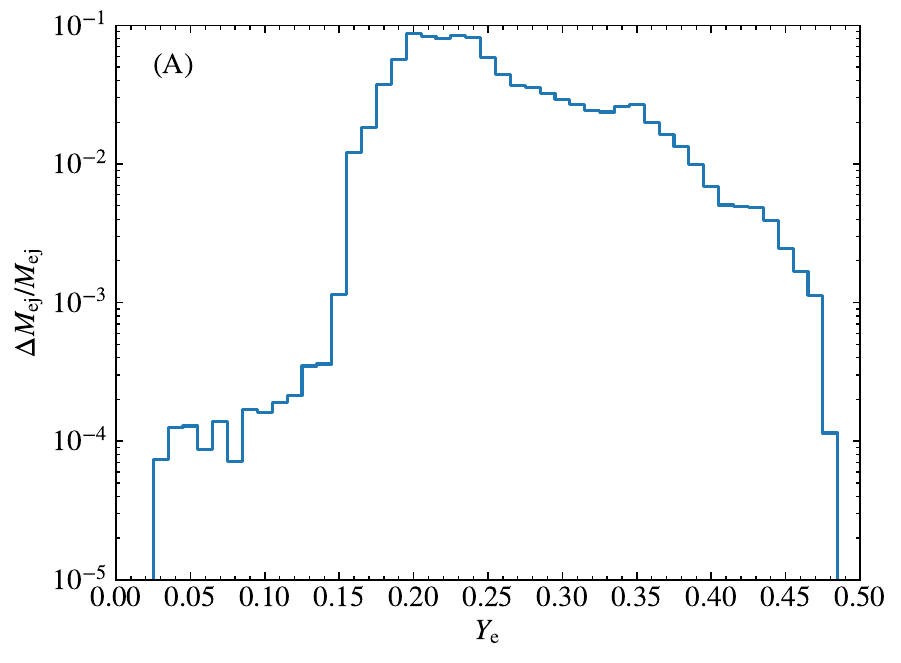}     
	\includegraphics[width=0.495\textwidth]{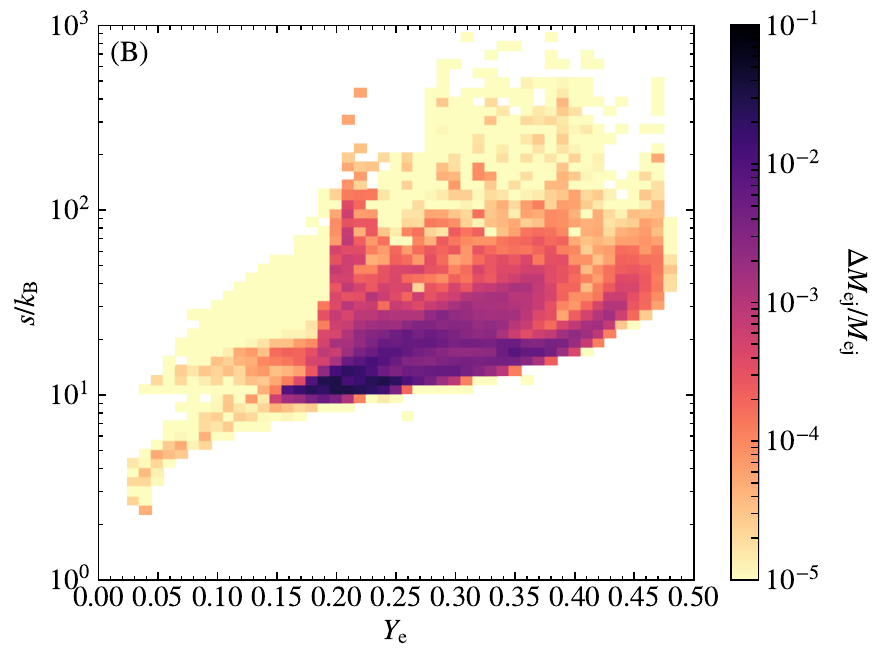}    
	\caption{\textbf{Nucleosynthesis-relevant ejecta properties.} (\textbf{A}) Ejecta-mass histogram of electron fraction evaluated at which the temperature in each particle decreases to 5~GK. (\textbf{B}) Two-dimensional ejecta-mass histogram of electron fraction and entropy per nucleon (in units of Boltzmann constant $k_\mathrm{B}$), both evaluated at 5~GK.
    }
	\label{fig:SM_hist} 
\end{figure}
Figure~\ref{fig:SM_hist} (\textbf{A}) shows the ejecta-mass histogram of electron fraction $Y_\mathrm{e}$ evaluated at which the temperature in each tracer particle decreases to 5~GK. We find a distribution of $Y_\mathrm{e} = 0.03$--0.48 with a peak at $\sim 0.2$. This is different from our previous results 
based on the axisymmetric viscous hydrodynamics simulation of the merger remnant in which the distribution peaks at $Y_\mathrm{e} \sim 0.3$ and extends over a slightly proton-rich region \cite{Fujibayashi:2020qda}. Figure~\ref{fig:SM_hist} (\textbf{B}) displays the two-dimensional ejecta-mass histogram of electron fraction and entropy per nucleon $s$ (in units of Boltzmann constant $k_\mathrm{B}$), both evaluated at which the temperature in each tracer particle decreases to 5~GK. We find a broad correlation between electron fraction and entropy, since both are increased by the heating, either due to 
the magneto-turbulent viscosity or neutrinos, as similarly found in our previous results 
based on the axisymmetric viscous hydrodynamics simulations~\cite{Fujibayashi:2020qda}. However, we also find a branch of $Y_\mathrm{e} \sim 0.2$ extending from $s/k_\mathrm{B} \sim 10$ to a few 100 owing to the mass ejection induced by the Lorentz force exerted by the global magnetic field, which is absent in the previous models~\cite{Fujibayashi:2020qda}.

\begin{figure} 
	\centering
	\includegraphics[width=0.5\textwidth]{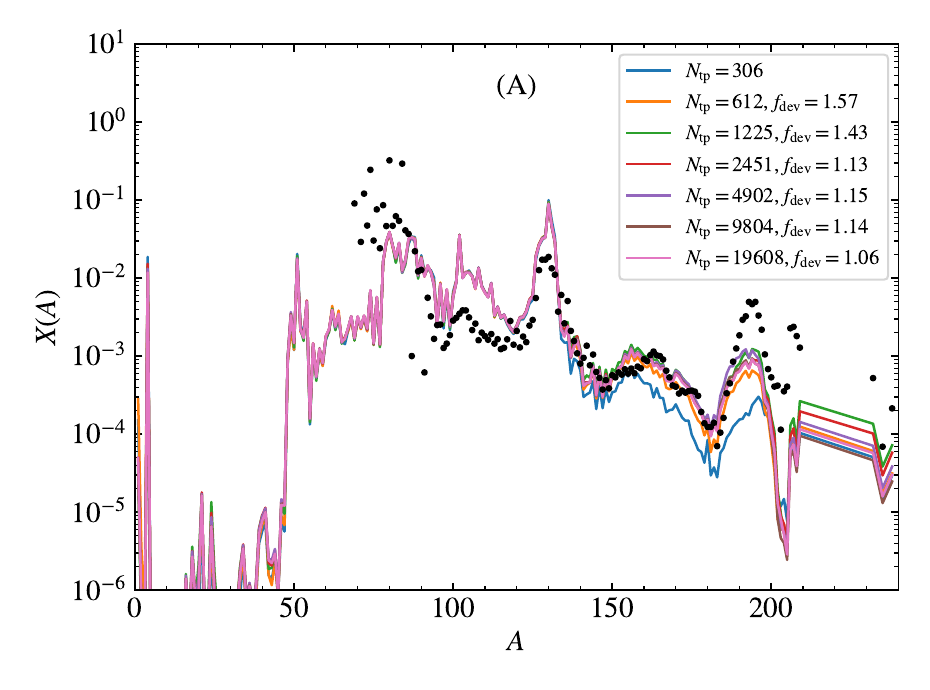}
	\includegraphics[width=0.48\textwidth]{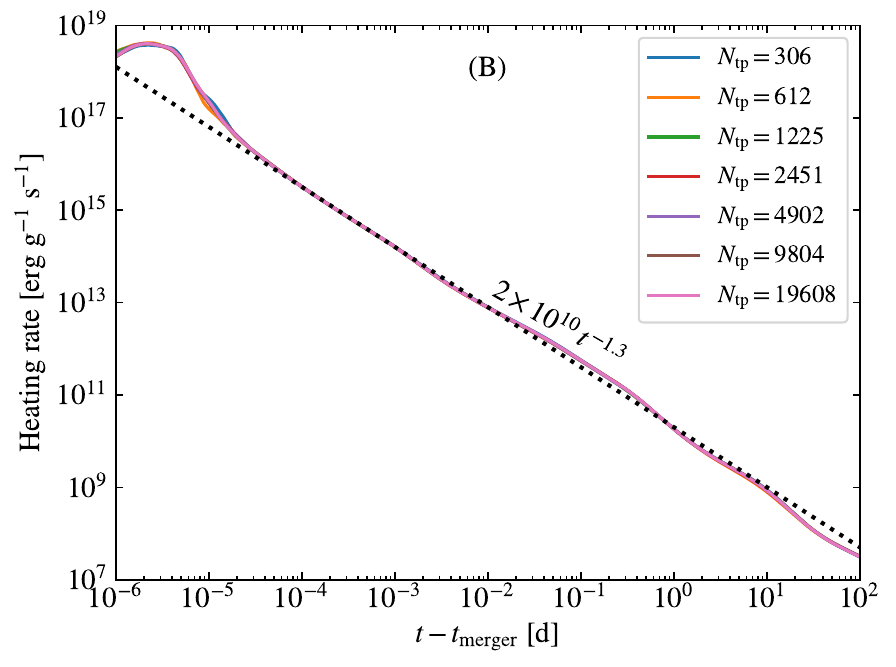} 
	\caption{\textbf{Nucleosynthetic yields and radioactive heating rates.} (\textbf{A}) Mass fractions averaged for the ejecta as a function of atomic mass number $A$ at the end of nucleosynthesis calculations (1~Gyr), which are compared with the solar $r$-process distribution \cite{Prantzos2020} (circles) scaled to match the result ($N_\mathrm{tp} = 19608$) at $A = 151$. The unstable nuclei, except for Th and U, are assumed to have $\alpha$-decayed. Each curve indicates the result with selected tracer particles of $N_\mathrm{tp}$ specified in the legend. (\textbf{B}) Radioactive heating rates averaged for the ejecta as a function of time. The resulting curves are nearly independent of $N_\mathrm{tp}$ and thus overlapped, which approximately decay as $\propto t^{-1.3}$ (dotted line). 
    }
	\label{fig:SM_nucl_conv} 
\end{figure}

The main peak formation at $Y_\mathrm{e}\approx 0.2$ in the mass-histogram of the electron fraction in fig. \ref{fig:SM_hist}~(\textbf{A}) reflects the freeze-out 
condition of $Y_\mathrm{e}$. At high temperatures, $Y_\mathrm{e}$ approximately follows the local equilibrium established by electron and positron captures. As the ejecta expand and these reactions become too slow to maintain equilibrium, $Y_\mathrm{e}$ freezes out. Similar behavior has been discussed in previous studies (e.g., \cite{2008MNRAS.390..781M, Fujibayashi:2020qda, 2022MNRAS.509.1377J}). The systematically lower electron fraction in the present simulation than in models employing an effective turbulent viscosity description can be attributed, at least in part, to the more rapid expansion of the post-merger ejecta owing to the global magnetic field. Freeze-out therefore occurs while the density is still high enough for the electrons to remain degenerate, so that the equilibrium value of $Y_\mathrm{e}$ is still low when it is imprinted on the ejecta.

The effects of neutrino absorption are not taken into account in the above interpretation of the freeze-out of the main $Y_\mathrm{e}\approx 0.2$ component. Nevertheless, this approximation is expected to be reasonable around the freeze-out of this component because the ejecta have already reached distances of $\sim100$\,km from the central remnant, where the neutrino-absorption timescale is longer than the relevant expansion timescale. Neutrino absorption may instead have a more significant impact during the earlier evolution, closer to the remnant. In particular, the ejecta with $Y_\mathrm{e}\gtrsim 0.35$, which constitute a subdominant component distinct from the main $Y_\mathrm{e}\approx0.2$ ejecta, are more strongly affected by neutrino irradiation.


The tracer-particle analysis indicates that the branch of $Y_\mathrm{e}\approx0.2$ and $s/k_\mathrm{B}\approx10$--300 in fig.~\ref{fig:SM_hist}~(\textbf{B}) is a component of the ejecta toward the polar direction with a high velocity with $(0.1-0.5)c$ (see fig.~\ref{fig:SM_ejeprof}). However, this component is not initially ejected along the polar axis. Instead, it is first expelled away from the polar axis during the early post-merger phase, when its electron fraction is largely set to $\approx 0.2$ by the freeze-out process described above. The material subsequently falls back toward the central remnant and is re-ejected at high velocity toward the polar region by magnetohydrodynamic processes. Therefore, a further detailed analysis of the thermodynamic histories of fluid elements is required for a comprehensive understanding of how $Y_\mathrm{e}$ is determined.

Figure~\ref{fig:SM_nucl_conv} (\textbf{A}) shows mass fractions as a function of atomic mass number $A$ at the end of nucleosynthesis calculations (1~Gry), which is mass-averaged for the ejecta. The unstable nuclei of $A > 209$, except for long-lived Th ($A = 232$) and U ($A = 235$ and 238), are assumed to have $\alpha$-decayed. Each color indicates the result with the number of trajectories $N_\mathrm{tp}$ specified in the legend, in which $N_\mathrm{tp}$ increases by approximately a multiplier 2 from 306 (step~0) to 19608 (step~6). 
In the legend, the deviation factor for each step ($n$) with respect to the previous step ($n-1$), defined as $f_\mathrm{dev} = 10^\sigma$; $\sigma^2 = \sum_A \left[\log_{10} X^{(n)}(A) - \log_{10} X^{(n-1)}(A)   \right]^2/N_\mathrm{tot}$, is provided,
where $X^{(n)}(A)$ is the calculated mass fraction of isobar $A$ at step~$n$ and $N_\mathrm{tot}$ the total number of isobars with $X(A) > 10^{-5}$. We find that the abundance distributions are well converged for $N_\mathrm{tr} \ge 2451$ (step~3). For the subsequent HD and kilonova light curve simulations, we adopt the result with $N_\mathrm{tp} = 19608$ (step~6), in which the deviation from step~5 is sufficiently small ($f_\mathrm{dev} = 1.06$). We find that the adopted yields reasonably reproduce the distribution of the solar $r$-process abundances \cite{Prantzos2020} (circles; shifted to match the calculation at $A = 151$) around the first peak ($A \sim 80$--90) and rare-earth ($A \sim 140$--180) regions, although the regions around $A \sim 100$/130 and 195 are somewhat overproduced and underproduced, respectively. 

The mass fraction of He is 0.012, which is 4 times smaller than that in our previous result based on the axisymmetric viscous hydrodynamics simulations for the case with a long-lived remnant massive neutron star~\cite{Fujibayashi:2020qda}. The mass of $^{56}$Ni is negligible in our present model. This is due to the fact that the amount of ejecta with $Y_\mathrm{e} > 0.45$ and $Y_\mathrm{e} > 0.48$ required for the production of He and $^{56}$Ni, respectively, is small and absent in our result. 

Figure~\ref{fig:SM_nucl_conv} (\textbf{B}) displays the radioactive heating rates as a function of time, which are mass-averaged for the ejecta. The heating is due to $\beta$-decay, $\alpha$-decay, and nuclear fission, although the contributions from the latter two are subdominant because of the small production of trans-lead elements. The heating rate already has converged with a small number of tracer particles, which justifies the use of nucleosynthesis yields with $N_\mathrm{tp} = 19608$ for the subsequent HD and kilonova light curve simulations.

\subsubsection*{Mapping errors}



\begin{figure} 
	\centering
	\includegraphics[width=0.48\textwidth]{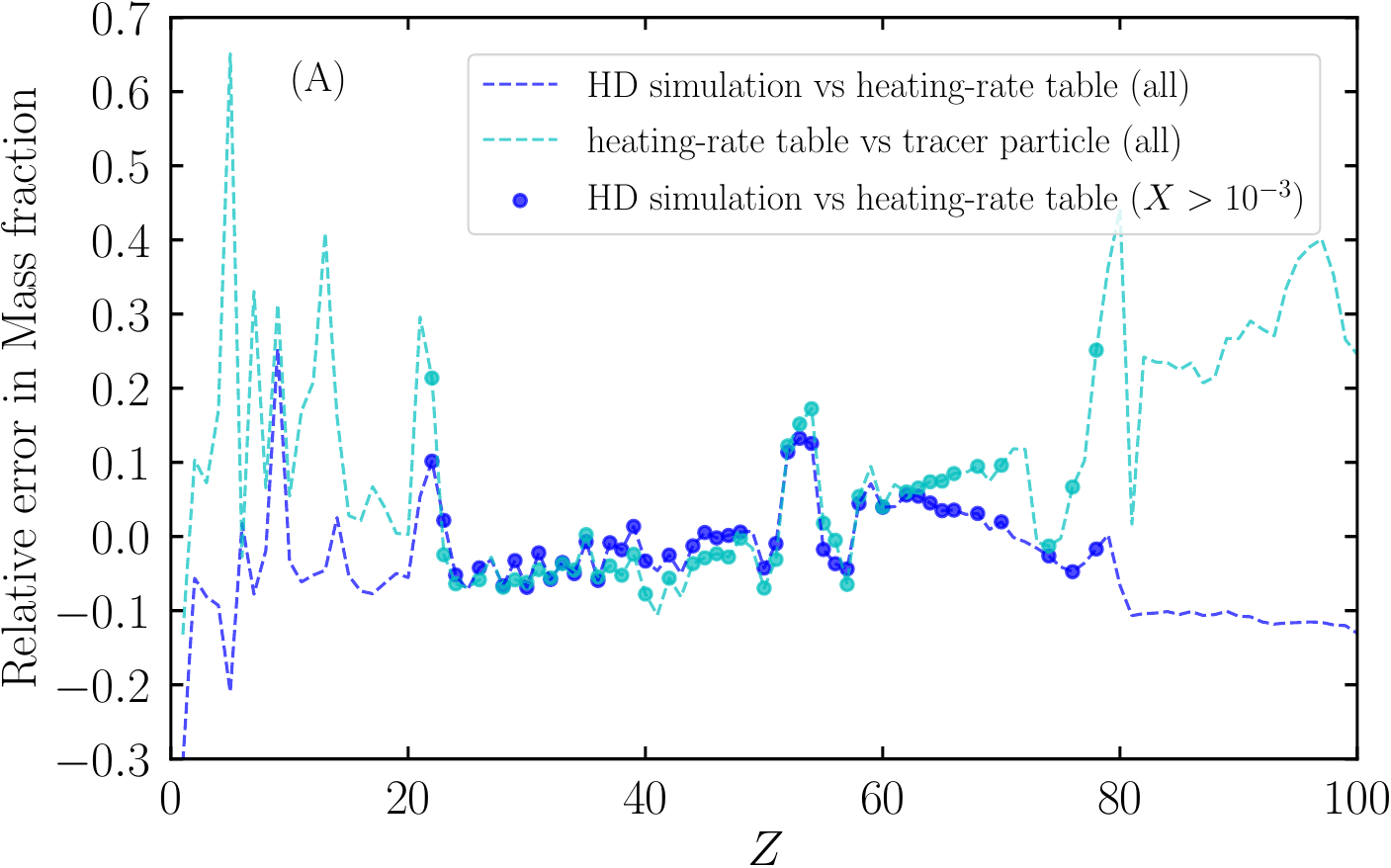}
	\includegraphics[width=0.48\textwidth]{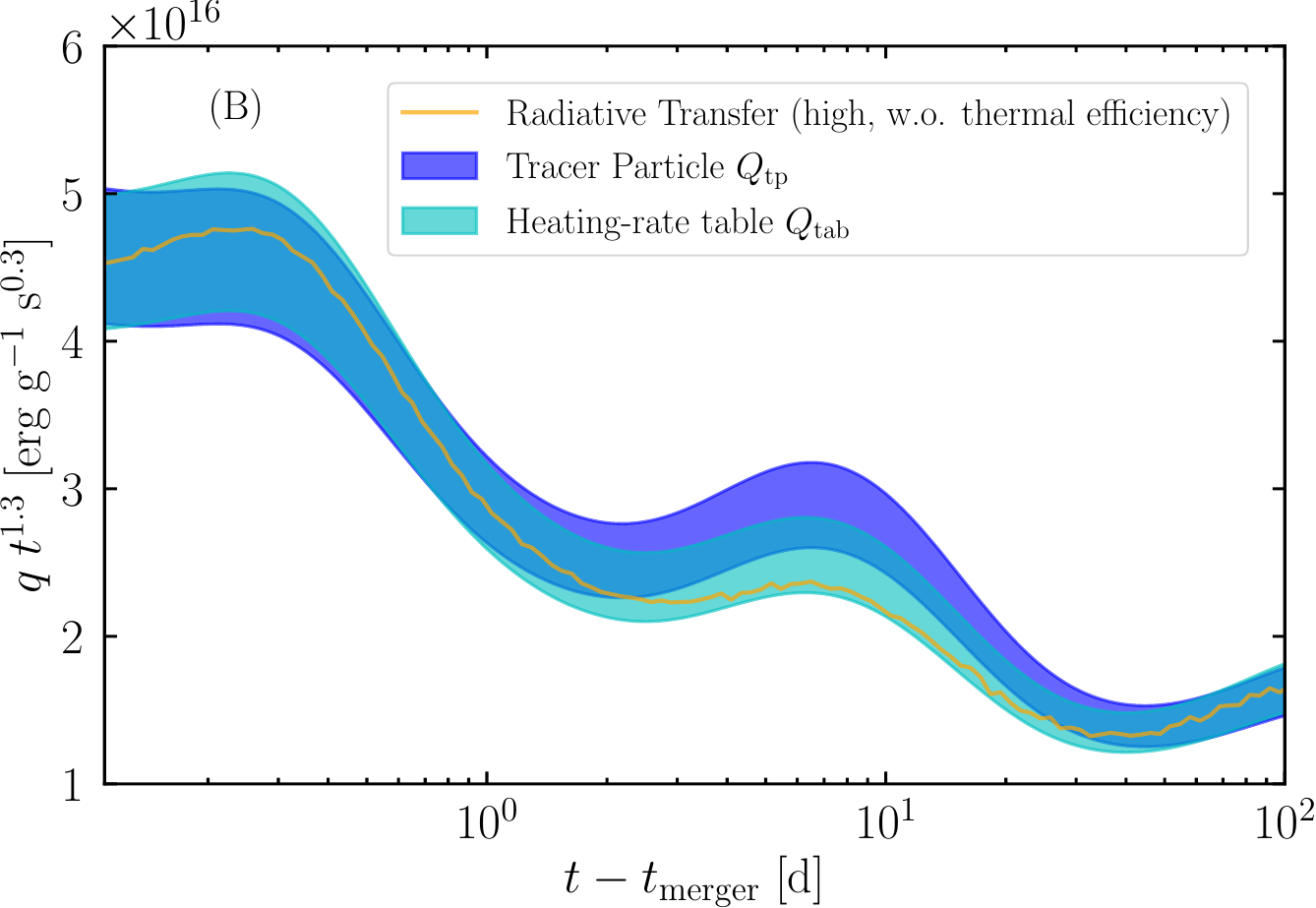} 
	\caption{\textbf{Verification of the mapping procedure for elemental abundances and heating rates.} 
(\textbf{A}) Relative differences in the mass-weighted elemental abundance patterns. The cyan curve shows the relative difference between the original tracer-particle data and the table-reconstructed abundances. 
The latter is calculated from the lookup table augmented by numerical relativity data.
The blue curve shows the relative difference between the table-reconstructed abundances 
and the abundances of the final profile used in the radiative transfer simulation. The filled circles denote the elements whose abundances are greater than $10^{-3}$, and the dashed curves are for all the elements. (\textbf{B}) Specific radioactive heating rate as a function of the post-merger time. Results directly taking the mass average from tracer particles (blue) and reconstructed from the numerical relativity simulation data (cyan) 
are plotted with $\pm 10\%$ uncertainty (shaded regions). The orange curve represents the mass-averaged specific heating rate used in the radiative transfer simulation. Note that the heating rate going into neutrinos is not considered here, and, for a fair comparison with the other heating rates, the thermalization efficiency from~\cite{Barnes:2016umi} is not applied to the orange curve.
}
	\label{fig:SM_mapcheck} 
\end{figure}

As described in Materials and Method, the elemental abundances and radioactive heating-rate histories are interpolated from the tracer particles to construct the table used in the long-term HD simulation. The tabulated quantities are further interpolated in the HD simulation when they are assigned to fluid elements. These interpolation procedures can introduce mapping errors in the elemental abundances and heating rates. In addition, the subsequent HD evolution can modify the mass-weighted abundance pattern and heating-rate history that are eventually used for the radiative transfer simulation, for example, because a fraction of the matter may fall back as a result of hydrodynamic interactions. We assess the magnitude of these effects and their impact on the quantities used for the kilonova radiative transfer simulation.

Figure~\ref{fig:SM_mapcheck} shows a comparison of the elemental abundance pattern and radioactive heating rate obtained with different stages of the procedure. In the left panel, we compare the mass-weighted elemental abundances evaluated from the original tracer-particle data with those reconstructed using the elemental abundance table and the mass flux extracted from the numerical relativity simulation data at $r=r_\mathrm{ext}$, which is shown by the cyan curves. The latter quantity is calculated by
\begin{align}
X(Z) &= \frac{1}{M_\mathrm{input}}\int \oint X(Z,t_{\rm ext},\theta_{\rm ext},\phi_{\rm ext}) \rho u^r \sqrt{-g} {r_\mathrm{ext}}^2 d\Omega_{\rm ext} dt_{\rm ext}, \\
M_\mathrm{input} &= \int \oint \rho u^r \sqrt{-g}  {r_\mathrm{ext}}^2 d\Omega_{\rm ext} dt_{\rm ext},
\end{align}
where ${r_\mathrm{ext}}^2 d\Omega_{\rm ext}$ is the surface area element on the extraction sphere, $X(Z,t_{\rm ext},\theta_{\rm ext},\phi_{\rm ext})$ is the elemental mass fraction of the matter extracted at time $t_{\rm ext}$ and angles $(\theta_{\rm ext},\phi_{\rm ext})$ on the extraction sphere, which is interpolated from the table, and $\rho u^r \sqrt{-g}$ is the conserved radial mass flux extracted from the numerical relativity simulation data.

The relative difference is typically within $\approx$10\% for most relevant elements. Larger deviations are found for some elements, in particular at low atomic numbers and in a few narrow ranges of $Z$. These deviations mainly reflect the sparse sampling of the nucleosynthesis calculation and the interpolation from discrete tracer-particle results to the grid of extraction time and angular coordinates. However, the overall abundance pattern is well reproduced, especially for elements with high abundances ($X(Z)>10^{-3}$), and the dominant abundance features relevant for the opacity calculation are not significantly altered. 

We also compare the elemental abundances reconstructed from the table and extracted mass flux with the mass-averaged elemental abundance of the ejecta profile used in the radiative transfer simulation, which is displayed as the blue curves. This comparison quantifies the change in the long-term HD evolution after the mapping procedure. The relative difference is within $\approx$10\% for the dominant elements with $X(Z)>10^{-3}$. This indicates that the abundance pattern relevant for the radiative transfer simulation is affected by these mapping procedures by at most $\sim 10\%$. 

The right panel of fig.~\ref{fig:SM_mapcheck} compares three specific radioactive heating rates as a function of time: $Q_{\rm tp}(\tau)$ constructed by taking the mass average of the contributions of each tracer particle (blue), $Q_{\rm tab}(\tau)$ constructed from the heating-rate table and mass flux extracted from the numerical relativity simulation data (cyan), and $Q(\tau)$, which is the mass average of the specific heating rate used in the radiative transfer simulation (orange). $Q_{\rm tab}(\tau)$ is calculated as
\begin{align}
Q_{\rm tab}(\tau) &= \frac{1}{M_{\rm input}}\int \oint q(\tau;t_{\rm ext},\theta_{\rm ext},\phi_{\rm ext}) \rho u^r \sqrt{-g} {r_\mathrm{ext}}^2 d\Omega_{\rm ext} dt_{\rm ext}.
\end{align}
For the comparison, $Q_{\rm tp}$ and $Q_{\rm tab}$ are shown with$\pm 10\%$ uncertainty (shaded region) in the right panel of fig.~\ref{fig:SM_mapcheck}.
The heating rates used in the radiative transfer simulation agree well with the heating rate reconstructed from the table as well as those directly evaluated from the tracer-particle data over the timescale relevant for the kilonova emission. Indeed, we see that the differences between the heating rates are within $\sim10\%$.

We therefore conclude that the mapping from the selected nucleosynthesis tracer particles to the input table for the long-term HD simulation, as well as the mapping from the input table to the ejecta data for the radiative transfer simulation, does not introduce significant systematic errors in the global elemental abundance pattern or the radioactive heating-rate history. While these mapping errors may affect the detailed abundances of individual elements, especially those with small mass fractions or localized ejecta distributions, they are unlikely to alter the main kilonova light-curve results presented in this work.

\subsubsection*{Ejecta mass evolution in the HD simulation}

\begin{figure} 
	\centering
	\includegraphics[width=0.65\textwidth]{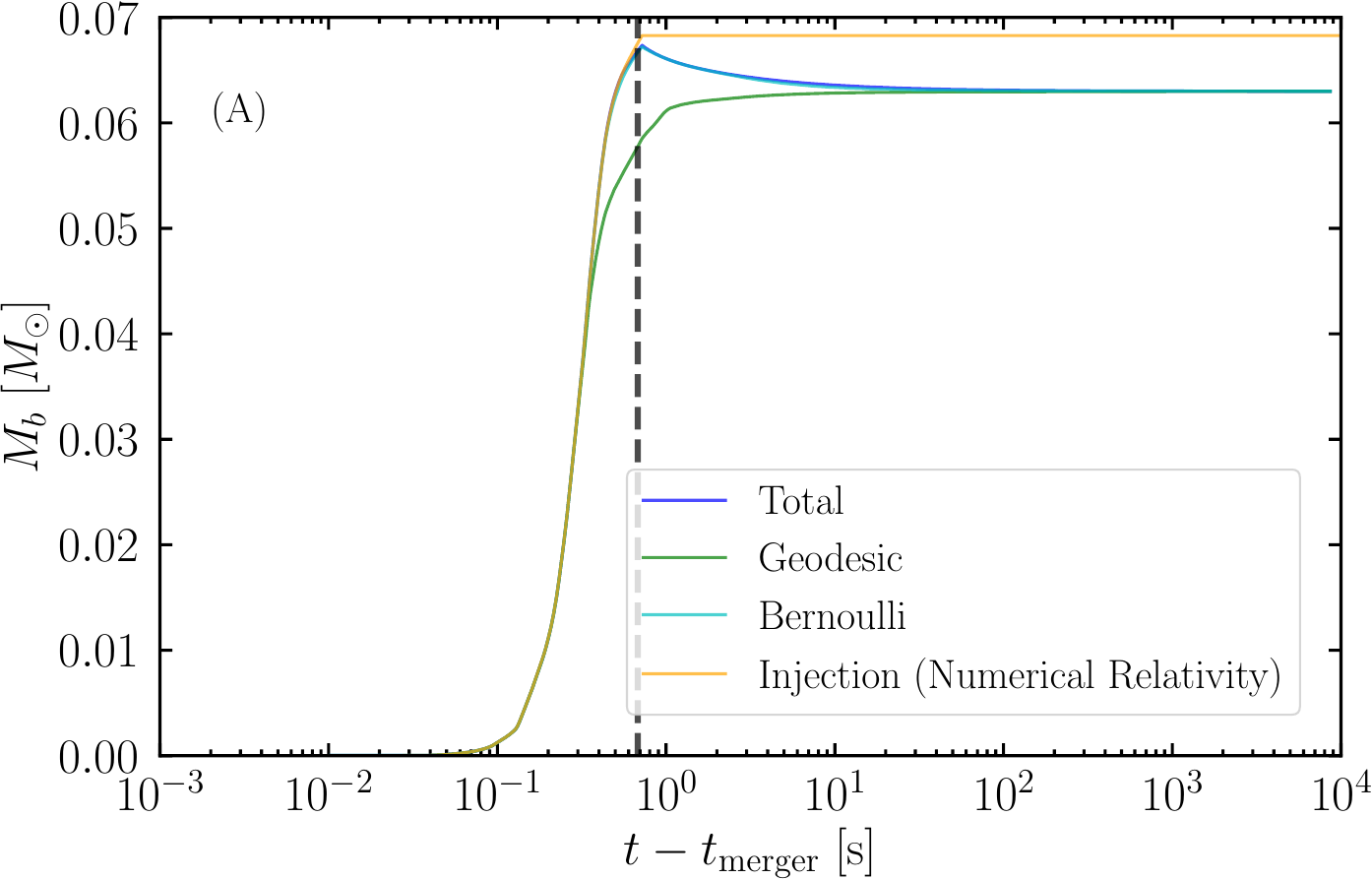}
	\caption{\textbf{Ejecta mass evolution in the HD simulation.} (\textbf{A}) The total baryonic mass contained within the computational domain is plotted as a function of time. The blue curve denotes the total baryonic mass in the HD simulation. The green and cyan curves represent the ejecta mass, with 
    the geodesic and Bernoulli criteria, respectively. The orange curve indicates the total injected baryonic mass calculated from the numerical relativity simulation. The black dashed vertical line marks the epoch at which the material injection from the inner boundary is truncated, i.e., the final moment of the numerical relativity simulation. 
    } 
	\label{fig:SM_hdmass} 
\end{figure}

Figure~\ref{fig:SM_hdmass} shows the time evolution of the rest mass in the computational domain for the HD simulation. The total mass within the domain up to the end of the numerical relativity simulation approximately agrees with the mass estimated by time-integrating the mass flux at the extraction radius, which is used as the inner boundary condition. Initially, the ejecta mass satisfying the geodesic condition is smaller than the total mass but gradually increases over time, whereas the mass satisfying the Bernoulli condition remains approximately equal to the total mass in the domain. This indicates that a fraction of the matter, 
whose internal energy has yet to be converted to kinetic energy, becomes unbound as it is accelerated by the pressure gradient. As time evolves, the mass of the matter satisfying the geodesic condition and that satisfying the Bernoulli condition both converge to the mass remaining in the computational domain. The total ejecta mass and total ejecta kinetic energy at $t-t_{\rm merger}=0.1\,{\rm d}$ are $0.063\,M_\odot$ and $1.3\times 10^{51}\,{\rm erg}$, respectively. Matter falls back after the vacuum inner boundary condition is applied in the HD simulation due to the termination of the numerical relativity data. The total amount of this fallback matter is $\approx 5\times 10^{-3}\,M_\odot$. In addition to the matter that artificially falls back, the ejecta components that have not yet reached the extraction radius and are therefore not accounted for in the kilonova prediction have a similar amount of mass. Nevertheless, since such a missing component has a mass only up to $\approx 16\%$ of the total ejecta mass and has a long diffusion time scale due to its low velocity, we expect that it introduces only a minor error to the light curve prediction compared to the other sources of errors and uncertainties, as also discussed in~\cite{Kawaguchi:2020vbf,Kawaguchi:2024hdk}.

\subsubsection*{Ejecta morphology in the HD simulation}
\begin{figure} 
	\centering
	\includegraphics[width=0.48\textwidth]{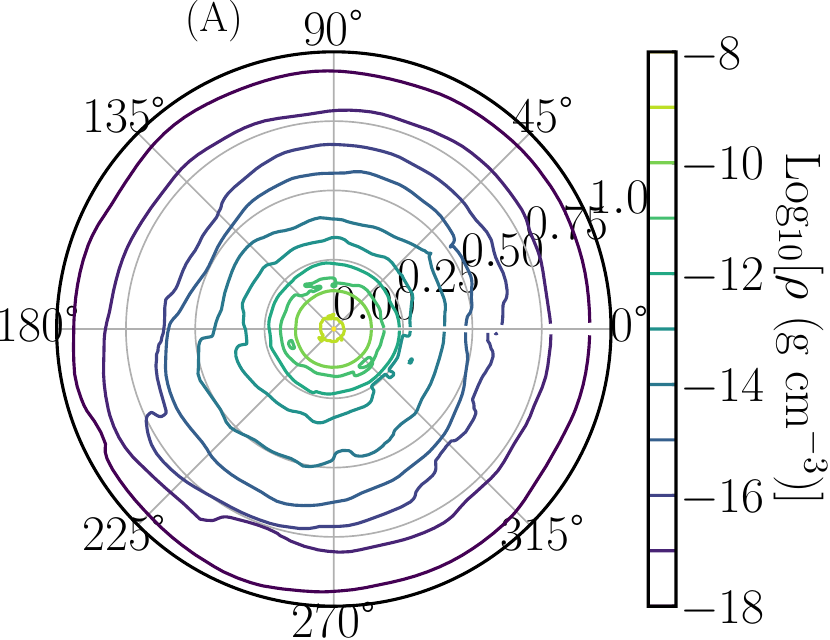}
	\includegraphics[width=0.48\textwidth]{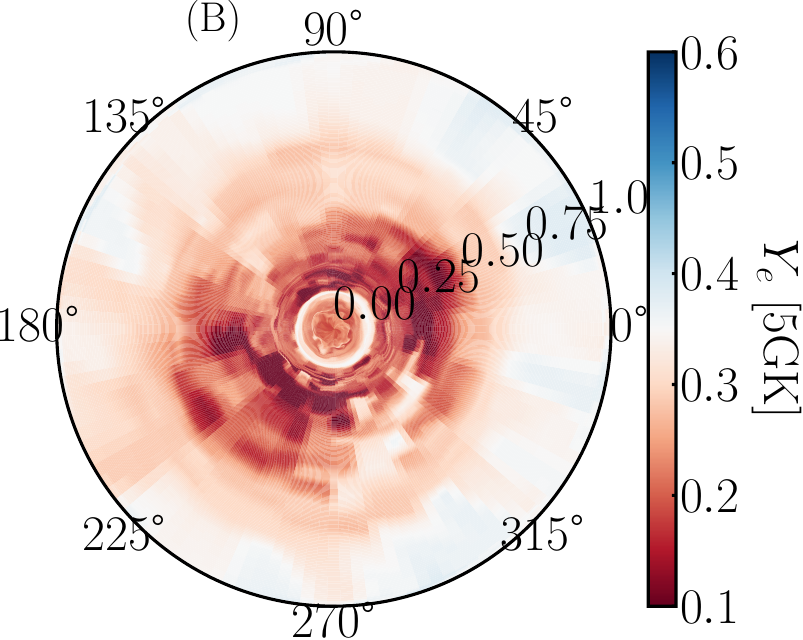}\\
    \includegraphics[width=0.48\textwidth]{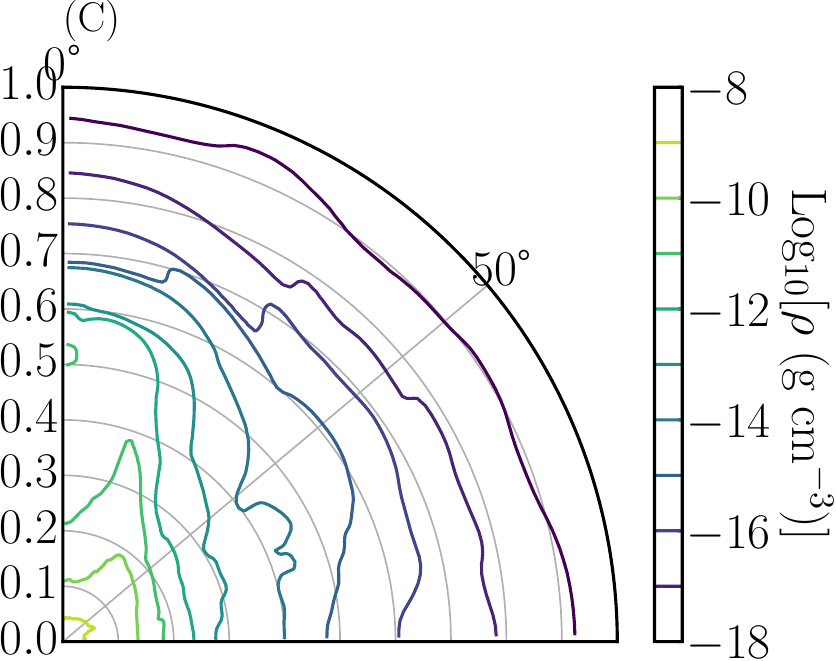}
	\includegraphics[width=0.48\textwidth]{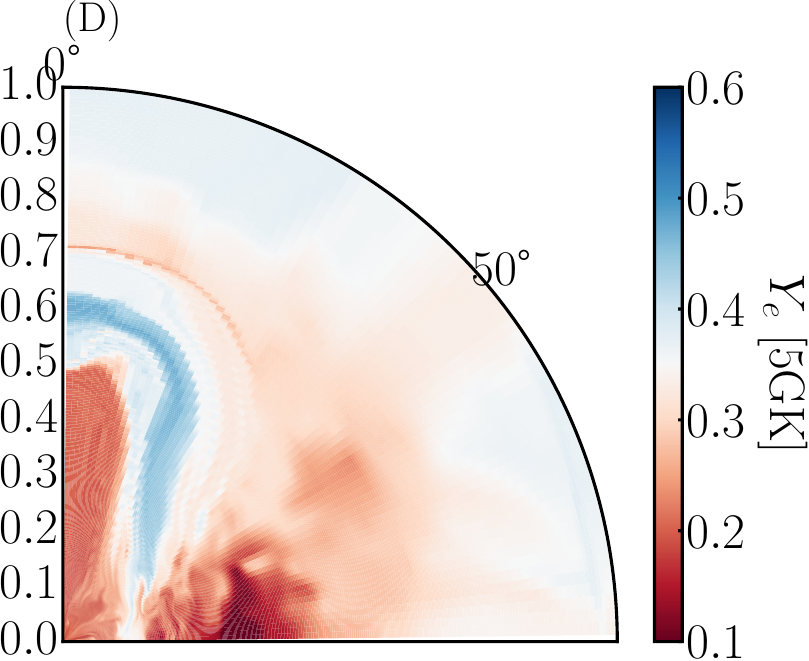} 
	\caption{\textbf{Ejecta rest-mass density and electron fraction profiles.} (\textbf{A}) Rest-mass density and (\textbf{B}) electron fraction ($Y_e$) profiles on the orbital plane at $t-t_{\rm merger}=0.1\,{\rm d}$. (\textbf{C}) and (\textbf{D}) are the same, but on a meridional plane along $\phi=0^\circ$. The value at the temperature of $5\times 10^9\,{\rm K}$ is shown for the electron fraction profile. 
    The radial coordinate is in units of $ct$.}
	\label{fig:SM_ejeprof} 
\end{figure}

Figure~\ref{fig:SM_ejeprof} displays the spatial profiles of the rest-mass density and electron fraction ($Y_{\rm e}$) of the ejecta at $0.1\,{\rm d}$ obtained by the ejecta HD simulation. The rest-mass density profile exhibits an approximately axisymmetric morphology, with the post-merger component developing a distinct prolate shape. The $Y_{\rm e}$ profile also reveals an approximately axisymmetric morphology, except for the fast-moving dynamical ejecta ($v > 0.1\,c$), which reveals moderate non-axisymmetric features.
Within the prolate post-merger ejecta, the typical electron fraction is found to be $Y_{\rm e} \sim 0.2$--$0.3$. Notably, this $Y_{\rm e}$ value tends to be systematically smaller than those reported in previous numerical models, particularly employing the viscous description, e.g.,~\cite{Fujibayashi:2020qda,Just:2023wtj}, suggesting a more lanthanide-rich environment for the subsequent {\it r}-process nucleosynthesis in our model. 


\subsubsection*{Viewing angle dependence of kilonova light curves}

\begin{figure} 
	\centering
	\includegraphics[width=0.48\textwidth]{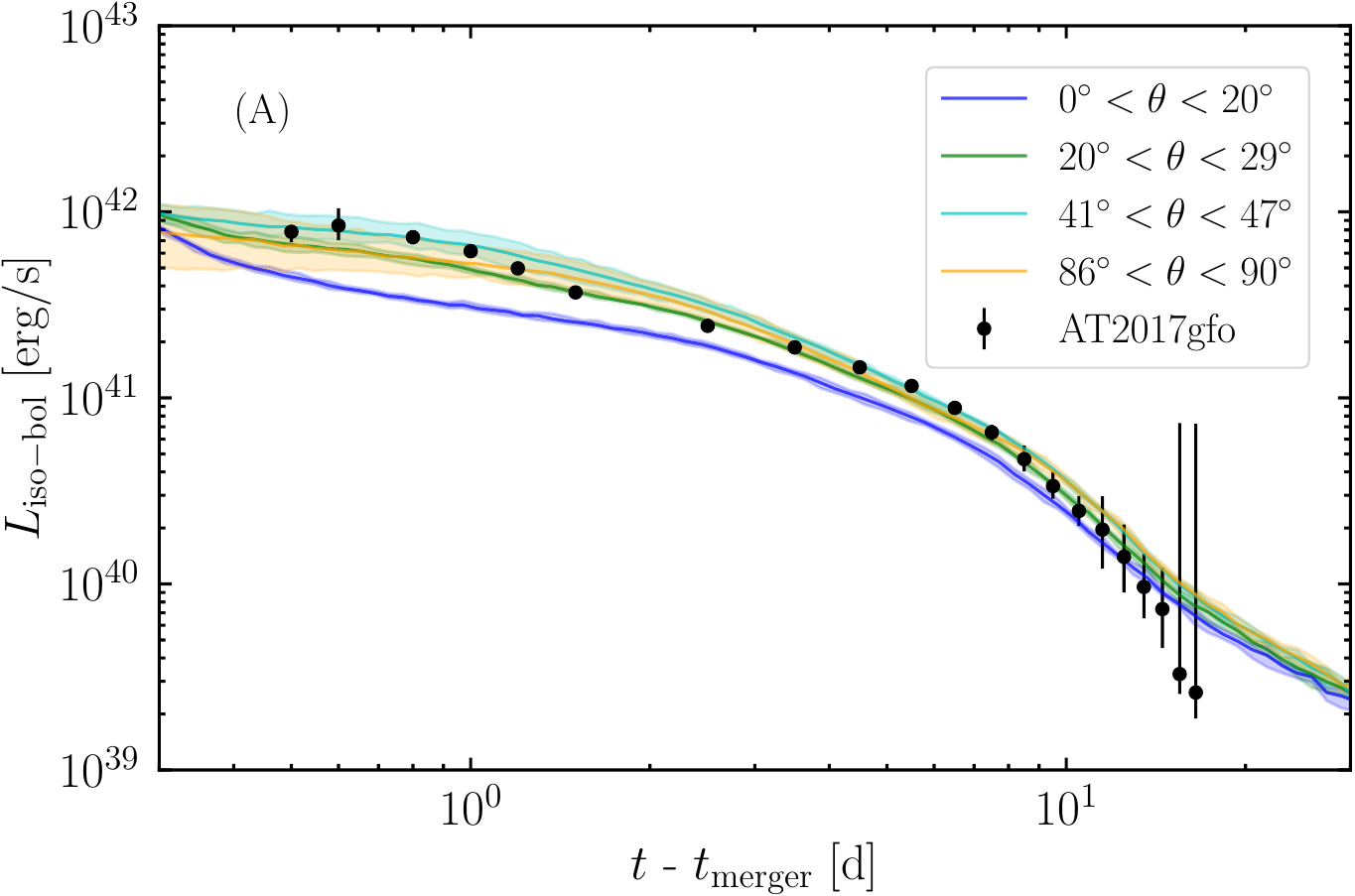}
	\includegraphics[width=0.48\textwidth]{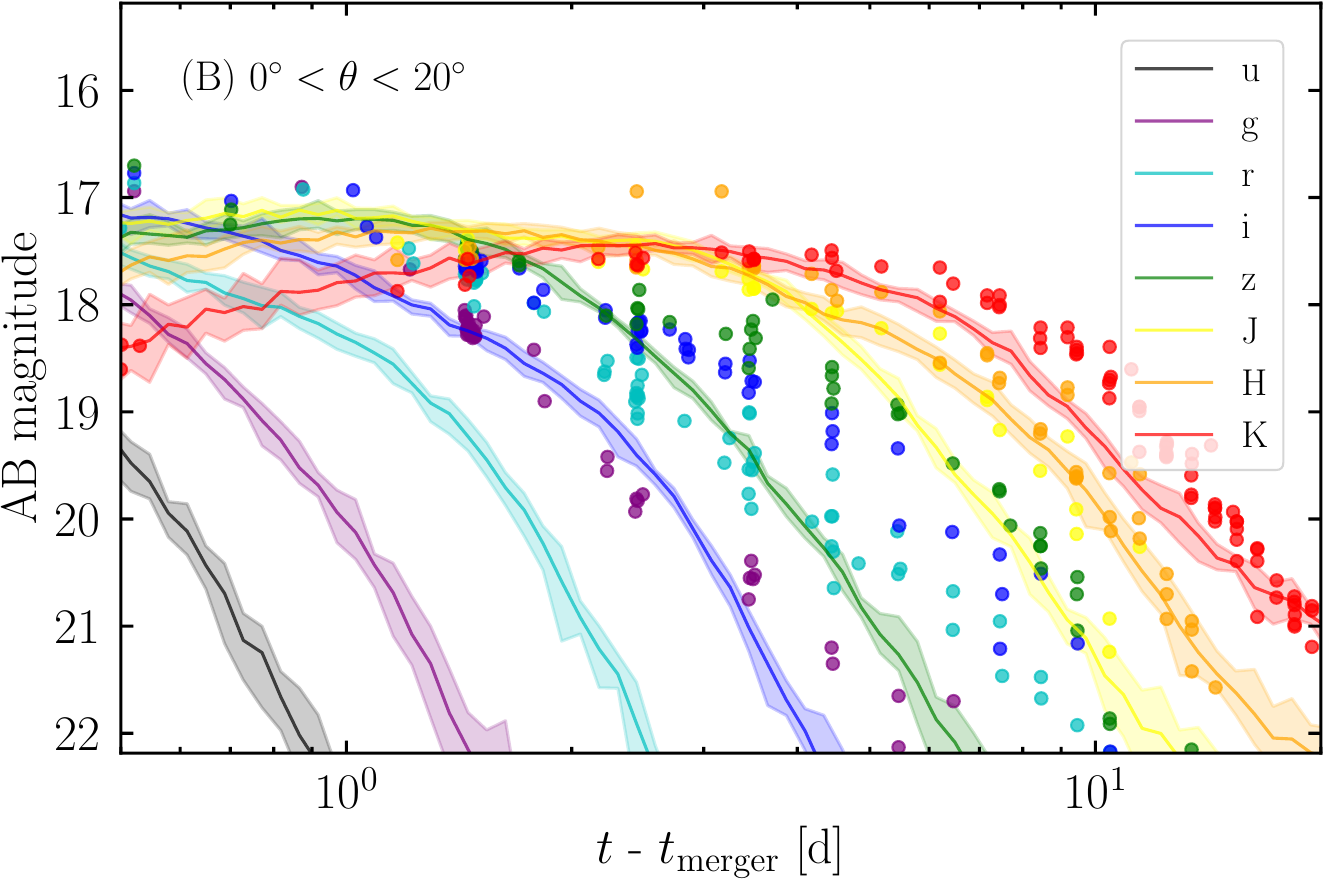}
	\includegraphics[width=0.48\textwidth]{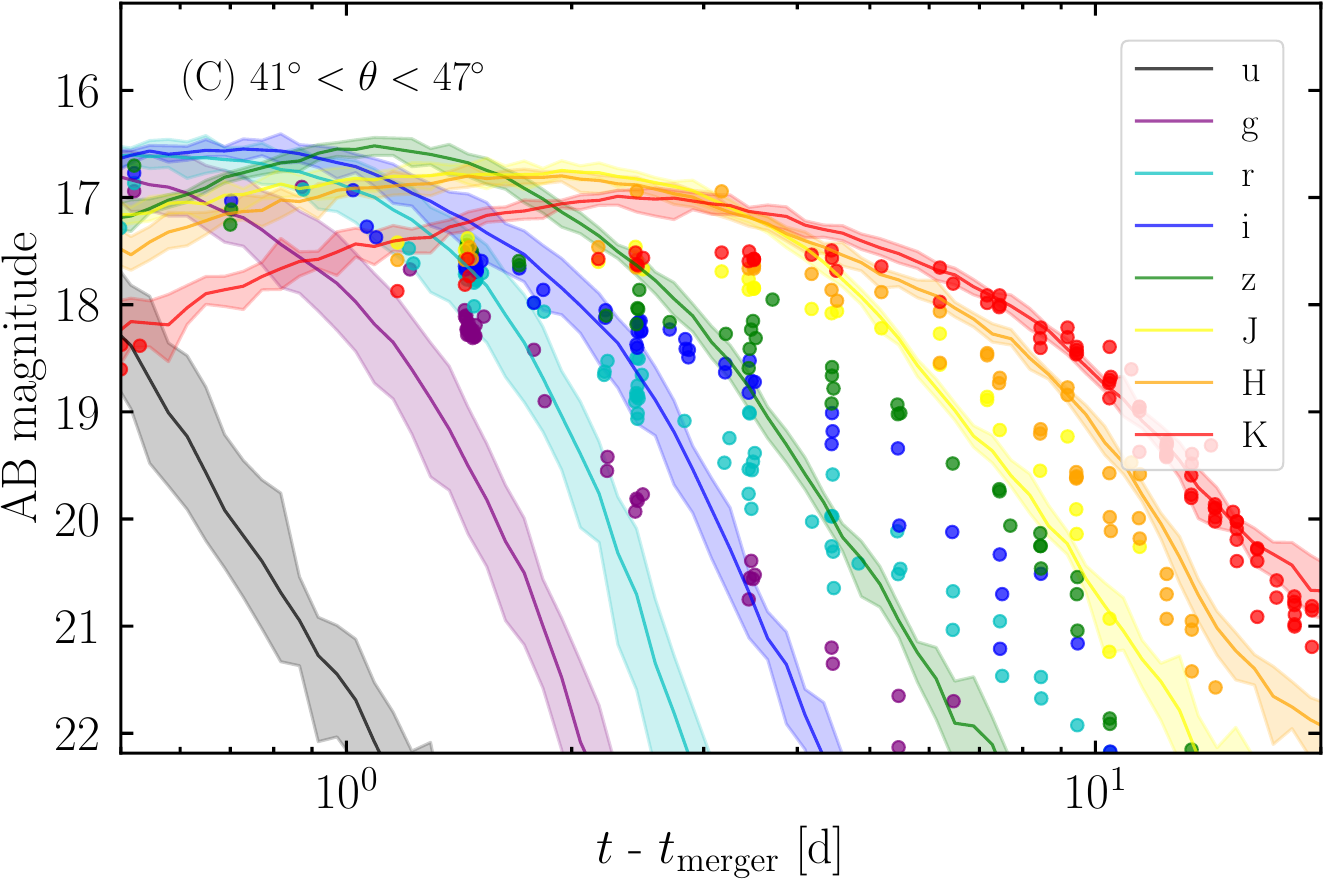}
	\includegraphics[width=0.48\textwidth]{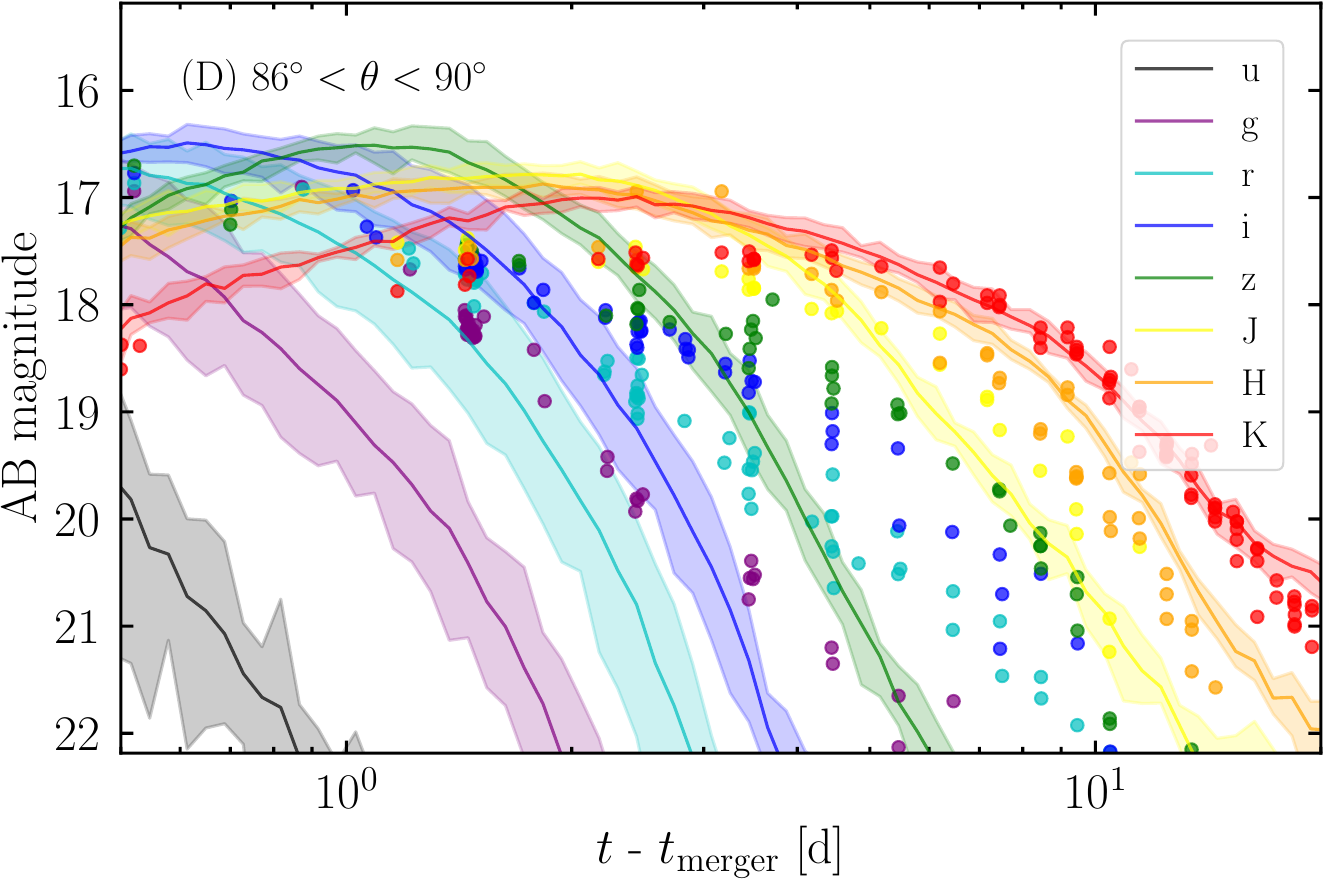}
	\caption{\textbf{Viewing angle dependence of kilonova light curves.} (\textbf{A}) The isotropic bolometric light curves and $ugrizJHK$-band light curves for various latitudinal viewing angles (\textbf{B}) for $0^\circ \le\theta \le 20^\circ$, (\textbf{C}) for $41^\circ \le\theta \le 47^\circ$, and (\textbf{D}) for $86^\circ \le\theta \le 90^\circ$. The shaded regions denote the variation ranges with respect to the longitudinal viewing angles. The $ugrizJHK$-band light curves are computed assuming a hypothetical distance of $40\,{\rm Mpc}$. The bolometric luminosity observed in AT2017gfo is shown by using the data in~\cite{Waxman:2017sqv}. The ugrizJHK data points denote the observation data of AT2017gfo taken from~\cite{Villar:2017wcc}.}
	\label{fig:SM_lcang} 
\end{figure}

Figure~\ref{fig:SM_lcang} shows the viewing angle dependence of the predicted kilonova light curves with the fiducial setup. Focusing on the latitudinal angle dependence, the bolometric luminosity is smallest when observed along the polar axis and largest when viewed from $\approx45^\circ$. This is because the projected area of the prolate post-merger ejecta depends on the viewing angle, whereas the emission viewed from the orbital plane is suppressed due to the presence of the opaque dynamical ejecta (see fig.~\ref{fig:SM_ejeprof}). The same relationship between viewing angle and brightness is found in broadband magnitude light curves. Note that the light curves for a viewing angle of $20^\circ \le \theta \le 29^\circ$, which is consistent with the constraints from GW170817 and its associated jet radio afterglow~\cite{LIGOScientific:2017vwq,Mooley:2018qfh}, agree approximately with the spherically averaged light curves.

The variation in the light curves for different longitudinal angles is not strong, reflecting the fact that the ejecta morphology exhibits an approximately axisymmetric shape. This variation is more pronounced for an orbital plane viewer and less significant for a polar viewer. Furthermore, the variation is more pronounced in the early optical emission and less significant in the later near-infrared emission. This indicates that the longitudinal angle dependence of the light curves is primarily caused by that of the lanthanide-rich dynamical ejecta, which are primarily distributed around the orbital plane (see fig.~\ref{fig:SM_ejeprof}); their radiative transfer effects become less important in the later phase due to the decrease in density and optical depth.
\subsubsection*{Uncertainties and systematic errors in kilonova light curve modeling} \label{sec:SM_uncertain}

\begin{figure} 
	\centering	
	\includegraphics[width=0.48\textwidth]{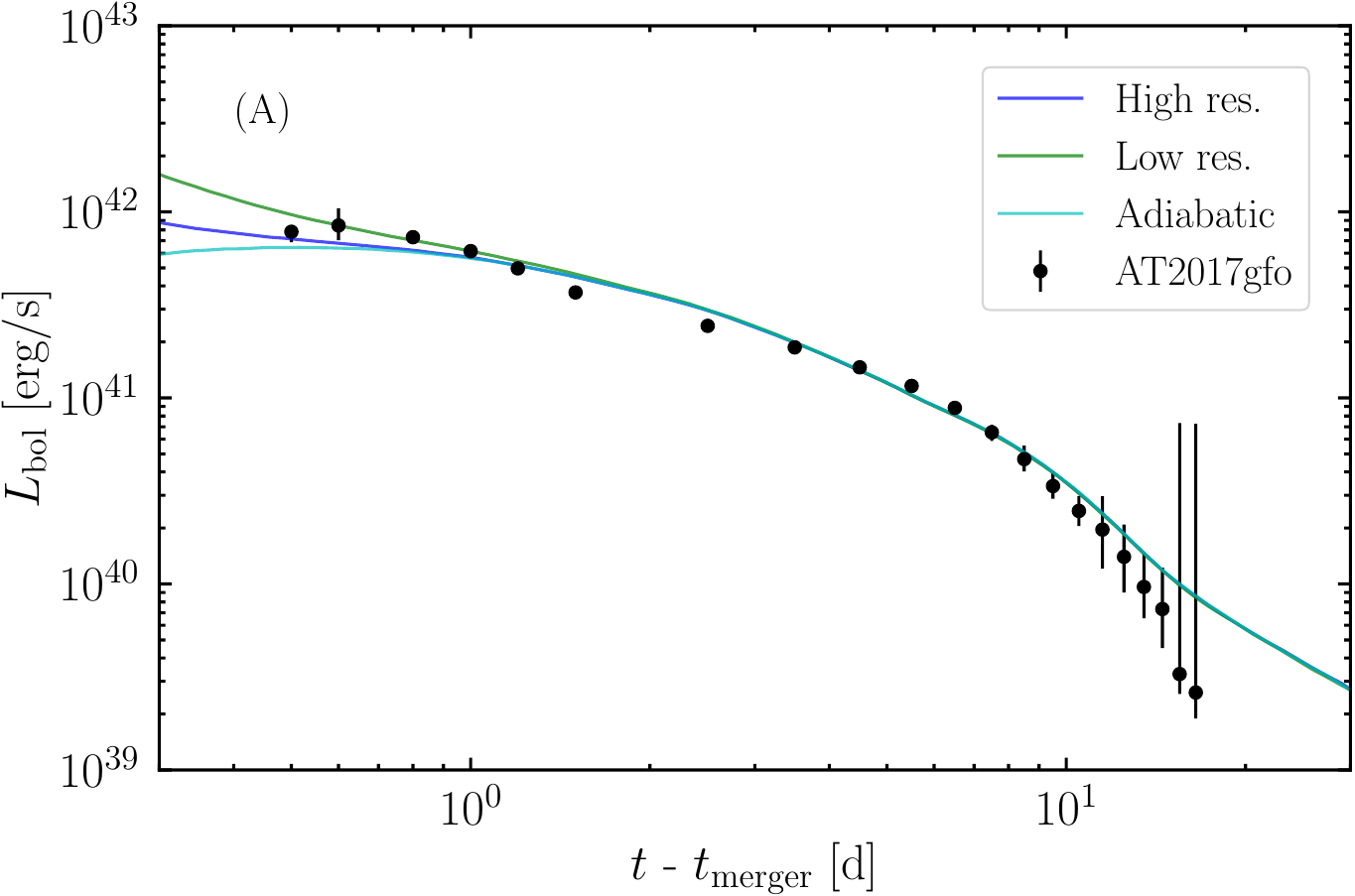} 
    \includegraphics[width=0.48\textwidth]{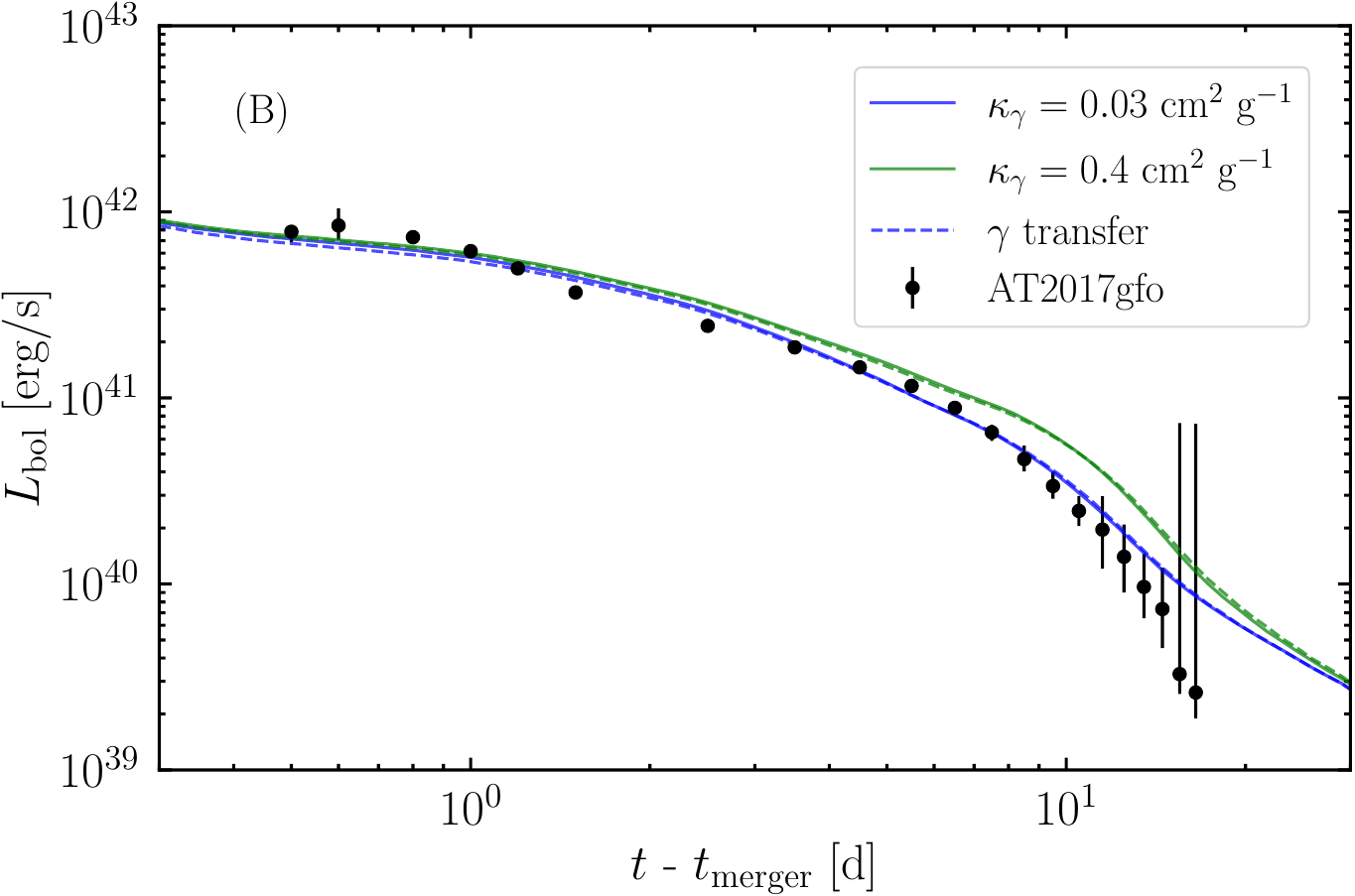}
	\caption{\textbf{Systematic uncertainties in the predicted bolometric light curves.} (\textbf{A}) An impact of the grid resolution in HD simulation on the bolometric luminosity. The blue, green, and cyan solid curves represent the models with high resolution HD profile, low resolution HD profile, and high resolution HD profile but with the initial radiation profile estimated by assuming adiabatic evolution, respectively. (\textbf{B}) Dependence on the gray gamma-ray effective opacity ($\kappa_\gamma$) and thermalization treatment. Solid curves use the analytic formula from~\cite{Barnes:2016umi} with $0.03\ {\rm cm^2\ g^{-1}}$ (blue) and $\kappa_\gamma = 0.4\ {\rm cm^2\ g^{-1}}$ (green). The dashed curves denote the results explicitly solving the gamma-ray transport ($\gamma$ transfer). 
    The bolometric luminosity observed in AT2017gfo is shown by using the data in~\cite{Waxman:2017sqv}.}
	\label{fig:SM_lcsys} 
\end{figure}

Figure~\ref{fig:SM_lcsys} (\textbf{A}) illustrates the impact of the grid resolution in the HD simulation on the predicted kilonova light curves. The comparison reveals that the model with a lower HD grid resolution yields a higher kilonova brightness during the very early phase ($t-t_{\rm merger} \lesssim 0.5\ {\rm d}$). On the other hand, the light curves after $\approx 1\ {\rm d}$ are approximately identical among the models, indicating that the predicted emission in the later epochs is robust against the choice of grid resolution. This early-phase discrepancy stems from the unphysical numerical dissipation of kinetic energy into internal energy within the HD simulation due to the truncation error. 
Because the internal energy profile at the end of the HD simulation serves as the initial radiation profile for the subsequent radiative transfer simulation, this enhanced numerical heating artificially boosts the early-time flux. 

Consequently, we must bear in mind that the kilonova brightness in the first few hours can be systematically overestimated due to this resolution effect. To gauge the extent of this overestimation, we perform an additional radiative transfer simulation where the initial radiation profile is instead estimated by assuming adiabatic cooling balanced by radioactive heating. Indeed, the bolometric luminosity calculated from the high-resolution HD profile agrees with this adiabatic-evolution model within $\approx 10\%$ for $t-t_{\rm merger} > 0.5\ {\rm d}$. Therefore, we conclude that the predicted kilonova light curves after $0.5\ {\rm d}$ are robust and free from significant systematic errors induced by the finite grid resolution of the HD simulation.

The thermalization efficiency of radioactive gamma-rays is a critical factor in kilonova modeling, yet its determination relies heavily on the gamma-ray effective opacity, which carries significant uncertainties. This opacity parameterizes complex, energy-dependent interactions into a simplified gray value, making it highly sensitive to the radioactive decay gamma ray spectra and elemental abundances. Reflecting these physical complexities, previous literature has proposed widely disparate values: an analytic value of $0.4\ {\rm cm^2\ g^{-1}}$~\cite{Barnes:2016umi}, a composition-dependent range of $0.07$--$0.4\ {\rm cm^2\ g^{-1}}$ ~\cite{Hotokezaka:2019uwo}, and more recent lower estimates of $0.03$--$0.05\ {\rm cm^2\ g^{-1}}$~\cite{Guttman:2024bxl}.

Figure~\ref{fig:SM_lcsys} (\textbf{B}) illustrates the impact of the uncertainty in the gamma-ray effective gray opacity on the predicted bolometric luminosity, comparing our fiducial value of $0.03\ {\rm cm^2\ g^{-1}}$ (e.g.,~\cite{Guttman:2024bxl}) with a higher value of $0.4\ {\rm cm^2\ g^{-1}}$ (e.g.,~\cite{Barnes:2016umi}). While the two models show similar evolution in the very early phase, the discrepancy becomes increasingly noticeable from $\approx 3$ days post-merger and reaches up to a factor of two at around 10 days. This baseline discrepancy stems from the large variation in the proposed values of the gamma-ray effective opacity in the literature. Our results emphasize that such a systematic uncertainty in the gamma-ray thermalization efficiency must be carefully taken into account when interpreting kilonova light curves.

To evaluate the validity of the gamma-ray thermalization prescription itself, we adopt the analytic formula from~\cite{Barnes:2016umi} to determine the thermalization efficiency from the effective opacity. We also perform a radiative transfer simulation that explicitly solves the gamma-ray transport using the same gray gamma-ray effective opacity. As shown in fig.~\ref{fig:SM_lcsys}, the resulting bolometric light curves are nearly identical as long as the same effective opacity is employed. This agreement indicates that the choice of the analytic formula itself does not introduce significant systematic errors, confirming that the primary source of uncertainty lies in the choice of the effective opacity value rather than the formalization of the thermalization process.

\begin{figure} 
	\centering	
	\includegraphics[width=0.48\textwidth]{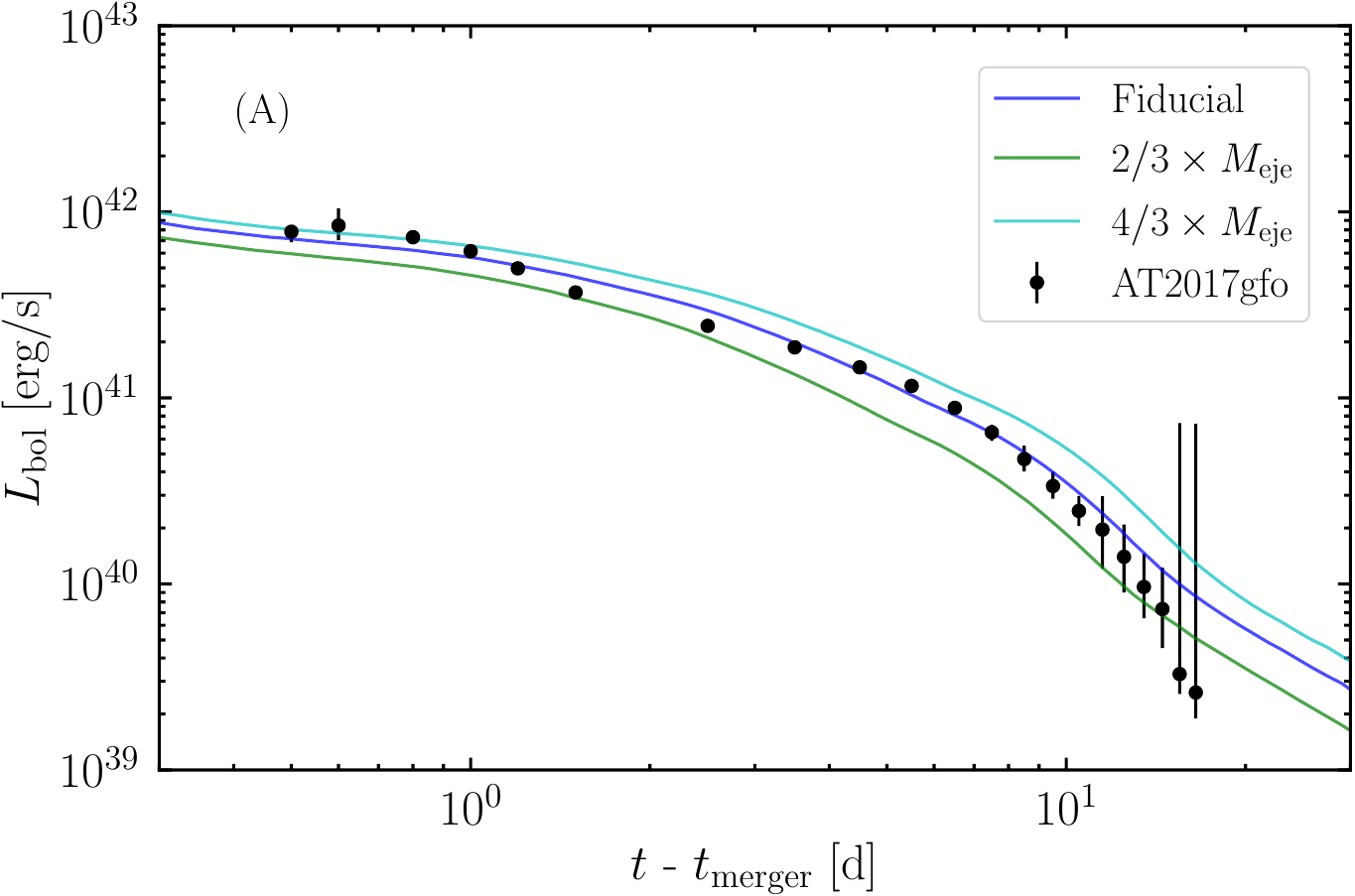}
	\includegraphics[width=0.48\textwidth]{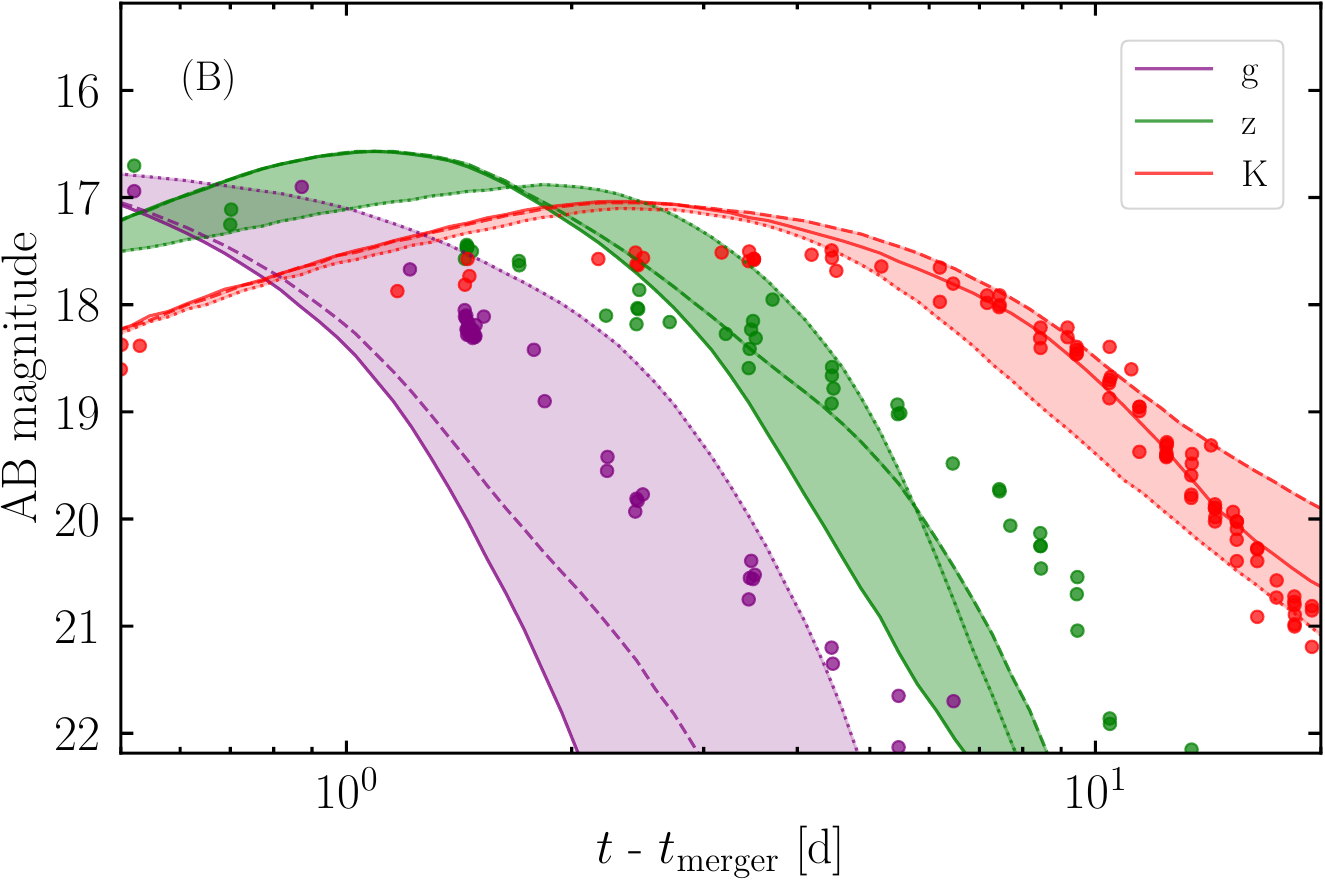} \\
	\caption{\textbf{Light curve possible uncertainty.}
    (\textbf{A}) Total bolometric light curve of the fiducial setup and those obtained by radiative transfer simulations using the same spatial profile as the fiducial model, but with the density scaled uniformly by a factor of $2/3$ and $4/3$. (\textbf{B}) Comparison of the {\it gzK}-band light curves for models in which the contributions from the neutral (dashed curves) or both neutral and first ionized atoms (dotted curves) are switched off. The light curves shown in the solid curves are the results with the fiducial setup and are the same as in Fig.~\ref{fig:main_knlc}.}
	\label{fig:SM_uncert} 
\end{figure}

In addition to gamma-rays, energy is deposited in the merger ejecta by massive decay products such as $\beta$-decay electrons/positrons, alpha-particles, and fission fragments, which are generally expected to continue thermalizing efficiently on timescales longer than those of gamma-rays. For consistency, we investigate whether uncertainties in the thermalization of massive decay products have a significant impact on our model light-curves.

The thermalization efficiency of $\beta$-decay electrons has been studied in great detail \cite{Barnes:2016umi,Hotokezaka:16,Kasen.Barnes:2019,Waxman:2019,Hotokezaka.Nakar:2020,Shenhar:24,Andalman:2026}. These works find broadly similar values for the typical thermalization efficiency timescale of $\beta$-decay electrons of $\sim$ 10 days. Applying the various thermalization formulae to bulk ejecta properties, we find a maximum variation of $\lesssim 20$\% for the timescales of $t-t_{\rm merger} \sim 1$--$30$ days relevant to the light-curves presented in this work, much smaller than the expected uncertainty in the intrinsic energy generation from radioactive decay \cite{Barnes:21,Zhu:21}. Thermalization of alpha-particles and fission fragments has been less studied, with greater variation found between existing works \cite{Barnes:2016umi,Hotokezaka.Nakar:2020,vdBerg:2026}. However, for the model studied here, we find a negligible amount of energy going into alpha-decay particles and fission fragments compared to gamma-rays and $\beta$-decay electrons ($\lesssim 1\%$ of the total heating rate from 0.1 day to 30 days), such that the thermalization efficiency has no impact on the total ejecta heating rate. Therefore, we find that uncertainties in the thermalization efficiencies of massive decay products are negligible in the present work.

The ejecta mass is highly sensitive to the binary configuration. In the case of an asymmetric binary, the total mass slightly increases given the chirp mass. Therefore, we expect the slightly earlier black hole formation, and consequently slightly less massive post-merger ejecta in the asymmetric binary. To evaluate the impact of this mass variation on the resulting light curves, we perform an additional radiative transfer simulation using the same spatial profile as the fiducial model, but with the density scaled uniformly by a factor of $2/3$. While this reduction is our primary focus, we also investigate a scaled factor of $4/3$ for completeness. Figure~\ref{fig:SM_uncert} (\textbf{A}) illustrates the comparison. Due to the reduced total radioactive heating rate and the shorter photon diffusion timescale associated with a lower optical depth, the kilonova light curves in this low-mass model exhibit a lower peak luminosity and a more rapid temporal evolution compared to the fiducial case. For the case with the increased ejecta mass, the opposite behavior is observed; the kilonova light curves exhibit a higher peak luminosity and a slower temporal evolution compared to the fiducial model.

In our radiative transfer code, we assume local thermodynamic equilibrium (LTE) for the ionization and excitation populations, as well as a common temperature for the gas and radiation fields. However, this assumption can break down in low-density regions. Indeed, non-LTE calculations by~\cite{Hotokezaka:2021ofe,Pognan2022MNRAS} demonstrated that atoms can remain more highly ionized than predicted under LTE. To broadly evaluate the impact of this non-LTE effect on the resulting light curves, we adopt the prescriptive approach of~\cite{Kawaguchi:2020vbf,Kawaguchi:2022bub} and perform a test simulation by artificially prohibiting the presence of neutral and singly ionized states. As shown in fig.~\ref{fig:SM_uncert} (\textbf{B}), this suppression leads to an enhancement in the optical emission. This brightening trend is broadly consistent with the recent detailed non-LTE modeling by~\cite{Brethauer:2025plw}, which could significantly improve the agreement between our kilonova model and the observational data of AT2017gfo. Hence, incorporating these non-LTE effects will be a crucial ingredient for accurately modeling kilonova light curves and performing direct, reliable comparisons with observations.

Additionally, the expansion opacity method commonly employed in radiative transfer simulations is potentially formally invalid at early times at blue wavelengths due to the overlap of many strong lines (e.g. \cite{Eastman.Pinto:1993,Baron:1996,Kasen:2006ce}). This may lead to a substantial underestimation of photon emissivity \cite{Morag:2026}, which may further account for missing UV and blue flux in our light curves. 

\subsection*{Helium-Strontium profile}
\begin{figure} 
	\centering
    \includegraphics[width=0.49\textwidth]{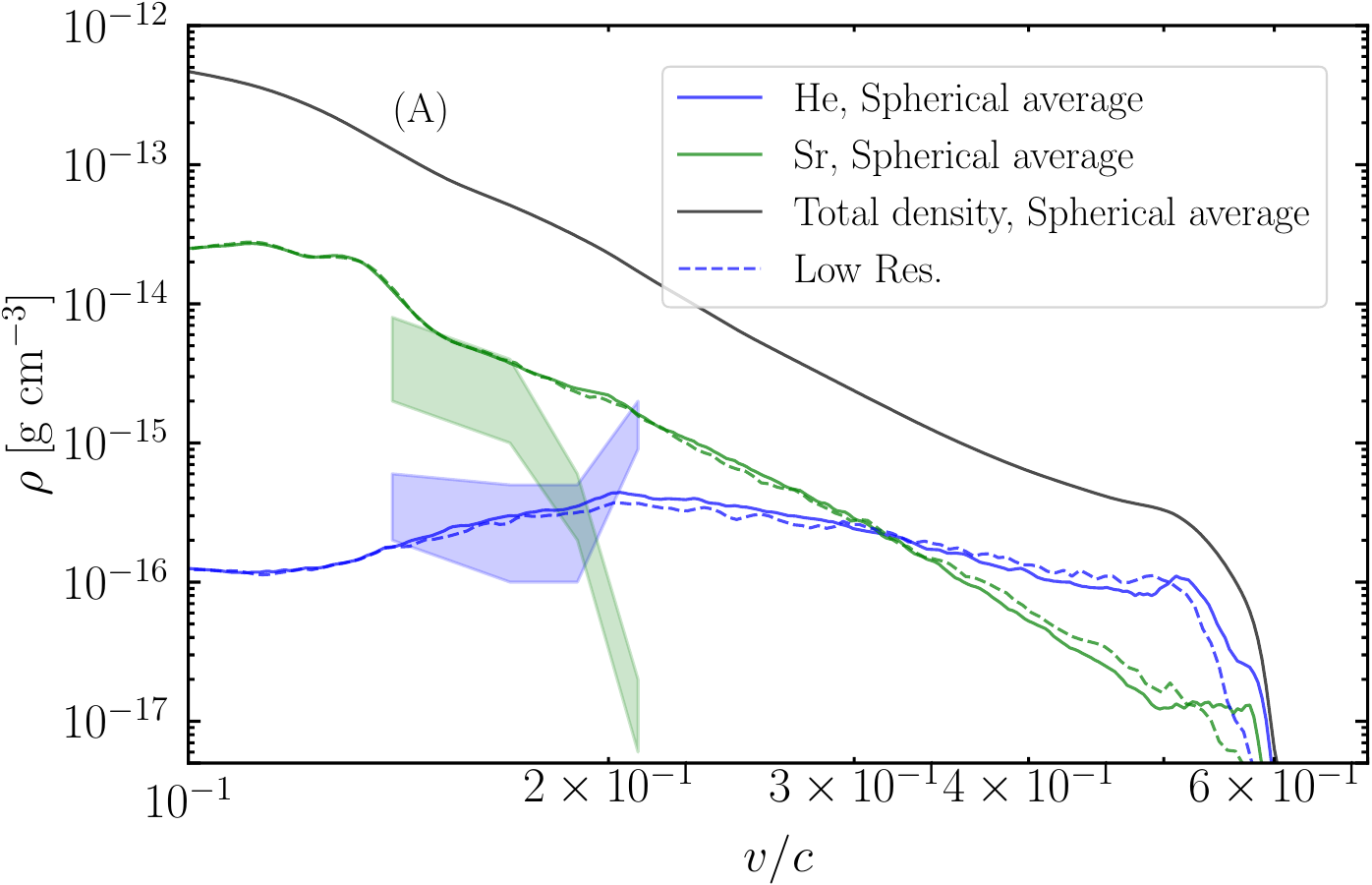}
	\caption{\textbf{Helium-Strontium profile.}
    (\textbf{A}) The density profiles of helium (He; blue) and strontium (Sr; green) as a function of radial velocity $v/c$ at 1~day since the merger. 
    The colored regions indicate the He and Sr densities, either of which reproduces the absorption feature at $\sim 8000$~{\AA} in AT2017gfo \cite{Chiba:2026flt}. The dashed curves denote the low-resolution result. 
    }
	\label{fig:SM_HeSr} 
\end{figure}

The absorption feature at $\sim 8000$~{\AA} in the photospheric spectra of AT2017gfo, which was first attributed to Sr, is suggested to be also explained by He~\cite{Perego:2020evn,Tarumi:2023apl}. Figure~\ref{fig:SM_HeSr} presents the radial density profiles of He and Sr in the ejecta at 1~d after the merger in our model. The calculated He profile is broadly consistent with the constraint (blue colored area) inferred from the prominent absorption feature observed in AT2017gfo~\cite{Sneppen:2024jch,Arya:2026jvs,Chiba:2026flt}.

On the other hand, the Sr density in the high-velocity outer region exceeds the inferred upper limit required to reproduce the observed spectral features of AT2017gfo. However, we note that this constraint is highly sensitive to the adopted total mass density profile of the ejecta. Moreover, the impact of the detailed spatial distributions of He and Sr in such high-velocity regions on the formation and evolution of the absorption features remains unclear, motivating future detailed spectral modeling. 



\clearpage 

\paragraph{Caption for Movie S1.}
\textbf{Visualization for the Kelvin-Helmholtz instability, the large-scale dynamo, and jet launching:}
The link~\url{http://www2.yukawa.kyoto-u.ac.jp/~kenta.kiuchi/anime/FUGAKU2025/out_yuv420p_3D.mp4} visualizes the volume rendering of the rest-mass density with the magnetic field lines (magenta curve) on the left, the magnetic-field strength in the center, and the electron fraction on the right, with the zoom-in/out scale. 




\end{document}